\documentclass[%
 reprint,
 amsmath,amssymb,
 aps,
]{revtex4-2}
\usepackage{natbib}
\usepackage{graphicx}% Include figure files
\usepackage{dcolumn}% Align table columns on decimal point
\usepackage{bm}% bold math
\usepackage{xcolor}
\begin{document}

\preprint{APS/123-QED}

\definecolor{ultramarine}{rgb}{0.07, 0.04, 0.56}
\newcommand{\is}[1]{\textcolor{ultramarine}{\textbf{(IS: #1)}}}
\newcommand{\GIZMO}{{\small GIZMO}}
\newcommand{\gizmourl}{\href{http://www.tapir.caltech.edu/~phopkins/Site/GIZMO.html}{\url{http://www.tapir.caltech.edu/~phopkins/Site/GIZMO.html}}}
\newcommand{\FIREurl}{\href{http://fire.northwestern.edu}{\url{http://fire.northwestern.edu}}}

\title{Galactic Center Gamma-Ray Emission in MHD Galaxy Formation Simulations with Full Cosmic Ray Spectra}% Force line breaks with \\

\author{Isabel S. Sands}
\affiliation{TAPIR, California Institute of Technology, Mailcode 350-17, Pasadena, CA 91125, USA}
\author{Philip F. Hopkins}%
\affiliation{TAPIR, California Institute of Technology, Mailcode 350-17, Pasadena, CA 91125, USA}%
\author{Sam B. Ponnada}%
\affiliation{TAPIR, California Institute of Technology, Mailcode 350-17, Pasadena, CA 91125, USA}%
\author{Lina Necib}
\affiliation{Department of Physics and Kavli Institute for Astrophysics and Space Research, MIT, Cambridge, MA 02139, USA}

\author{Dušan Kereš}
\affiliation{Department of Astronomy and Astrophysics, University of California, San Diego, La Jolla, CA 92093, USA}
\affiliation{Department of Physics, University of California, San Diego, La Jolla, CA 92093, USA}

\author{Yen-Hsing Julius Lin}
\affiliation{Department of Astronomy and Astrophysics, University of California, San Diego, La Jolla, CA 92093, USA}

\date{\today}% It is always \today, today,

         
\begin{abstract}
    The Milky Way's galactic center is a highly dynamical, crowded environment. $\gamma$-ray observations of this region,  such as the excess of GeV-scale $\gamma$-rays observed by Fermi-LAT, have been of tremendous interest to both the high-energy astrophysics and particle physics communities. However, nearly all past studies of $\gamma$-ray emission make simplifying assumptions about cosmic ray (CR) propagation that may not be valid in the galactic center. Recent numerical breakthroughs now enable fully time-dependent dynamical evolution of CRs in magnetohydrodynamic simulations with resolved, multi-phase small-scale structure in the interstellar medium (ISM), allowing self-consistent comparisons to the Milky Way cosmic ray spectrum. We model $\gamma$-ray emission from cosmic ray interactions for a set of Feedback in Realistic Environments (FIRE) simulations of Milky Way-mass galaxies run with spectrally-resolved cosmic ray spectra for multiple species at MeV-TeV energies. We find that the galactic center $\gamma$-ray spectrum can vary by order-of-magnitude amounts in normalization, and by $\sim 10\%$ in spectral slope at high energies, driven by both injection from highly variable star formation and losses from variable structure in the turbulent ISM. $\gamma$-ray emission from inverse Compton scattering and relativistic non-thermal Bremsstrahlung is particularly variable on Myr timescales. We argue that features of the observed Milky Way $\gamma$-ray spectrum may arise from such transient phenomena in $\gamma$-rays produced from CR interactions. 
\end{abstract}

\maketitle

%\tableofcontents

\section{Introduction}

Modeling $\gamma$-ray emission from the Milky Way's galactic center (GC) is a complex, multi-dimensional problem, with many parameters associated with the diverse sources and processes that produce $\gamma$-ray photons \cite{Strong_2000}. The bulk of $\gamma$-ray emission in this region is produced by interactions between cosmic rays (CRs) and gas in the interstellar medium (ISM), predominantly CR nucleon- ISM nucleon collisions, and relativistic bremsstrahlung and inverse Compton scattering (ICS) of CR electrons and positrons \cite{BlumenthalRevModPhys.42.237, 1994A&A...286..983M, Ackermann_2017}. However, models for CR propagation and transport are not well constrained, with order-of-magnitude uncertainties in values like CR diffusivity \cite{Genolini_2015}. Moreover, many models for CR transport make simplistic assumptions for cosmic ray propagation and losses, such that CRs are in steady-state equilibrium, have an axisymmetric distribution, no dynamic coupling with gas, and are not subject to time-dependent phenomena such as galactic winds, magnetic fields, or turbulence \cite{Amato_2018}. 

All of these effects become especially important in the highly active, densely-populated astrophysical environment of the galactic center. The Central Molecular Zone (CMZ), a dense region of molecular gas estimated to make up as much as 10\% of the gas mass of the Milky Way, has been found to be highly turbulent \cite{BRYANT2021101630}. While the star formation rate in the CMZ is thought to be lower than in the disk of the Milky Way, the overall star formation history of the galactic center is uncertain: previously assumed to be continuous, more recent work has suggested that there were late bursts of star formation in the last 1 Gyr \cite{Figer_2004, 2020NatAs...4..377N, Nogueras_Lara_2020}. In addition, the Sgr A* supermassive black hole may have had periods of activity in the past that emitted CRs and regulated star formation on scales from the GC out to the circumgalactic medium (CGM) \cite{2010RvMP...82.3121G}. Because cosmic rays provide a source of feedback that influences star formation within the galaxy, and thus impacts future CR and $\gamma$-ray emission, this problem is inherently time-dependent \cite{Chan_2019, Hopkins_2019, Hopkins_2020, Chan_2022}. 

It is imperative to understand these underlying dynamical effects in CR and $\gamma$-ray emission in order to interpret observations of $\gamma$-rays in the GC, several features of which are hotly debated. The Fermi Large Area Telescope (Fermi-LAT) has detected an excess of GeV $\gamma$-rays within the central 10 degrees of the Milky Way \cite{Ackermann_2017}. While some studies have attributed the signal to a population of unresolved point sources (e.g. millisecond pulsars) \cite{Lee_2016, Bartels_2016}, others have claimed that it is a smoking gun signal of annihilating WIMP dark matter \cite{Hooper_2011, 2016PDU....12....1D, Hooper:2022bec}. However, this is a percent-level excess, which could be explained by insufficient modeling of astrophysical sources \cite{Carlson_2014, Petrovi__2014, Gaggero_2015}. Similarly, Fermi-LAT has detected two $\gamma$-ray super-bubbles above and below the plane of the Milky Way, the origin of which is unknown \cite{Su_2010}. Various production mechanisms for the Fermi bubbles have been suggested, including AGN jets and CR accelerated by episodes of supernova feedback and galactic winds; there is also disagreement as to whether they are hadronic or leptonic \cite{sarkar2024fermierositabubbleslooknuclear}. Understanding the time-dependent interplay of stars, gas, cosmic rays, and AGN is essential to elucidating the origin of the Fermi bubbles. 

Cosmological simulations in which cosmic rays are evolved with full magnetohydrodynamics (MHD) have only recently become possible, owing to numerical breakthroughs in the simulation of plasma physics alongside stellar physics  \cite{Booth_2013, Salem_2013, 2016ApJ...816L..19G, Butsky_2018, Chan_2019, Buck_2020, Werhahn_2021, Pfrommer_2022, Farcy_2022, thomas2022cosmic}. More recent work has extended these numerical techniques to evolve multiple CR species in multiple energy bins in  galaxy formation simulations \cite{Hopkins_CR}. This set of simulations, in which the resolved, multiphase ISM can affect CRs, have enabled self-consistent comparisons to the Milky Way's CR spectrum. So far, these new numerical methods have been used to constrain CR transport by comparing simulations to CR observations in the local ISM, predict radio synchrotron emission \cite{Ponnada_2023}, and explore the origin of the far infrared-radio correlation \cite{ponnada2024hookslinessinkersagn}. In this work, we extend our analysis of these simulations to $\gamma$-ray emission from cosmic ray interactions with the ISM. We model galactic center $\gamma$-ray emission from neutral pion decay, relativistic non-thermal bremsstrahlung, and inverse Compton scattering for the three simulations of Milky Way-mass spiral galaxies in our sample. 

This paper is structured as follows. In section \ref{sec:sims}, we provide a brief description of the simulations analyzed in this work, and describe our method for modeling $\gamma$-ray emission in post-processing. In section \ref{sec:results}, we present the re-constructed $\gamma$-ray spectra, and describe the galactic ``weather" that leads to order-of-magnitude variation in the spectrum normalization and slope over Myr timescales. We then discuss the specific phenomena that lead to variations in the galactic center $\gamma$-ray emission, and their implications for interpreting observations of the Milky Way's galactic center in section \ref{sec:discussion}. Finally, we conclude and outline future work that we plan to undertake to extend our dynamical modeling of galactic $\gamma$-rays in simulations. 

This is the first in a series of three papers that will utilize the $\gamma$-ray modeling introduced in this work for simulations with full CR spectra. The second paper will explore the formation of extended $\gamma$-ray bubbles and halos that arise from the inverse Compton scattering of CR leptons in these simulations. The third paper will focus on a more detailed exploration of CR physics as an explanation for the galactic center $\gamma$-ray excess detected by Fermi-LAT.

% Outline paper 

\section{Simulations}
\label{sec:sims}

\subsection{FIRE methods}

The simulations analyzed in this work were previously run and presented in \cite{Hopkins_CR}; all of the numerical methods used are taken directly from that work, and summarized here to provide context. We consider a set of three Milky Way analog galaxies, labeled as \textbf{m12f}, \textbf{m12i}, and \textbf{m12m}, respectively. All three simulated galaxies are comparable to the Milky Way in stellar and halo mass. \textbf{m12f} and \textbf{m12m} are morphologically similar to the Milky Way: they are spiral galaxies with a thin disk, and \textbf{m12m} is barred. \textbf{m12i}, however, has a polar disk: the inner disk is nearly perpendicular to the outer disk, as a result of its merger history. While none of these simulations are a perfect match to the Milky Way, they are all still good analogs in different respects. The galaxy-to-galaxy variation in $\gamma$-ray emission can inform the range of behavior we may expect from the Milky Way and other nearby galaxies, such as M31.

\subsubsection{Fiducial FIRE physics}
The simulations analyzed in this work were run with GIZMO, a massively-parallel multi-physics solver, using the Lagrangian meshless finite-mass method. We utilize the FIRE-2 implementation of Feedback in Realistic Environments (FIRE) physics for star formation, stellar feedback, and cooling \cite{Hopkins_2018}.
In the simulations, star formation occurs in dense, self-gravitating, self-shielding, Jeans-unstable gas. The simulations explicitly follow stellar feedback from type Ia and type II supernovae, AGB and OB mass loss from stellar winds, and multi-wavelength photo-heating and radiation pressure. Cooling and heating from $10-10^{10}$K is driven by the metagalactic background and stellar sources. Magnetic fields are evolved self-consistently from seed magnetic fields at redshift $z = 100$.

\subsubsection{Cosmic Ray Transport}

% CR methods summary
These simulations are a ``controlled restart'' of a set of three cosmological simulations of galaxies  run with the single-bin CR physics introduced in \cite{Chan_2019, Hopkins_2019, Hopkins_2020} identified as Milky Way-like. Supernovae and fast winds inject a fixed fraction of pre-shock ejecta kinetic energy $\epsilon^{\rm{inj}}_{CR} \sim 0.1$ into the CRs, of which a fraction $\epsilon_{e} \sim 0.02$ goes into the leptonic species. A universal injection spectrum that is a power-law function of rigidity/ momentum is assumed: $j(R) \propto R^{-4.2}$. The CR transport follows the methods detailed in \cite{Hopkins_CR}, and includes adiabatic, streaming, and diffusive re-acceleration loss terms, in addition to Coulomb and ionization losses, catastrophic losses due to hadronic collisions, radioactive decay, annihilation, Bremsstrahlung, inverse Compton scattering, and synchrotron radiation. In addition, there is a  model for secondary losses, such as electrons and positrons produced from pion decay, etc.

The CR species simulated include electrons, positrons, protons, anti-protons, and heavier nuclei including B, CNO, stable Be ($^7\rm{Be}$ and $^9\rm{Be}$) and unstable $^{10}\rm{Be}$ in MeV to TeV energy bins. The relative normalization of the CR spectrum for each species is set according to local ISM measurements \cite{Bisschoff_2019}. The distribution function  $f(p, \mu, ...)$ of each species is evolved according to the two-moment formalism derived in \cite{Hopkins_2021a}, which allows for arbitrary, non time-steady, anisotropic momentum distribution, free streaming at speed $c$, and  alfavenic streaming (as distinct from the more simplified Fokker-Planck that is traditionally solved). Additionally, CR transport is anisotropic along magnetic field lines, dynamically coupled to the evolution of gas and magnetic fields, and can affect/ is affected by the multi-phase ISM, with snapshots saved approximately every Myr in order to study the dynamically effects of CR feedback on a relatively short timescale. The CR spectrum of each species is discretized in piecewise power-law fashion with coarse rigidity intervals.

In these simulations, all parameters that determine the evolution of the distribution function are explicitly evolved and predicted, except for the CR scattering rate, which presumably arises from sub-AU gyroscopic radii, and is therefore unresolved. Following standard practice, we fit the pitch angle-weighted scattering rate $\bar{\nu}$ to a simple power law function of rigidity, $\bar{\nu} = \bar{\nu}_0\beta \,  (R/R_0)^{-\delta}$ by reference to local ISM observables \cite{Zwei13, thomas.pfrommer.18:alfven.reg.cr.transport}; here, $\bar{\nu}_0 = 10^{-9} \, \rm{s}^{-1}$, $R_0 = 1 \, \rm{GV}$, and $\delta = 0.5$. For phenomenological galactic CR propagation codes, which typically use the isotropic Fokker-Planck equation and fit to $D_{xx} = \beta D_0 (R/R_0)^{\delta}$, these values are equivalent to $D_0 = 10^{29} \, \rm{cm}^2/s$, which is similar to the best-fit ISM value found by codes like GALPROP and DRAGON2 \cite{De_La_Torre_Luque_2021, Korsmeier_2021}. Both these phenomenological codes and these simulations have been Both this and those models have been vetted against a wealth of local ISM models, and show to reproduces observables such as the CR spectrum, $e^+/e^-$ ratio, etc. \cite{Hopkins_CR}.

\subsection{Modeling $\gamma$-ray emission}

 During post-processing, we model $\gamma$-ray emission from relevant cosmic ray interactions with the ISM. We consider the three most relevant processes: neutral pion decay, inverse Compton scattering, and relativistic non-thermal Bremsstrahlung. This first requires us to interpolate the coarse-binned CR spectra onto a more fine energy grid using the power-law slopes saved in each coarse bin. We extrapolate to higher energy bins as necessary by assuming the final coarse bin power-law slope. 
 
 The dominant source of $\gamma$-ray emission in the galactic disk is diffuse emission from neutral pion decay, which occurs when a CR proton collides with a nucleus in the ISM \cite{1994A&A...286..983M}. We use the parametrization of the proton-pion-$\gamma$-ray cross section presented in \cite{Kafexhiu_2014}, which is fit to the Geant4 simulations \cite{GEANT4:2002zbu, Allison_06, Allison:2016lfl} above 2 GeV, and experimental data from 0.28-2 GeV. The $\gamma$-ray luminosity from neutral pion decay is
\begin{equation}
    L_{\rm{pion}}= n_{\rm{CR}}n_{\rm{nuc,ISM}}  c \beta V E_{\gamma}\int  \langle \frac{\rm{d}\sigma(T_{\rm{p}},E_{\gamma} )}{d E_{\gamma}} d T_p\rangle ,
\end{equation}
where $E_{\gamma}$ is the energy of the $\gamma$-ray photon, $n_{\rm{CR}} $ is CR proton number density, $n_{\rm{nuc, ISM}}$ is the number density of ISM nucleons, $T_{\rm{p}}$ is the CR proton kinetic energy, and $V$ is the volume of each CR element in the simulation \cite{1994A&A...286..983M}.

The second-most significant contribution to GC $\gamma$-ray flux is relativistic non-thermal Bremsstrahlung, following Blumenthal (1970) \cite{BlumenthalRevModPhys.42.237}: 
\begin{equation}
    L_{\rm{Brem}} = \frac{3}{\pi}\alpha_{\rm{fs}} \sigma_{\rm{T}} V n_{\rm{H}} n_{e\pm} \left( \ln{2 \gamma_{e}} - \frac{1}{3} \right) 
\end{equation}
where $\alpha_{\rm{fs}}$ is the fine-structure constant, $\sigma_T$ is the Thompson cross-section, $\gamma_e$ is the electron/ positron Lorentz factor, and the $\gamma$-ray energy scales as $E_{\gamma} \approx \frac{1}{3}E_{0}$. 

Finally, we calculate inverse Compton scattering of CR electrons from the Cosmic Microwave Background (CMB), IR, optical, and UV radiation bands.

\begin{equation}
    L_{\rm{ICS}} =\sum_b \frac{4}{3} c \sigma_T  U_{\rm{rad, b}} n_{\rm{CR}, e^{\pm}} (\beta)^2 (\gamma_e)^2
\end{equation}
where $b$ is the radiation band (CMB, IR, optical, and UV) \cite{Longair:1992ze}. The energy density for the latter three bands is evolved in the simulation and output, while we treat CMB as a uniform background with a peak at $6 \times 10^{-4}$ eV and density $0.26 \times 10^{-9}$ GeV cm$^{-3}$ \cite{2020MNRAS.491.3702H, Dodelson:2003ft}. Because there is a far higher density of CMB photons than IR, optical, or UV photons, ICS contributes significantly to $\gamma$-ray emission below 1 GeV. For the other three bands, we pick a characteristic photon energy for the entire band, $E_0$, and estimate the energy of the IC-scattered photon as $E_\gamma \approx (\gamma_e)^2 E_0$ \cite{Longair:1992ze}. 

We do not explicitly model photon production from processes that are expected to produce a negligible amount of $\gamma$-rays, i.e. synchrotron and secondary synchrotron radiation, electron-positron annihilation, and non-relativistic thermal Bremsstrahlung. Additionally, we do not inject a mock source spectrum from pulsars, although this option will be explored in future work.

\section{Results}
\label{sec:results}
\subsection{Comparison with the Local ISM as Calibration Test}
\label{sec:calibration}
\begin{figure}
    \centering
    \includegraphics[width=\linewidth]{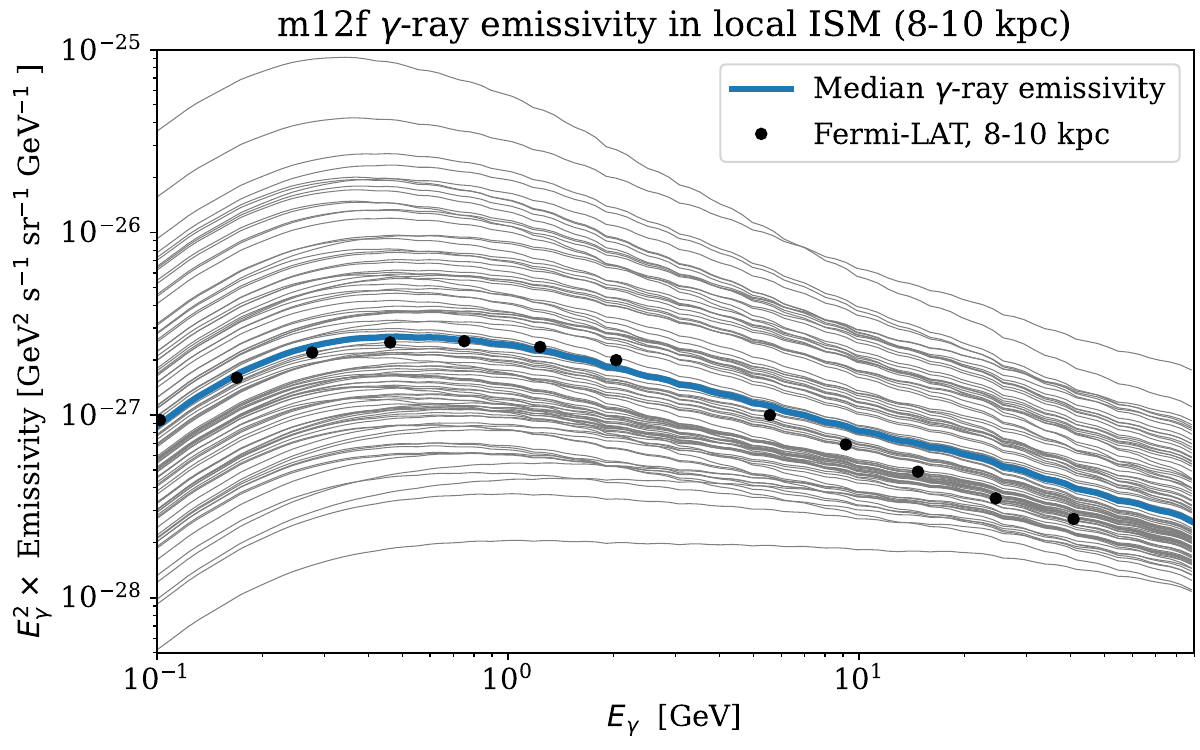}
    \caption{$\gamma$-ray emissivity per H atom for galaxy \textbf{m12f} within a 8-10 kpc annulus, compared to Fermi-LAT measurements of $\gamma$-ray emissivity in the local ISM \cite{Acero_2016}; the median emissivity is shown in the thick blue line, with thinner lines representing the emissivity in randomly-selected gas cells within the annulus.  The median emissivity for \textbf{m12f} (and \textbf{m12i} and \textbf{m12m} as well) is in good agreement with the local ISM Fermi-LAT measurement, which serves as an important validation of the CR transport model and the simulated Milky Way analogs used in this work. There is considerable variation in spectral shape for different cells, ranging from flattened profiles to steeper slopes, driven in part by differences in the gas density, as discussed in \cite{Hopkins_CR}.}
    \label{fig:GR_emiss}
\end{figure}

Previous work has shown that these simulations predict CR spectra and properties that are consistent with observations in the local ISM  \cite{Hopkins_CR}. Since the simulated galaxies in our sample have been established as good Milky Way analogs in the local ISM, we test our $\gamma$-ray modeling in this environment first. 

We calculate the $\gamma$-ray emissivity per H atom  in a galactocentric annulus from 8-10 kpc, selecting only cells with a gas number density of $0.1- 1 \, \rm{cm}^{-3}$ in order to compare with local ISM-like environments. Comparing with the $\gamma$-ray emissivity per H atom measured by Fermi-LAT in the same annulus \cite{Acero_2016}, we find that the calculated $\gamma$-ray emissivity for all three simulated galaxies in our sample is a good match to the observations within a factor of $\sim 2$. In Figure \ref{fig:GR_emiss}, we show the median $\gamma$-ray emissivity for galaxy \textbf{m12f}, along with the emissivities from 100 randomly-selected gas cells within this annulus. While the median emissivity is strikingly similar to that observed by Fermi-LAT, we see that the emissivities can vary by more than 2 orders of magnitude in amplitude, as well as shape; the emissivity for some cells is sharply peaked just below 1 GeV, while for others it is flatter throughout. Even with minimal variation in HI density, the small-scale structure of the ISM can have a significant effect on CR transport and $\gamma$-ray emission. 

\begin{figure}
    \centering
    \includegraphics[width=\linewidth]{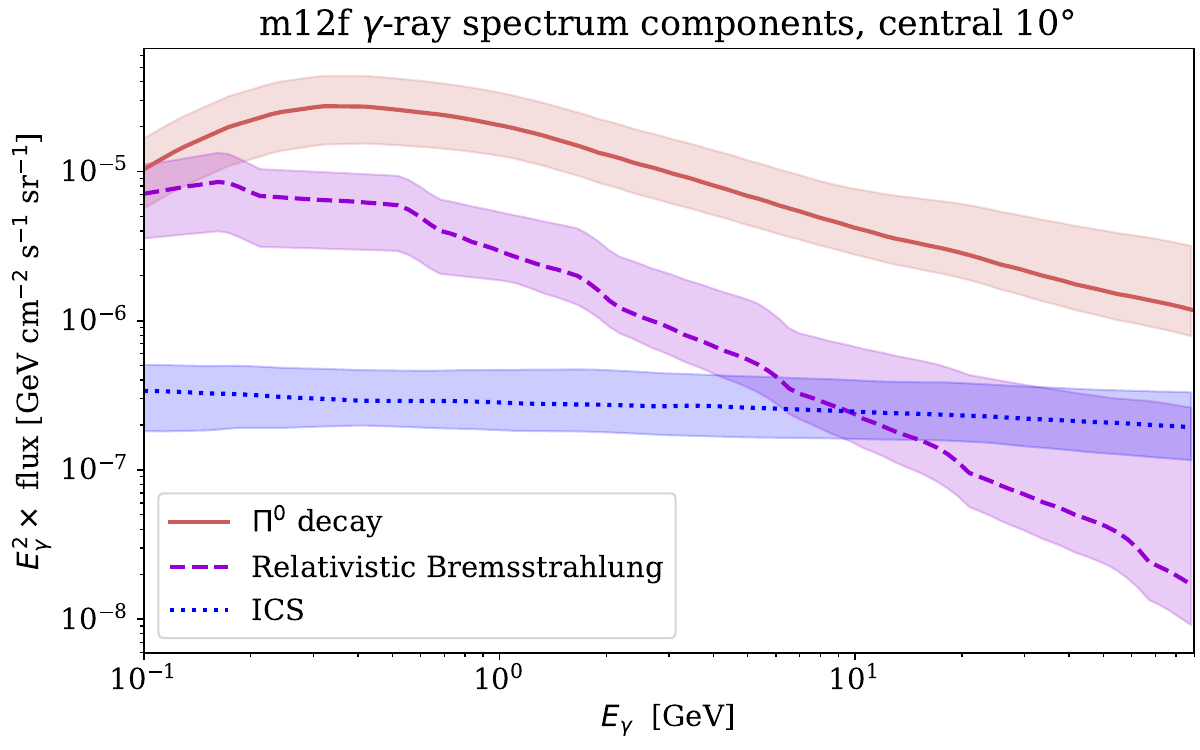}
    \caption{$\gamma$-ray emission in the galactic center of galaxy \textbf{m12f} from the three most significant $\gamma$-ray producing CR interactions: neutral pion decay, relativistic non-thermal bremsstrahlung, and inverse Compton scattering, as viewed by an observer positioned within the solar circle. The median value of each component over snapshots spanning 460 Myr is plotted, with the full range of values shaded. Neutral pion decay produces most of the $\gamma$-ray emission, with bremsstrahlung important below 10 GeV. The higher-energy tail of the bremsstrahlung is highly variable. ICS is nearly constant from 0.1 to 100 GeV. Note that the ``steps" in the bremsstrahlung are a numerical artifact from the coarse-binned rigidities.}
    \label{fig:m12m_spread}
\end{figure}

\begin{figure}
    \centering
    \includegraphics[width=\linewidth]{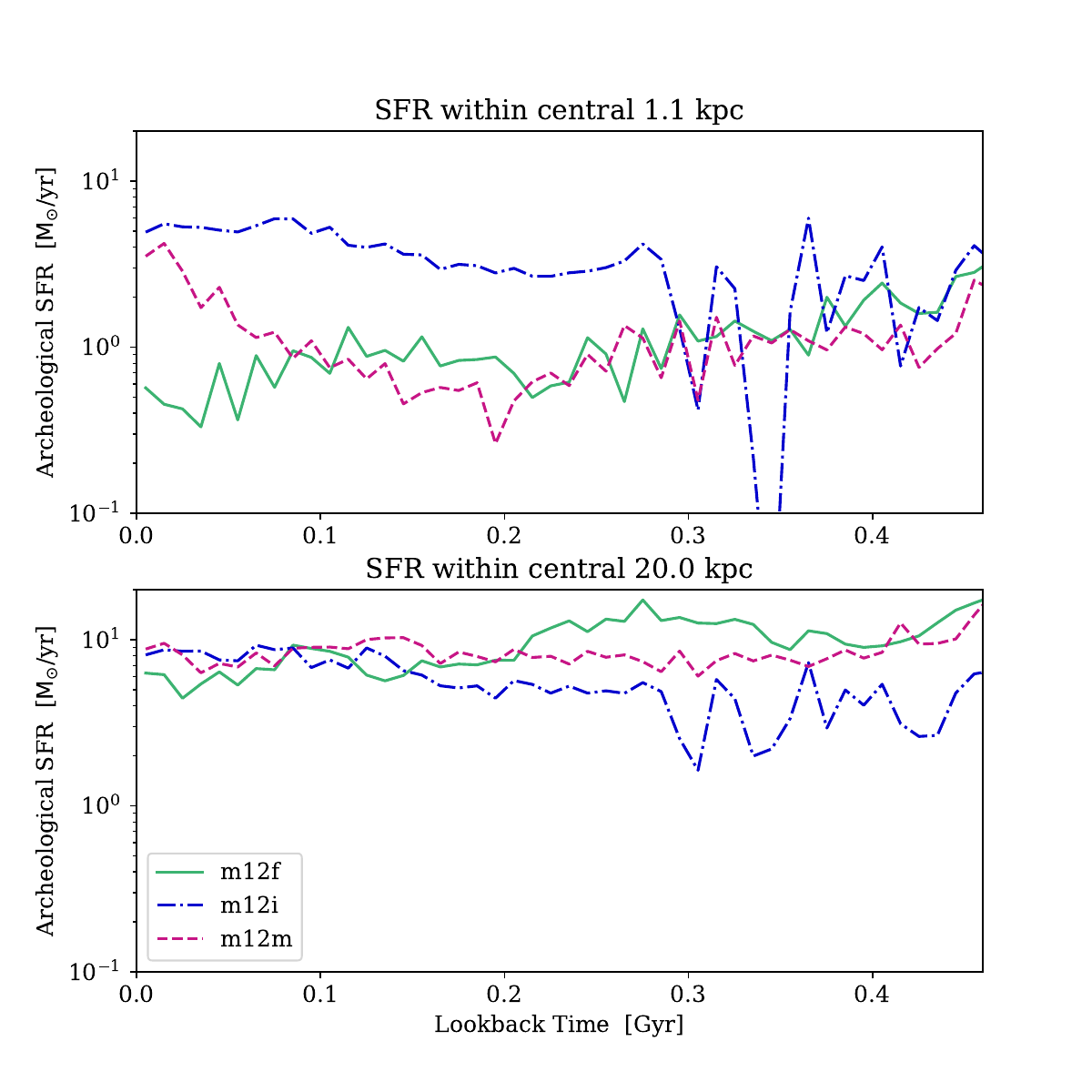}
    \caption{ `Archaeological' star formation rate (SFR) averaged over 10 Myr for the three galaxies in our sample, within the galactic center (top) and the disk (bottom), during the last 460 Myr of the simulation (i.e. when the full multi-species CR spectra are evolved). SFR is mutually correlated with supernova rate, gas density, and radiation fields, all of which affect $\gamma$-ray emission \cite{2017MNRAS.467.2301T, 2021ApJ...908L..31O, Hopkins_CR}.  Within the galactic center, \textbf{m12i} maintains a comparatively high SFR over the last 300 Myr, while \textbf{m12m} gradually increases over the last 100 Myr. \textbf{m12f} slowly declines in SFR throughout the simulation. All three galaxies show more burstiness throughout the stellar disk.}
    \label{fig:sfr}
\end{figure}

\begin{figure}[h!]
    \centering
    \includegraphics[width=\linewidth]
    {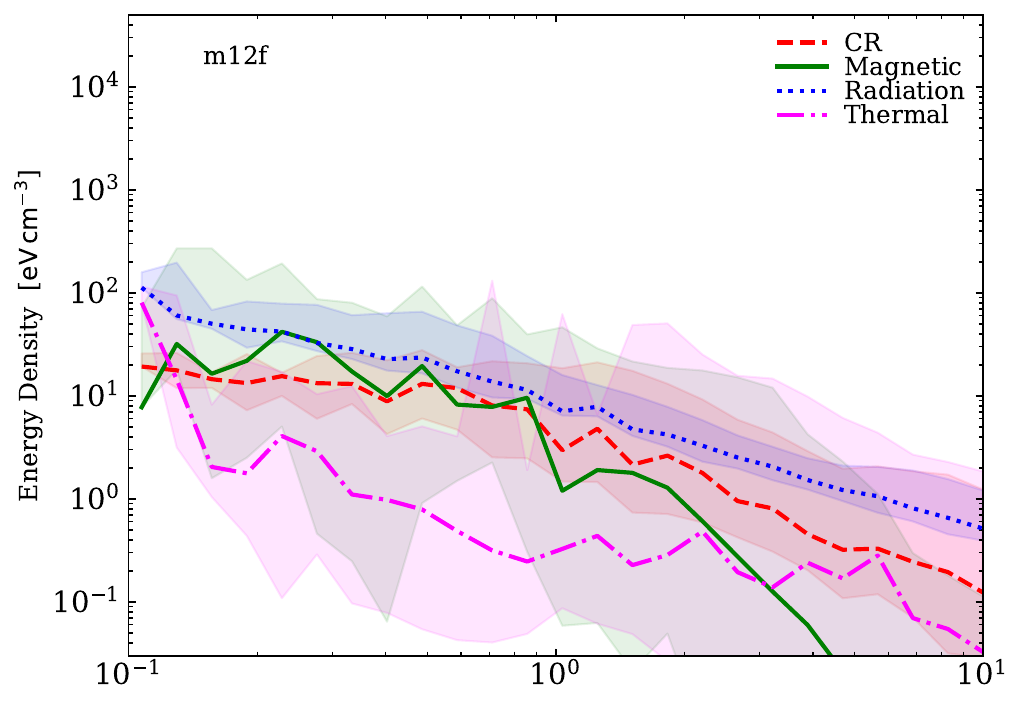}
    \includegraphics[width=\linewidth]
    {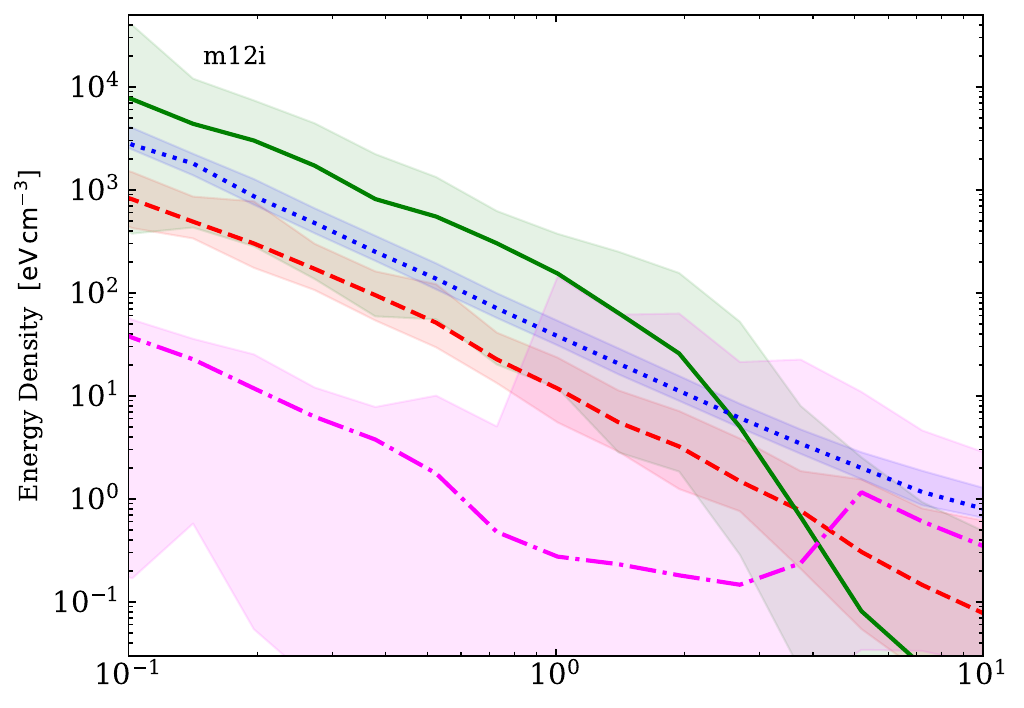}
    \includegraphics[width=\linewidth]
    {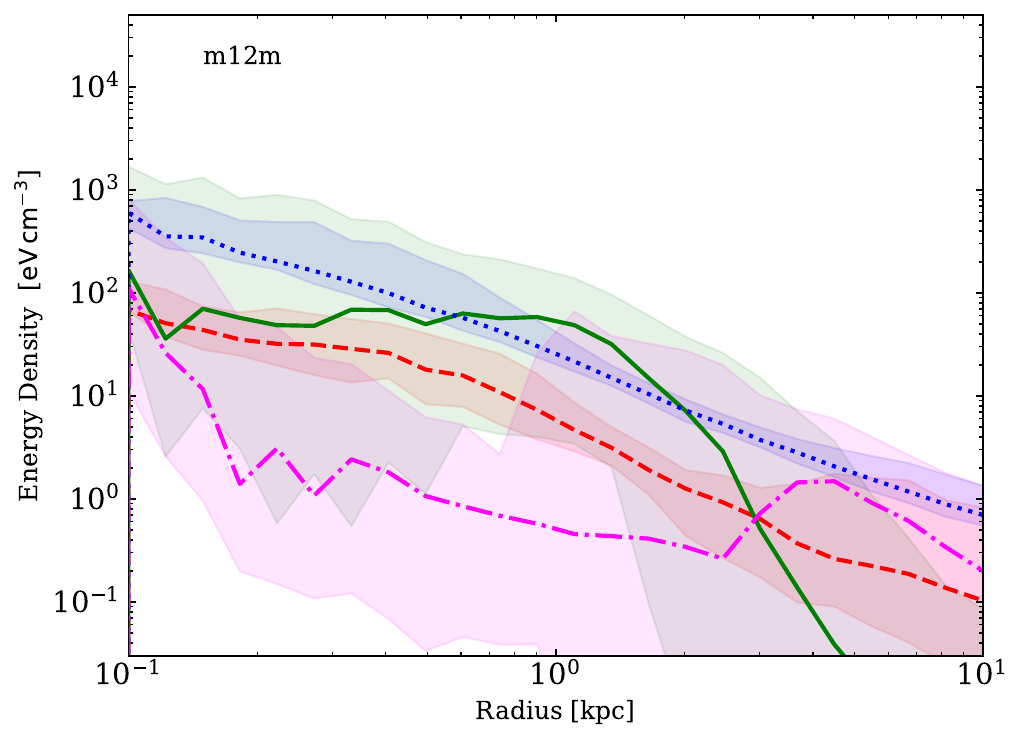}
    \caption{Energy density of the radiation field, magnetic field, thermal pressure, and CRs in galaxies \textbf{m12f} (top), \textbf{m12i} (middle), and \textbf{m12m} (bottom) vs galactocentric radius. Within the inner 1-3 kpc, the radiation and magnetic fields dominate, resulting in strong synchrotron and ICS losses in the CR leptons. The radiation, CR, and magnetic field energy density for galaxy \textbf{m12i} is roughly 2 orders of magnitude larger in the galactic center than those of \textbf{m12f} and \textbf{m12m}; it is also highly variable over time.}
    \label{fig:sim_stress}
\end{figure}

However, despite this similarity to the Milky Way in the local ISM, we find that all three of the simulated galaxies in our sample predict $\gamma$-ray emission roughly an order of magnitude larger than that measured by Fermi-LAT in the galactic center (Figure \ref{fig:m12m_spread}). This is largely due to the fact that, while these simulations are globally similar to the Milky Way in stellar and halo mass, they are more compact in their central regions. Since there are more stars (especially young stars sourced by recent star formation), there are more CR sources. In Figure \ref{fig:sfr}, we show `archaeological' star formation rates (SFR) for the simulations, both within the galactic center, and within the full galactic disk; the archaeological SFR infers when stars were formed based on the stellar population present at redshift $z = 0$. The higher SFR within the galactic center of \textbf{m12i} and \textbf{m12m} at late times results in particularly strong magnetic and radiation fields within the central kpc (Figure \ref{fig:sim_stress}). 

\begin{figure*}[t]
    \centering
    \includegraphics[width=\linewidth]{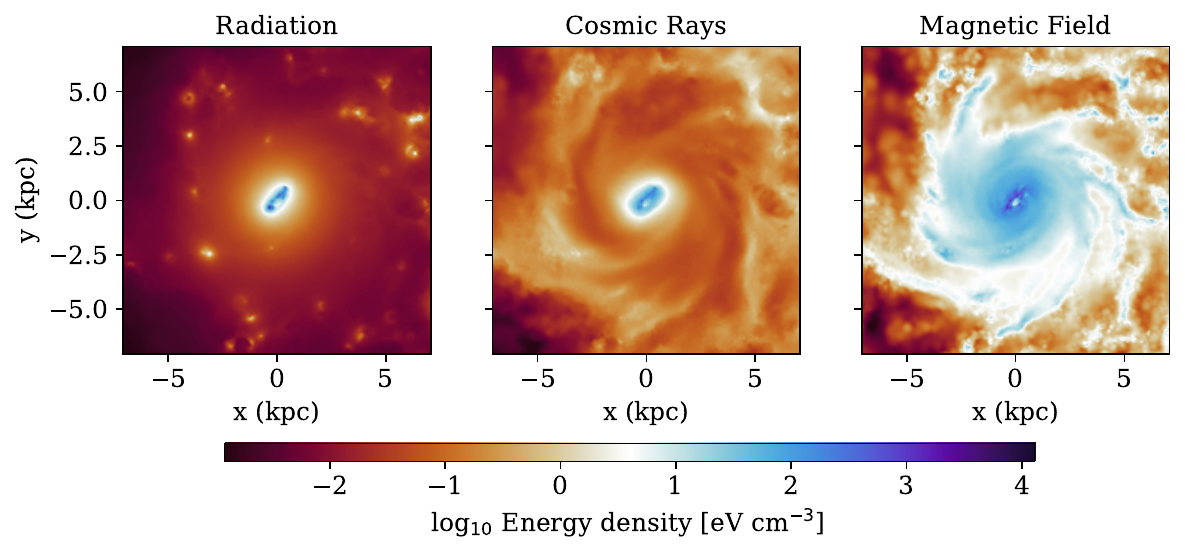}
    \caption{A spatial visualization of the quantities shown in Figure \ref{fig:sim_stress} for galaxy \textbf{m12m}: mass-weighted mean energy densities along the line-of-sight for the radiation field, cosmic rays, and magnetic field at redshift $z = 0$, within a $\pm 1$ kpc slice. All three density fields rise towards the galactic center, as well as in other dense regions like spiral arms and giant molecular clouds. Local variation in these fields results in sub-kpc spatial fluctuation in the total CR flux and the ratio of e.g. hadrons to leptons. The high energy density of the radiation and magnetic fields  in the central $\sim 1$ kpc results in significant synchrotron and inverse Compton losses for the CR electrons and positrons.}
    \label{fig:edens}
\end{figure*}

Another notable feature of our simulations is that, relative to the Fermi-LAT model presented in \cite{Ackermann_2017}, $\gamma$-ray emission from inverse Compton scattering makes up a considerably smaller fraction of the total $\gamma$-ray spectrum (Fig. \ref{fig:m12m_spread}). This relative deficiency of ICS-produced $\gamma$-rays is due to a deficiency of CR electrons; the region within approximately 2 kpc of the galactic center is dominated by magnetic and radiation pressure (Fig. \ref{fig:sim_stress} and Fig. \ref{fig:edens}), which, over a period of Myrs, leads to significant losses in the CR leptons via synchrotron radiation and ICS, respectively.

\begin{figure}
    \centering
    \includegraphics[width=\linewidth]{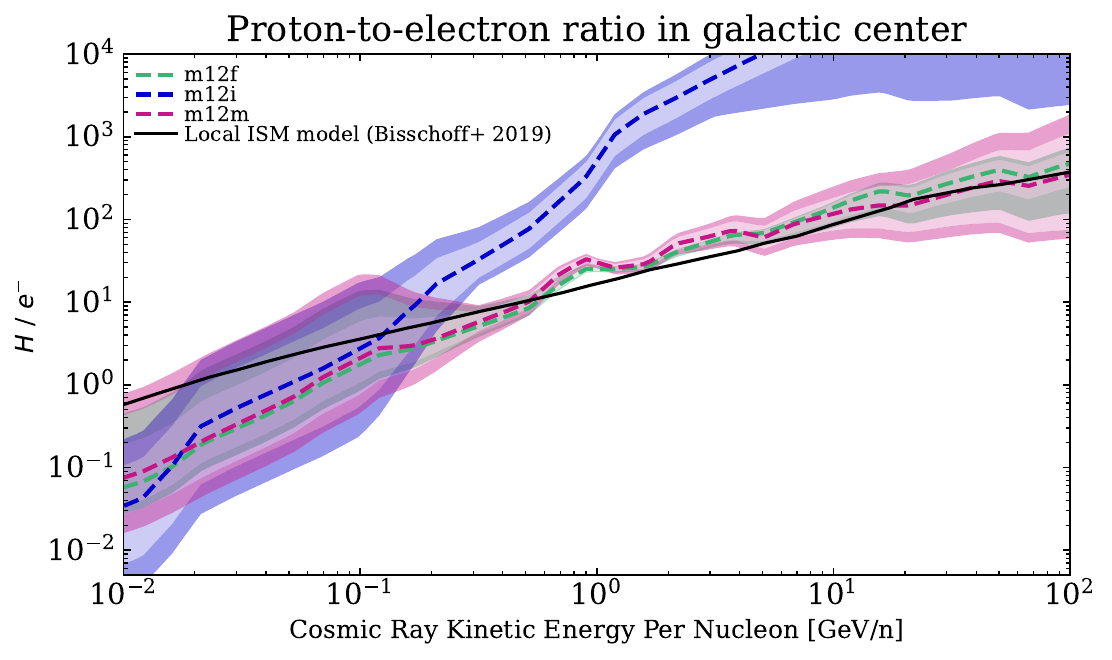}
    \includegraphics[width=\linewidth]{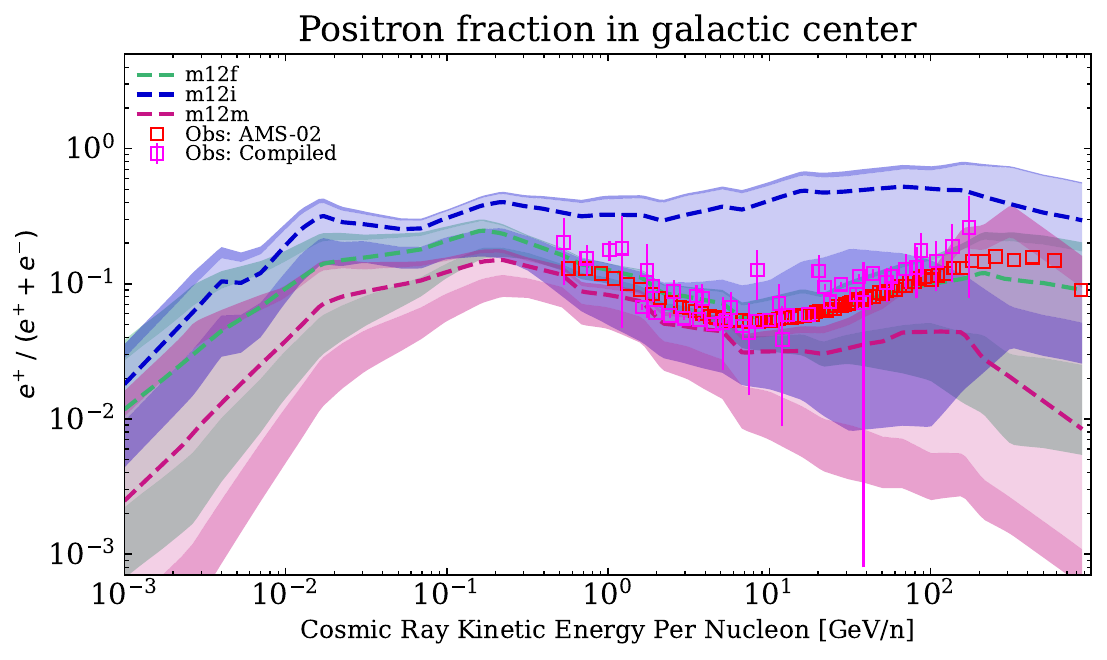}
    \caption{Top: Cosmic ray proton-to-electron ratio within a 1.1 kpc galactocentric radius, compared to the local ISM model derived in \cite{Bisschoff_2019}. The dashed line represents the median value within this region, and the light (dark) shaded area shows the $\pm1 \sigma$ ($\pm 2 \sigma$) range. \textbf{m12i}, which has particularly strong magnetic and radiation fields in its galactic center, has a sharp increase in proton-to-electron ratio above 0.1 GeV because CR electrons are lost to inverse Compton scattering and synchrotron radiation. \textbf{m12f} and \textbf{m12m} track the local ISM model closely on average, but have $\sim 2$ dex of scatter at high and low CR energies. Bottom: positron fraction  $e^+/(e^+ + e^-)$ within a 1.1 kpc galactocentric radius, compared to local ISM observations; red boxes represent data from the AMS-02 experiment \cite{2018PhRvL.120b1101A, 2019PhRvL.122d1102A, 2019PhRvL.122j1101A}, while pink boxes are compiled from multiple CR observatories \cite{1990A&A...233...96E,2007APh....28..154S, 2000ApJ...532..653B, 2011ApJ...742...14O, 2014ApJ...791...93A, 2017PhRvD..95h2007A, 2017arXiv170906442H, 2017ApJ...839....5Y, 2017Natur.552...63D, 2018PhRvL.120z1102A, 2019ARep...63...66A}. The simulations are broadly consistent with the observations, although there is $\sim$ 2 dex of scatter between 10-100 GeV for this quantity as well. The overall higher positron fraction for \textbf{m12i} is attributable to the high CR radiative losses, which preferentially affect CR leptons more than the positrons, which are produced as secondaries from CR proton interactions.}
    \label{fig:p_to_e}
\end{figure}

The synchrotron and ICS losses lead to variation and enhancement of the CR proton-to-electron ratio in the galactic center. In the top panel of Figure \ref{fig:p_to_e}, we compare the proton-to-electron ratio from the galactic center in our simulations to the model derived from local ISM data in \cite{Bisschoff_2019}. While the proton-to-electron ratio is in close agreement with the model when restricted to the local ISM-like environments described above, particularly above 1 GeV CRs, we see significant deviation from the model in the galactic center. \textbf{m12i} shows, by far, the most enhancement in proton-to-electron ratio in the GC, due to its strong magnetic and radiation fields (Fig. \ref{fig:sim_stress}). \textbf{m12f} and \textbf{m12m} show up to 2 orders-of-magnitude spread in the proton-to-electron ratio across all gas cells within 1.1 kpc of the galactic center for 10-100 GeV CRs, which contribute most to $\gamma$-ray production in the regime to which Fermi-LAT has the greatest sensitivity. 

In addition, we show the positron fraction of total CR leptons, $e^+/(e^+ + e^-)$ within the galactic center, and compare to local ISM observations from AMS-02, PAMELA, and other CR experiments (Fig. \ref{fig:p_to_e}, bottom) \cite{1990A&A...233...96E,2007APh....28..154S, 2000ApJ...532..653B, 2011ApJ...742...14O, 2014ApJ...791...93A, 2017PhRvD..95h2007A, 2017arXiv170906442H, 2017ApJ...839....5Y, 2017Natur.552...63D, 2018PhRvL.120z1102A, 2019ARep...63...66A,2018PhRvL.120b1101A, 2019PhRvL.122d1102A, 2019PhRvL.122j1101A}. The positron fraction in the galactic center is broadly consistent with the local ISM data, but, like the proton-to-electron ratio, shows $\sim 2$ dex of variation between 10-100 GeV. The rise in the positron fraction between 10 and 100 GeV in the local ISM, first detected by PAMELA and then confirmed by subsequent experiments \cite{PAMELA:2008gwm}, disagrees with standard CR propagation models \cite{Moskalenko_1998}. As for the galactic center excess, both annihilating dark matter and pulsars have been proposed as an origin of the anomaly, while other work suggests that the increase in positron fraction may simply arise from local inhomogeneities in the ISM \cite{Shaviv:2009bu}. The wide spread in positron fraction for gas cells within 1.1 kpc of the galactic center in these simulations lends credence to the latter interpretation, which extends to other anomalies (e.g., the 511 keV emission line detected in the galactic center) as well \cite{2011RvMP...83.1001P}. 

\subsection{Variation in $\gamma$-ray Spectrum}

\subsubsection{Variation over Time}

\begin{figure*}
 \centering
 \includegraphics[width=0.95\columnwidth]{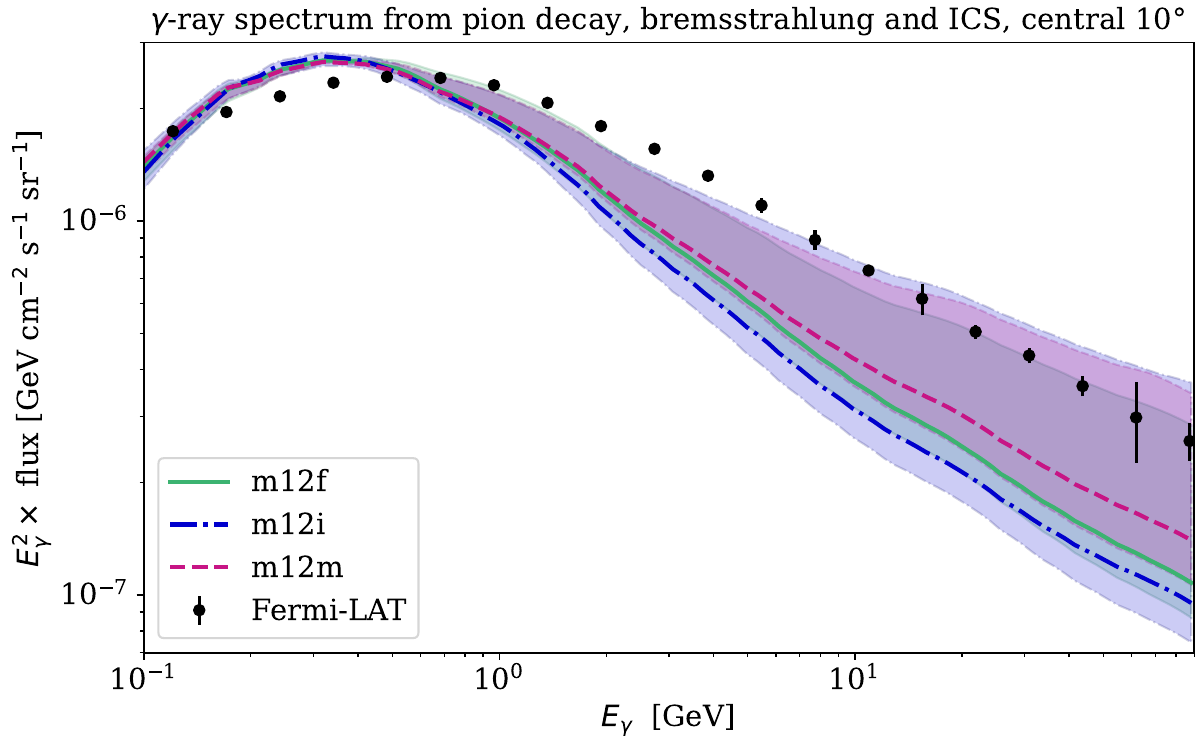}
 \includegraphics[width=0.95\columnwidth]{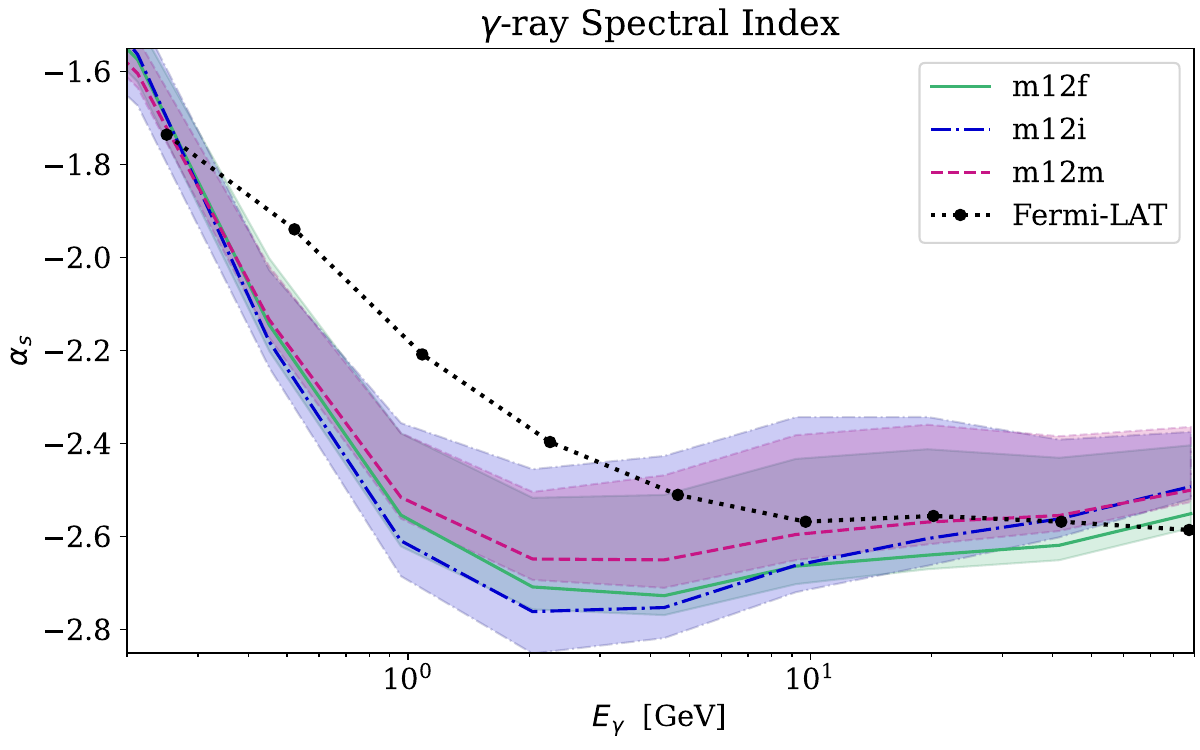}
    \caption{Left: Galactic center $\gamma$-ray spectrum for \textbf{m12f}, \textbf{m12i}, and \textbf{m12m} from pion decay, relativistic non-thermal bremsstrahlung, and inverse Compton scattering, calculated at snapshots over the last 460 Myr, and normalized to the measured Fermi-LAT galactic center $\gamma$-ray spectrum within the central 10 degrees of the Milky Way. These spectra arise from the CR transport implemented in the simulations, assuming a universal single power-law injection spectrum; unlike in phenomenological CR propagation codes, there is no variation over parameters to fit MW observables. Right: the $\gamma$-ray spectral index $\alpha_s \equiv  \frac{\ln F_{E}}{\ln E_{\gamma}}$ for the same set of snapshots. In general, the $\gamma$-ray spectra calculated from the simulations are more peaked and have a more rapid turnover than the spectrum measured by Fermi-LAT. However, the slope of the calculated spectra is consistent with the Fermi-LAT measurement above $E_{\gamma} \approx 4$ GeV.}
    \label{fig:GR_spec}
\end{figure*}

In order to determine the variation in $\gamma$-ray emission in the galactic center over time and along different lines-of-sight, we calculate the $\gamma$-ray spectrum from CR sources (pion decay, bremsstrahlung, and inverse Compton scattering) for each of the simulated galaxies. We sample from 460 Myr to 20 Myr in 10 Myr intervals, and from 20 Myr in 1 Myr intervals. We additionally sample from several different viewing angles within the galactic disk, each at a galactocentric distance of 8.1 kpc. As discussed in section \ref{sec:calibration}, nearly all of the $\gamma$-ray spectra from the simulations exceed the measured galactic center Fermi-LAT spectrum, in some cases by as much as an order of magnitude. In large part, this is due to the greater number of young stars leading to supernovae and stellar winds in the simulations versus the Milky Way. For the comparison of spectral slopes, we normalize the $\gamma$-ray spectrum from the simulations to match the Fermi-LAT $\gamma$-ray spectrum within the central 10 degrees of the Milky Way by integrating flux from 0.1 to 100 GeV.  In Figure \ref{fig:GR_spec}, we show the range of calculated $\gamma$-ray spectra from our sample of simulations \cite{Ackermann_2017}. We find that, while our simulated $\gamma$-ray spectra tend to be more peaked than the Milky Way's spectrum, they are consistent with the Fermi-LAT $\gamma$-ray spectrum and spectral slope above $\sim$ 4 GeV. Galaxies \textbf{m12i} and \textbf{m12m} show more variation in $\gamma$-ray spectrum shape than \textbf{m12f}, when all spectra are normalized to match the Fermi-LAT observations.  

Given that we impose very few constraints on our CR transport model to match the Milky Way (aside from normalizing the CR spectra for individual species when they are initialized at $z = 0.05$), the agreement between our simulated GC $\gamma$-ray spectra and the Fermi-LAT measurement is surprisingly good, especially above 1 GeV. There are several factors in the simulations that likely contribute to the more peaked spectrum below 1 GeV. First, we have assumed that the CR injection spectrum follows a single power-law; a more realistic injection spectrum may have a break at lower energies due to non-relativistic corrections, as is commonly adopted and fitted to in CR propagation codes \cite{Porter_2022}. In addition, the simulations use a constant cross-section for proton-to-pion production, while our analysis uses the cross-section parametrization described in \cite{Kafexhiu_2014}, potentially leading to a shift in the peak of the $\gamma$-ray spectrum. While the constant cross-section approximation used in the simulations is not a bad first-order approximation, it does under-predict pion production above $\sim 1$ GeV, and over-predicts it below a GeV \cite{2018A&A...615A.108Y}.

\begin{figure}[h!]
    \centering
    \includegraphics[width=\linewidth]
    {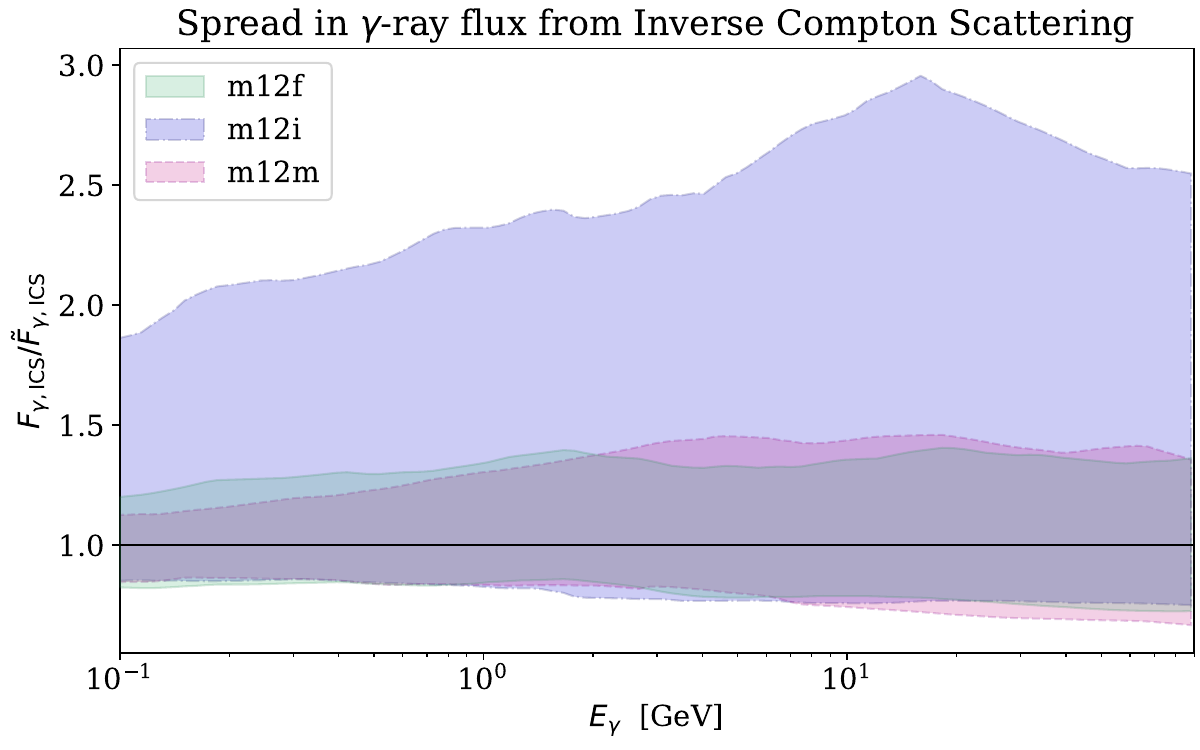}
    \includegraphics[width=\linewidth]
    {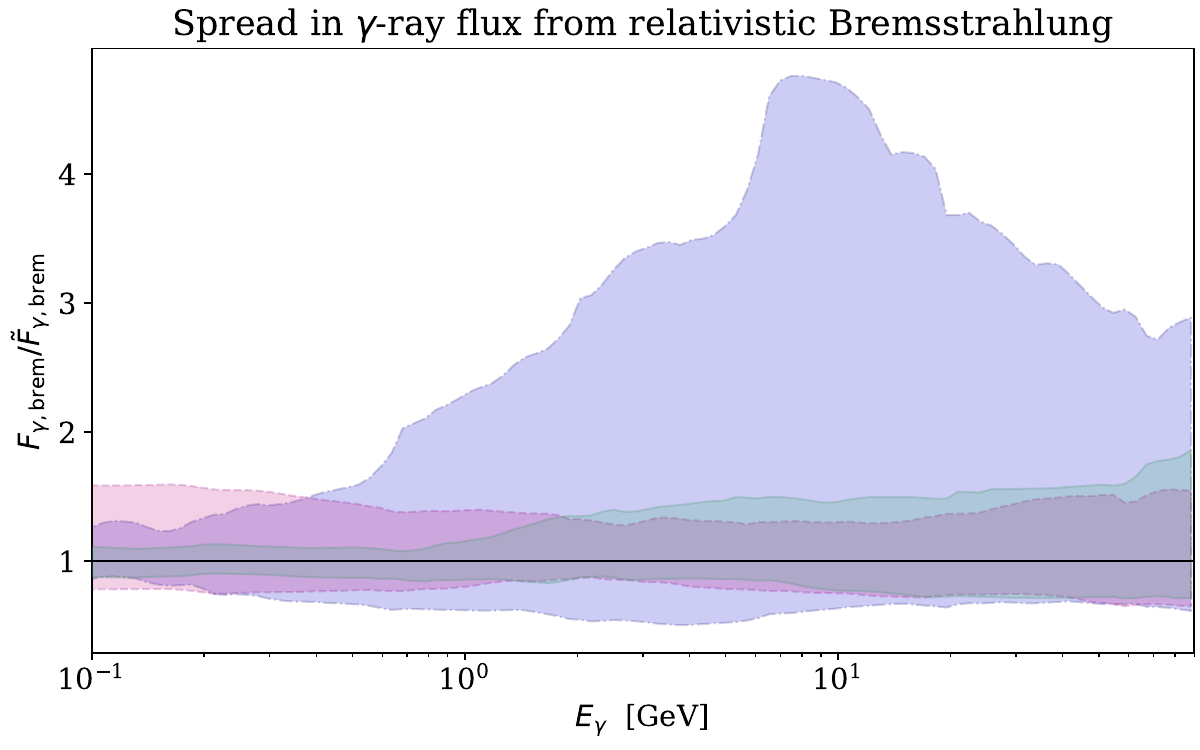}
    \includegraphics[width=\linewidth]
    {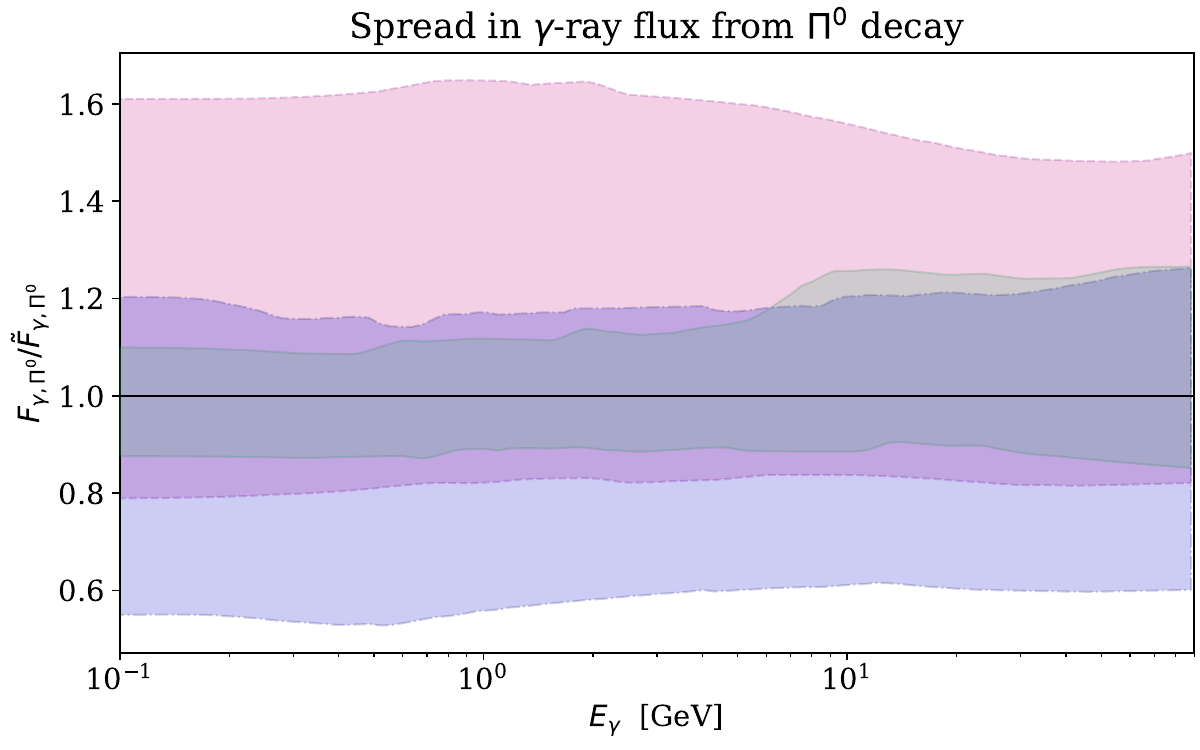}
    \caption{Spread in un-normalized $\gamma$-ray emission from inverse Compton (top), relativistic bremsstrahlung (middle), and pion decay (bottom), calculated by comparing the spectrum at each snapshot to the median value over all snapshots. \textbf{m12i} shows far more variation in its leptonic sources than both \textbf{m12f} and \textbf{m12m}, due to the galaxy's strong and fluctuating magnetic field. \textbf{m12m}, which has the greatest change in star formation rate at late times, has the largest spread in $\gamma$-ray from neutral pion decay. \textbf{m12f}, shows the least variation over time in all components of its $\gamma$-ray emission. (Note: 25th and 75th percentiles are shown here)}
    \label{fig:comp_spread}
\end{figure}

In Figure \ref{fig:comp_spread}, we consider each component of the $\gamma$-ray emission in order to determine what drives amplitude and shape variations in the GC $\gamma$-ray spectrum over time. For each of pion decay, Bremsstrahlung, and ICS, we compare the full spread of $\gamma$-ray spectra over time to its median value; unlike in Figure \ref{fig:GR_spec}, the spectra are not normalized to the same value (allowing us to see variation in amplitude). We shade the region between the 25th and 75th percentiles. The component of the GC $\gamma$-ray spectrum produced by pion decay shows that, while the two galaxies with more variable SFRs over the full duration of the simulation (\textbf{m12i}, which has a dip around 350 Myr, and \textbf{m12m}, which steadily increases in the last 100 Myr) can vary by a factor of a few in normalization, there is little variation in the shape of the pionic component (Figure \ref{fig:sfr}). In contrast, there is more variation in spectrum shape and amplitude for Bremsstrahlung and ICS-produced $\gamma$-rays; for \textbf{m12f} and \textbf{m12m}, the time variation in $\gamma$-ray spectrum amplitude from these two sources is comparable to the time variation from pion decay. What is most striking, however, is the large amplitude and shape variation for both Bremsstrahlung and ICS in \textbf{m12i}, which peaks at approximately 10 GeV. 

\begin{figure}
    \centering
    \includegraphics[width=\linewidth]
    {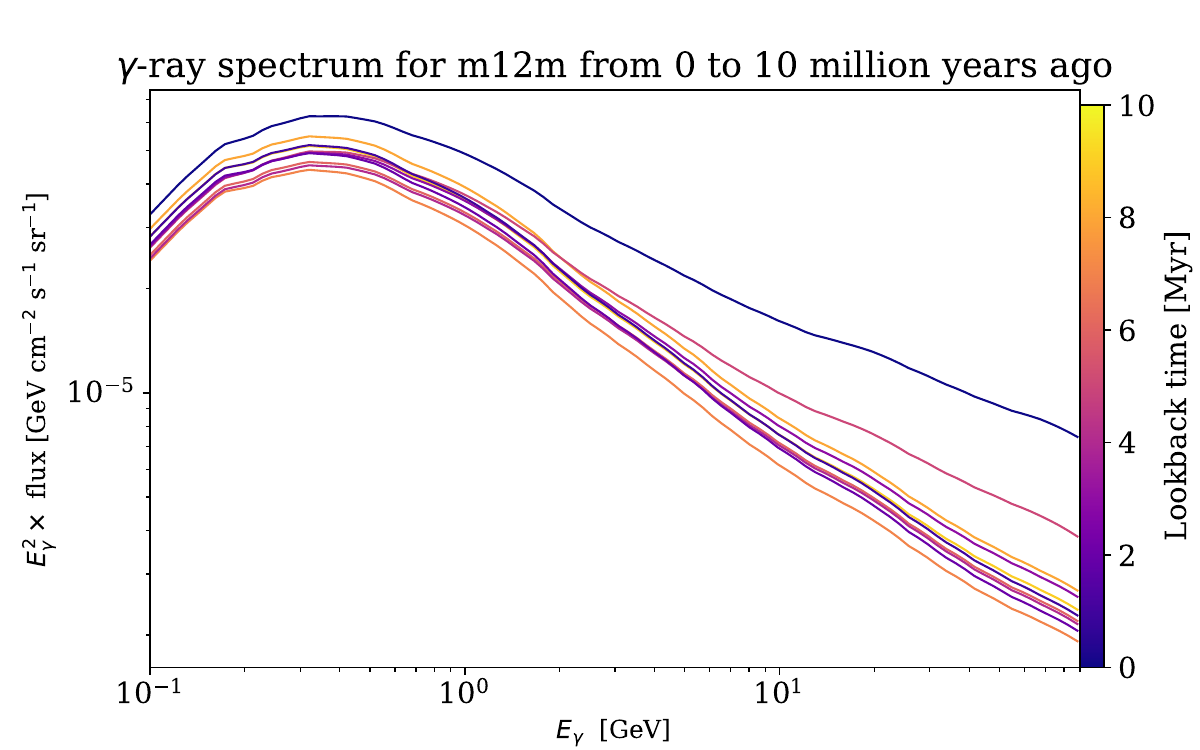}
    \caption{Spread in un-normalized $\gamma$-ray spectrum over the last 10 Myr of the simulation for galaxy \textbf{m12m}; the spectrum is calculated at 1 Myr intervals and color-coded by lookback time (more recent snapshots are darker). Both the shape and overall amplitude of the $\gamma$-ray spectrum can fluctuate considerably on a relatively short time scale of approximately 1 Myr. The fluctuations in spectrum shape and amplitude are not monotonic on $\sim 1$ Myr timescales.}
    \label{fig:m12m_spread_late}
\end{figure}

This variation in spectrum is typically stochastic, particularly on shorter time scales. At first order, it is driven by the CR injection rate (i.e. the supernova rate), which affects both the number density of CR protons and the distribution of HI to produce neutral pions, as well as magnetic and radiation field strength that affects CR leptons and thus $\gamma$-ray emission from Bremsstrahlung and ICS. While there may be an overall trend in the $\gamma$-ray spectrum amplitude or shape, on a $\sim 1$ Myr timescale, the spectrum can increase or decrease by a factor of a few (Figure \ref{fig:m12m_spread_late}). The clustering of CR sources (e.g., supernovae) in space and time can affect the makeup of the stellar population (young vs. old stars) and local gas density \cite{2017MNRAS.467.2301T,Moreno:2020bxn,2021ApJ...908L..31O,Li:2024aoq}, and thus alter the $\gamma$-ray spectrum shape.

\subsubsection{Variation in Line-of-Sight}

\begin{figure}
    \centering
    \includegraphics[width=\linewidth]{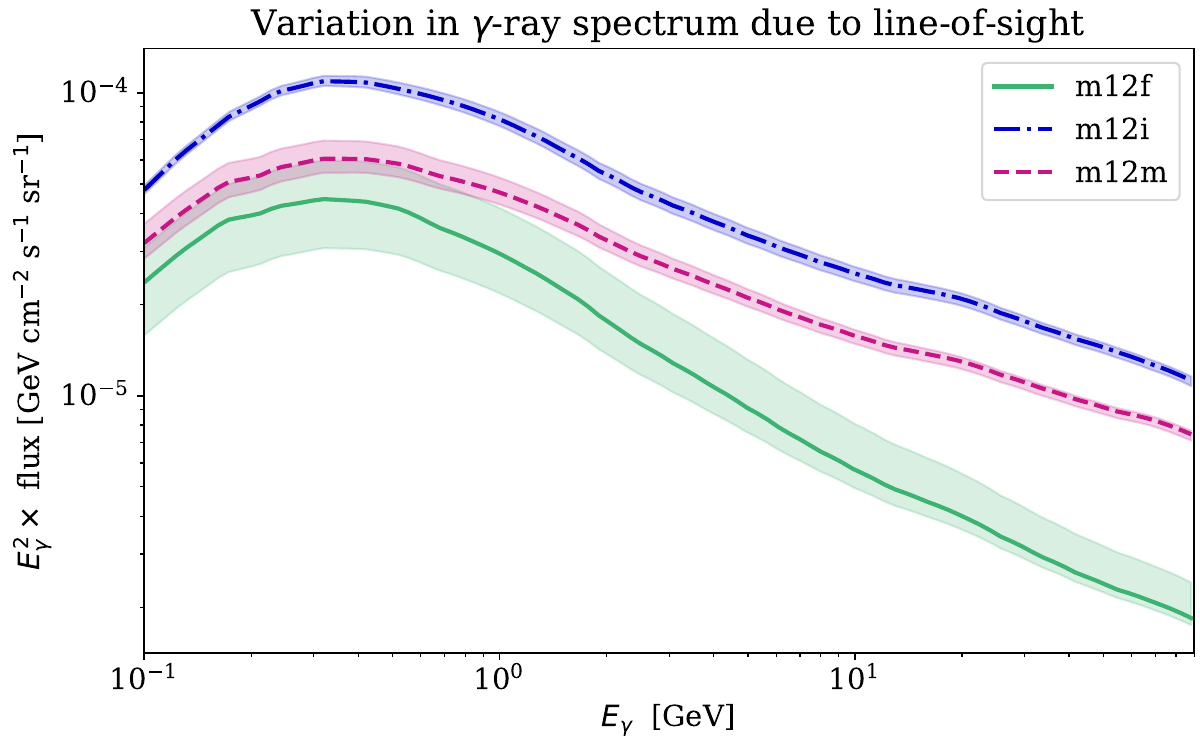}
    \caption{ Variation in un-normalized $\gamma$-ray spectra for \textbf{m12f}, \textbf{m12i}, and \textbf{m12m} along different lines-of-sight between the observer and the galactic center. The line-of-sight/ azimuthal position within the disk makes negligible difference for the spectra from \textbf{m12i} and \textbf{m12m}. However, it does affect the spectrum measured for \textbf{m12f}; for several azimuthal positions within the disk, there are dense gas clouds along the line-of-sight that contribute to the $\gamma$-ray spectrum by interacting with background CRs from star-forming regions. }
    \label{fig:spec_var_ang}
\end{figure}

We test the variation in $\gamma$-ray spectrum due to differing line-of-sights between observer in the galactic center by selecting several different azimuthal angles within the galactic disk at which to place the observer. In general, the simulated $\gamma$-ray spectra vary more with time than they do with line-of-sight. At $z = 0$, galaxies \textbf{m12i} and \textbf{m12m} show negligible variation with respect to line-of-sight; galaxy \textbf{m12f}, however, shows significant variation, particularly below 2 GeV (Figure \ref{fig:spec_var_ang}). This variation can be attributed to the presence of clumps of dense neutral gas along certain lines-of-sight for \textbf{m12f}. This result emphasizes that foreground removal and mitigation can be critical for the analysis of GC $\gamma$-ray emission.

\subsection{$\gamma$-ray emission maps}

Past work has used the angular distribution of the GC $\gamma$-ray emission measured by Fermi-LAT to argue for the existence of e.g. annihilating dark matter, pulsars, or emission from other astrophysical sources. Here, we show the emission from each CR process that contributes to GR emission. We find that there is drastic galaxy-to-galaxy variation in $\gamma$-ray emission maps.

\begin{figure*}
     \centering
     \includegraphics[width=0.9\linewidth]{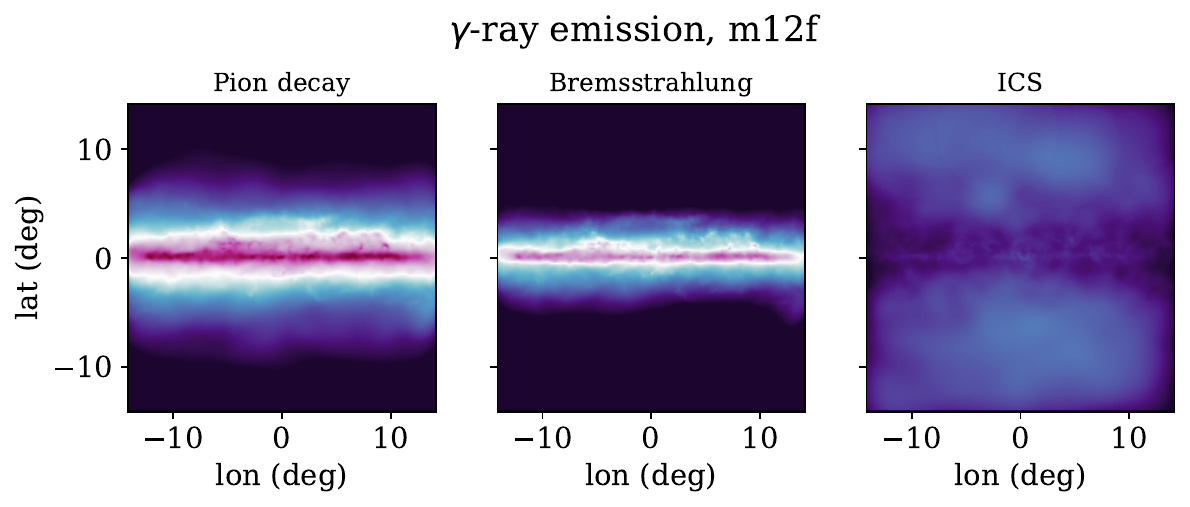}
     \includegraphics[width=0.9\linewidth]{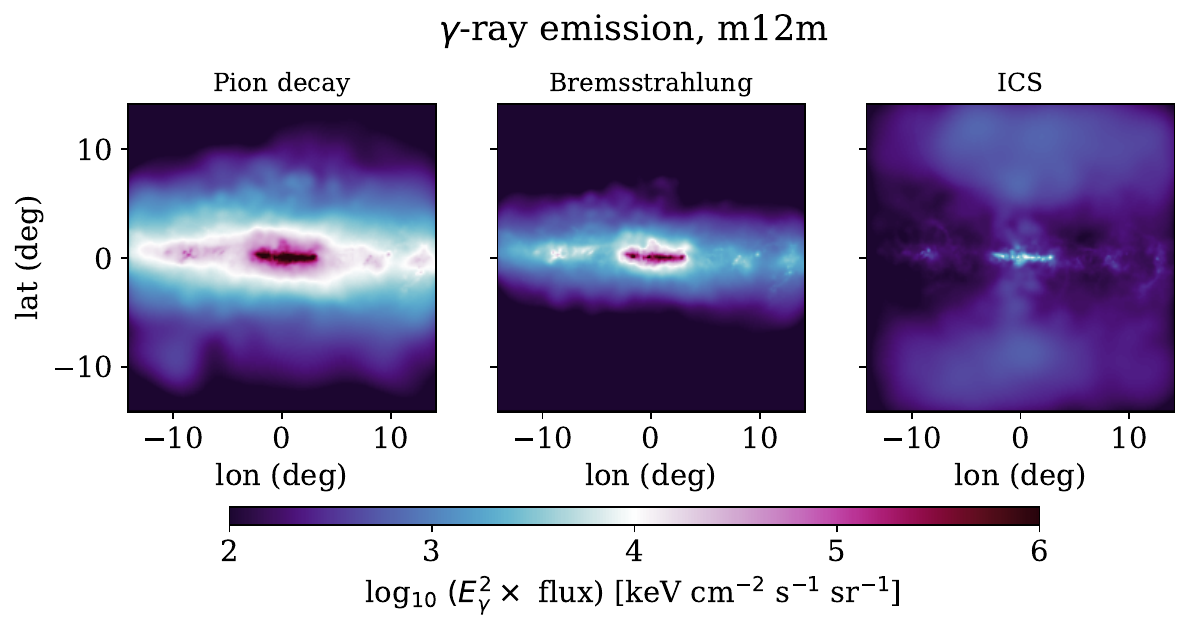}
     \caption{Maps of galactic center $\gamma$-ray emission, integrated over $E_{\gamma} = 0.1$ Gev to $100$ GeV, from neutral pion decay, brehmsstralung, and ICS for the two galaxies with more typical morphologies: \textbf{m12f} (top) and \textbf{m12m} (bottom). Neutral pion and bremsstrahlung emission are brightest within the the disk, while the ICS produces a pair of "clouds" above and below the plane of the disk. The $\gamma$-ray emission of \textbf{m12m} is rounder and more concentrated within the inner 5 degrees of the galaxy than the emission for \textbf{m12f}, which is more extended throughout the entire disk.}
     \label{fig:Gr_map_m12f_m12m}
 \end{figure*}

Two of the three galaxies in our sample, \textbf{m12f} and \textbf{m12m}, have typical geometries for spiral galaxies, with strong spiral arms and thin gas and stellar disks. Of the two, \textbf{m12f} has a thinner and more extended disk, while \textbf{m12m} has a late-forming bar and a more prominent bulge \cite{Debattista_2019}. The $\gamma$-ray emission from neutral pion decay and bremsstrahlung, when viewed by an observer 8.1 kpc from the galactic center, follows the geometry of the galaxies (Figure \ref{fig:Gr_map_m12f_m12m}). While neutral pion decay is the dominant source of $\gamma$-ray emission,  bremsstrahlung is an important contribution, particularly between latitudes of $\pm 2^{\circ}$.  $\gamma$-ray emission from inverse Compton is volume-filling, diffusing past the disk into a bi-lobed structure; although typically sub-dominant in the disk, this component dominates $\gamma$-ray emission far from the disk plane. The implications of this result for the structure and formation of the Fermi bubbles are discussed in a companion paper.

The $\gamma$-ray emission typically evolves on the scale of 10s to 100s of Myrs. Galaxies \textbf{m12f} and \textbf{m12m} both have less ordered, thicker disks shortly after the controlled restarts begin (approx. 450 Myr in lookback time). By 10 Myr in lookback time, these disks have become thinner and more ordered. Over the last 10 Myr of evolution, both galaxies have small, transient gas outflows that resolve on the scale of a few Myrs. These features lead to minor, but detectable changes in the pionic and bremsstrahlung $\gamma$-ray emission. The bi-lobed structure of the ICS emission is present throughout the evolution of the simulation, and tends to diffuse over time, unless there are strong injections of additional CR leptons (which occurs due to a late burst of star formation in \textbf{m12m}, but not in \textbf{m12f}).

\begin{figure*}
     \centering
     \includegraphics[width=0.9\linewidth]{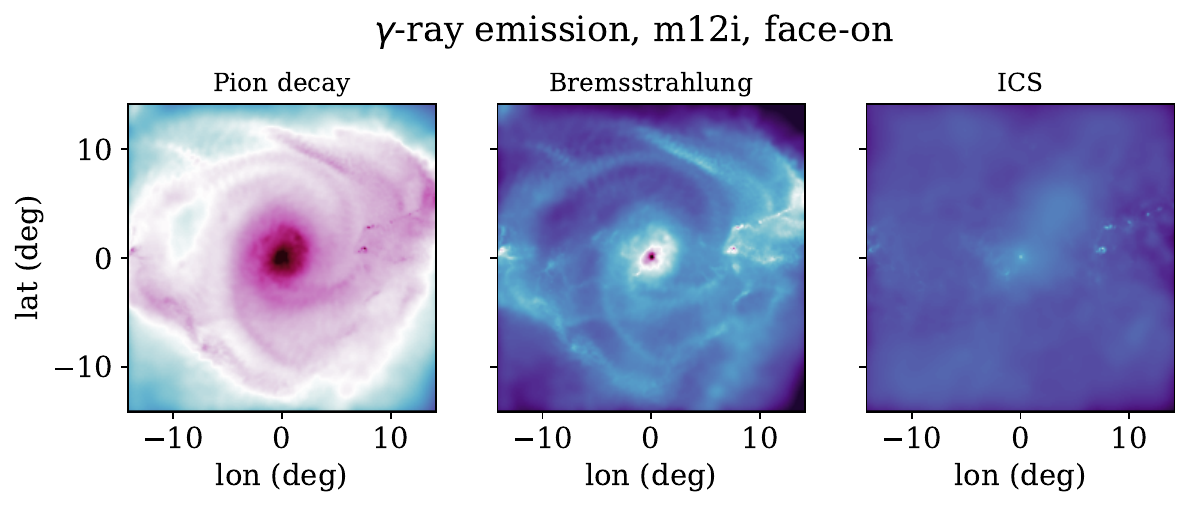}
     \includegraphics[width=0.9\linewidth]{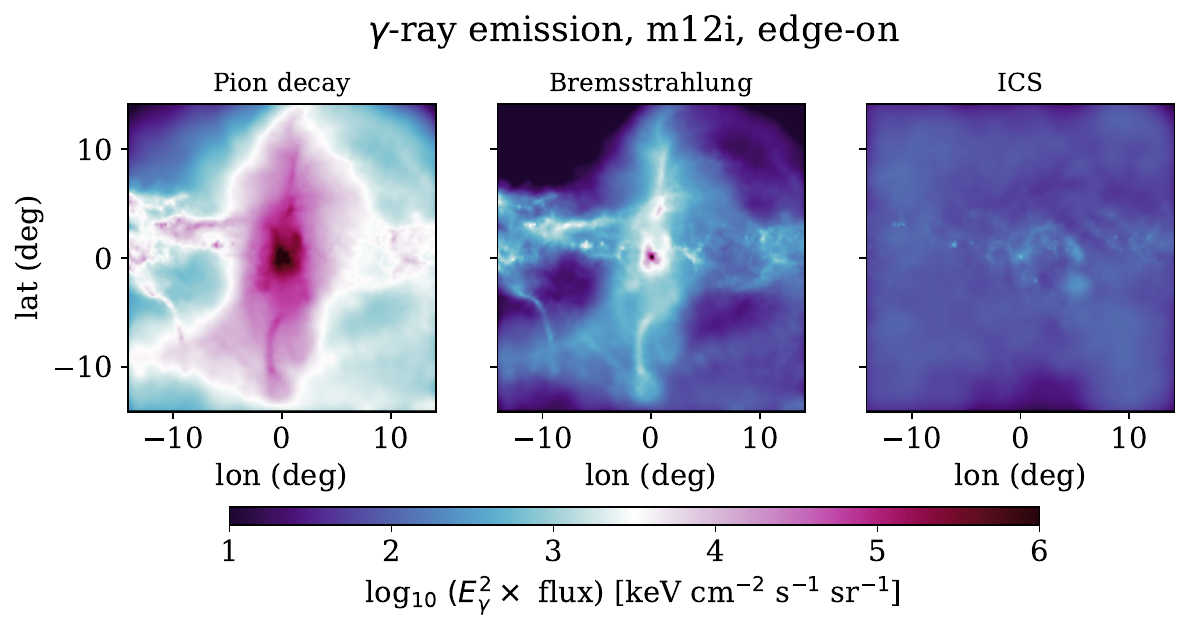}
     \caption{The same as figure \ref{fig:Gr_map_m12f_m12m}, but for galaxy \textbf{m12i}, which has a polar disk structure in which an inner disk is nearly perpendicular to the outer disk. Two different views of \textbf{m12i} are shown: one face-on (top) and the other edge-on (bottom) with the inner disk. Bright in $\gamma$-rays produced from neutral pion decay and bremsstrahlung, \textbf{m12i} is fainter in ICS $\gamma$-rays and does not show the same ICS-produced structures as \textbf{m12f} and \textbf{m12m}.}
     \label{fig:Gr_map_m12i}
 \end{figure*}

Galaxy \textbf{m12i} has a more complex geometry than either \textbf{m12f} or \textbf{m12m}: it has a slightly warped outer disk, a strong bar surrounded by a cavity with a lower SFR, and an inner disk of radius $\sim 3$ kpc that is perpendicular to the outer disk \cite{Hopkins_CR}. As a result, its $\gamma$-ray emission maps are highly dependent on azimuthal observer position and line-of-sight to the galactic center. In Figure \ref{fig:Gr_map_m12i}, we show two different views of the GC $\gamma$-ray emission for \textbf{m12i}: one face-on with the inner disk (top), and the other at an azimuthal angle of $90^{\circ}$ from the first for an edge-on view of the inner disk (bottom). 

Because the strong radiation and magnetic fields in the center of \textbf{m12i} lead to significant CR lepton losses through synchrotron and ICS, the $\gamma$-ray emission from ICS is faint compared to the the emission from pion decay. The synchrotron and ICS losses impact the high energy CR electrons more than the lower energy ($ < 1$ GeV) CR electrons. Since the low energy CRs contribute more to bremsstrahlung, there is still significant $\gamma$-ray emission from bremsstrahlung in \textbf{m12i}. The bi-lobed structure of the ICS is not as visually apparent as it is in \textbf{m12f} or \textbf{m12m}. It is present, but it is perpendicular to the inner disk and lies in the plane of the outer disk. \textbf{m12i} shows more dramatic evolution than \textbf{m12f} and \textbf{m12m} in its $\gamma$-ray emission over time, as the inner disk forms and stabilizes. 

\section{Discussion}
\label{sec:discussion}
\subsection{Factors affecting galactic center $\gamma$-ray emission}

The picture of galactic center $\gamma$-ray emission that we can gain from these simulations is highly dynamical and complex, and shaped by a multitude of correlated factors. These are, in large part, the same factors that affect CR propagation: magnetic field strength, radiation background, gas temperature and density, star formation/ supernova rates, etc. Previous studies of this set of simulations discusses these quantities and their relation to CR propagation in detail, with a focus on observables in the local ISM \cite{Hopkins_CR}. In this section, we focus on a subset of those factors that most impact $\gamma$-ray emission in the dense, highly turbulent environment of the galactic center, in which loss rates are universally higher due to high gas density and radiation and magnetic field energy densities. 

We can gain an intuitive understanding of what determines $\gamma$-ray flux from a given region at a moment in time by comparing the timescales for relevant loss rates (i.e. catastrophic pionic losses, bremsstrahlung losses, ICS and synchrotron losses) to the timescales for star formation and supernova rates to change significantly. From Fig. \ref{fig:sfr}, we see that the timescale for a starburst or drop in SFR is $\sim$ a few to 10 Myr. The loss times for the relevant $\gamma$-ray production processes are: 
\begin{equation}
    t_{\rm{\Pi_0}} \sim 0.02 \left( \frac{n}{10^4 \, \rm{cm}^{-3}}\right)^{-1} \left( \frac{E_{\rm{CR, p}}}{1 \,\rm{GeV}}\right)^{-0.28} \left(\frac{E_{\rm{CR, p}}}{200 \,\rm{GeV}} + 1 \right)^{0.2} \rm{Myr}
\end{equation}
for $\Pi_0$ decay \cite{1994A&A...286..983M}; 
\begin{equation}
     t_{\rm{Brem}} \sim 3\times10^{-3} \left( \frac{n}{10^4 \, \rm{cm}^{-3}}\right)^{-1} \frac{\ln(1 \, \rm{GeV} /E_0)}{\ln(E_{\rm{CR,e^{\pm}}}/E_0)} \,\rm{Myr}
\end{equation}
for relativistic non-thermal bremsstrahlung, where $E_0$ is the electron rest mass; and 
\begin{equation}
    t_{\rm{ICS, synch}} \sim 3 \left( \frac{E_{\rm{CR,e^{\pm}}}}{1 \,\rm{GeV}} \right)^{-1} \left( \frac{U_{\rm{rad, B}}}{100 \, \rm{eV \, cm^{-3}}} \right)^{-1} \rm{Myr}
\end{equation}
for ICS and synchrotron radiation, where $U_{\rm{rad, B}}$ refers to the radiation energy density for ICS, and magnetic field energy density for synchrotron, respectively. 

At fixed CR energy, the factors that determine these loss times are the gas number density $n$, radiation energy density $U_{\rm{rad}}$, and magnetic field energy density $U_{\rm{B}}$. Each of these three factors can vary locally in the galactic center-- number density by $\sim$ 3-5 dex, $U_{\rm{rad}}$ by $\sim$ 1-2 dex, and $U_{\rm{B}}$ by $\sim$ 2-3 dex (see Fig. \ref{fig:sim_stress}); they can also vary by order-of-magnitude amounts over 1-10 Myr timescales. Additionally, there are significant differences in the relative magnitude of these values between galaxies: for instance, $U_{\rm{B}}$ in the galactic center of $\textbf{m12i}$ is, on average, 3 dex larger than in $\textbf{m12f}$. 

Usually, these factors are correlated: the magnetic field and radiation energy densities are largest where the gas number density is highest. Thus, in an extreme environment typical of the galactic center in \textbf{m12i}, we could expect $n \sim 10^{3} \,\rm{cm}^{-3}$, $U_{\rm{rad}} \sim 10^3\, \rm{eV \,cm^{-3}}$, and $U_{\rm{B}} \sim 10^4 \,\rm{eV~cm^{-3}}$. Then the loss times for a 10 GeV CR are $t_{\Pi_0} \sim 1 \times 10^5$ years, $t_{\rm{brem}} \sim 2 \times 10^4$ years, $t_{\rm{ICS}} \sim 3 \times 10^4$ years, and $t_{\rm{synch}} \sim 3 \times 10^3$ years. Thus, in the galactic center, both CR protons and electrons at 10 GeV are in the calorimetric limit. In a less extreme GC environment, in a galaxy with a more standard morphology (e.g., \textbf{m12f}), where $n \sim 10^{2} \, \rm{cm}^{-3}$, $U_{\rm{rad}} \sim 10^2 \, \rm{eV \,cm^{-3}}$, and $U_{\rm{B}} \sim 10 \,\rm{eV \, cm^{-3}}$, ICS and bremsstrahlung loss times are $\sim 10^5$ years, compared to pionic loss time $\sim$ 1 Myr, so CR electrons can be in the calorimetric limit while CR protons are not. Notably, this is different from the local ISM and the galaxy overall, in which most CR protons and an $\mathcal{O}(1)$ function of CR leptons are expected to escape \cite{Chan_2019, Chan_2022, Hopkins_CR, Ponnada_2023}.

Considering these loss times relative to each other and SFR \& CR injection rates, we can better understand the amount of fluctuation in $\gamma$-ray emission from each galaxy in our sample as shown in Fig.~\ref{fig:sfr}. \textbf{m12i} has a nearly-constant bulk SFR in the GC in the last 300 Myr of the simulation; fluctuation in pionic $\gamma$-ray flux is driven by local variation in gas density and SFR (there is burstier star formation at earlier times, which is responsible for most of this galaxy's spread in pionic $\gamma$-ray flux). However, the strong magnetic and radiation fields, and their fluctuation over time and local variation, cause the drastic spread in $\gamma$-ray flux from bremsstrahlung and ICS. The strong CR lepton losses to ICS and synchrotron lead \textbf{m12i} to have a harder $\gamma$-ray spectrum than \textbf{m12f} and \textbf{m12m}. The latter two galaxies have burstier star formation in the galactic center; for \textbf{m12m} in particular, this leads to order-of-magnitude fluctuation in pionic $\gamma$-ray emission at late times. Both galaxies also have weaker radiation and magnetic fields than \textbf{m12i}, but it is still sufficiently strong to lead to CR lepton losses and drive variation in bremsstrahlung and ICS emission.

\subsection{Implications for Observations of the Milky Way}

While none of the galaxies in our sample are perfect Milky Way analogs, each of them are good analogs that reflect different attributes of the Milky Way. \textbf{m12i} has a warped disk, but a region of high gas density in the central few $\sim$ 100 pc similar to the Milky Way's CMZ.  \textbf{m12f} has the closest archaeological SFR to the Milky Way's estimated SFR of approximately 0.1 $\rm{M}_{\odot} \rm{yr}^{-1}$ within the galactic center \cite{Elia_2022}  \textbf{m12m} develops an x-shaped bulge at late times with some similarities to the Milky Way's \cite{Debattista_2019}. Each of these galaxy models allows us to consider which factors may be important for interpreting $\gamma$-ray observations of the Milky Way's galactic center. 

We find that local, turbulent ISM, and magnetic field strength and structure is important for determining $\gamma$-ray emission. This result is unsurprising, given that previous analysis of this set of simulations found that local structure affects CR spectra in the local ISM \cite{Hopkins_CR}. There are strong uncertainties in mapping the Milky Way's magnetic field due to the challenges of observing from within the plane of the galaxy \cite{Jaffe_2010}. While there are constraints on \textbf{B}-field strength in the local ISM, the magnetic field in the MW's galactic center is far less constrained, and many analytic models either assume a constant \textbf{B}-field or an axisymmetric one within the central few kpc \cite{Jaffe_2019}. Without strong upper bounds on the magnetic field strength in the galactic center, it is impossible to determine CR lepton loss rates and thus the $\gamma$-ray flux due to bremsstrahlung and ICS.

Additionally, the features of the ISM in the MW's galactic center can be radically different from the local ISM; in the CMZ, for instance, gas is much hotter, denser, and more turbulent \cite{BRYANT2021101630}. While there has been extensive progress in mapping the 3D structure of the CMZ , there are significant uncertainties in the placement of certain molecular clouds \cite{battersby20253dcmzicentral, walker20243dcmziiiconstraining}. Additionally, there are several sources of uncertainty in diffuse $\gamma$-ray emission that arise from the gas maps of the Milky Way used in modeling: there are potentially large error bars in the positioning of HI gas along the line-of-sight, as well as from dust and CO tracers of HII \cite{Ackermann_2017}. There has been extensive work to determine how these uncertainties propagate into measurements and modeling of the Milky Way's $\gamma$-ray emission, and to mitigate them \cite{2012ApJ...750....3A, Acero_2016}. Here, we explicitly show in Fig.~\ref{fig:spec_var_ang} that failure to remove foreground gas among the LOS can affect the observed GC $\gamma$-ray spectrum by $\mathcal{O}(1)$ factors, underpinning the need to cut down on these sources of errors. 

All three galaxies in our sample show stochastic, transient fluctuations in $\gamma$-ray emission on Myr timescales, indicating that the assumption of steady-state equilibrium used in many studies of the galactic center may be a poor approximation. Even for \textbf{m12f}, which shows the least amount of variation in $\gamma$-ray flux over time, the $\gamma$-ray spectrum amplitude and slope can vary by non-negligible $\mathcal{O}(1)$ factors. Fluctuations in pionic $\gamma$-ray flux usually traces SFR and results in a global enhancement or suppression of the $\gamma$-ray spectrum.  While there are significant uncertainties regarding star formation rates in the galactic center, time-varying pionic emission is therefore less consequential for interpreting features of the Fermi-LAT GC $\gamma$-ray spectrum such as an excess at certain energies. The variation in $\gamma$-ray flux from high-energy CR leptons, however, can have transient effects on the $\gamma$-ray spectrum for certain energy bins above 1 GeV (Fig. \ref{fig:comp_spread}). As discussed in the previous subsection, the magnitude of this effect and the energies at which it occurs are set by competing rates of star formation/cosmic ray injection, time and local variations in magnetic field strengths, and the corresponding synchrotron losses affecting the CR lepton spectra. Since this is an inherently dynamical effect, it would likely go undetected by any analysis of GC $\gamma$-ray emission that relies on fitting to static templates. However, this time-varying bremsstrahlung effect may be extremely important for interpretations of the MW's galactic center, since it is most prominent for 1-10 GeV $\gamma$-rays (corresponding to $\sim 3-30$ GeV CR electrons and positrons). Past work has argued that synchrotron-emitting CR leptons in the MW's galactic center can produce a sharply-peaked non-thermal bremsstrahlung component of diffuse $\gamma$-ray emission that can account for the GeV $\gamma$-ray excess detected by Fermi-LAT \cite{Yusef_Zadeh_2012}. We will use our simulations to explore this scenario in a companion paper. 

While our analysis shows that multiple assumptions made by models of galactic $\gamma$-ray can fail, we acknowledge that these simulations utilize a specific model of CR transport and a specific prescription for baryonic feedback, and are finite in resolution. Changing several of the modeling choices made in these simulations (e.g. the power-law CR injection spectrum, values for the fraction of ejecta kinetic energy that goes into CRs and leptons) could result in either stronger or weaker fluctuations in $\gamma$-ray emission. However, many of the implications for $\gamma$-ray observations of the Milky Way's GC are likely robust to these choices: we have considered three different realizations of Milky Way-mass spiral galaxies, with a range of e.g. gas densities and magnetic field strengths in their galactic centers. 
Additionally, many CR transport parameters are not well constrained, and we have selected values that are known to closely reproduce observed CR spectra in the local ISM, but first-principles CR transport models  can easily predict strong variation in galactic center plasma environments \cite{yan.lazarian.04:cr.scattering.fast.modes, yan.lazarian.2008:cr.propagation.with.streaming, lazarian:2016.cr.wave.damping, zweibel:cr.feedback.review, Rusz17, farber:decoupled.crs.in.neutral.gas, kempski:2020.cr.soundwave.instabilities.highbeta.plasmas.resemble.perseus.density.fluctuation.power.spectra, kempski:2023.large.amplitude.fluctuations.and.cr.scattering, kempski.li.2024:unified.cr.scattering.plasma.scattering.from.strong.field.curvature.intermittent.ism.structure.explained.together, krumholz:2020.cr.transport.starbursts.upper.limit.kappa.gamma.rays, hopkins:m1.cr.closure,ji:2021.cr.mhd.pic.dust.sims, hopkins:2021.sc.et.models.incompatible.obs, squire:2021.dust.cr.confinement.damping, butsky:2023.cosmic.ray.scattering.patchy.ism.structures, fitzaxen:2024.cr.transport.into.gmcs.suppressed.starforge, barretomota:2024.mirror.scattering.ism.crs}.

\section{Conclusion}

In this work, we have modeled $\gamma$-ray emission from CR interactions in simulations of Milky Way-mass spiral galaxies in which multi-species CR spectra are evolved alongside full MHD. This allows us, for the first time, to study in detail how the dynamical coupling of CRs to galactic magnetic fields and turbulence in the ISM affects $\gamma$-ray emission from the galactic center over time. We demonstrate that astrophysical effects on diffuse $\gamma$-ray emission can lead to large time-dependent, non-steady state equilibrium effects in the observed galactic center $\gamma$-ray spectrum: time-dependent and local variability in supernova rates, gas density, and magnetic and radiation field strengths can lead to order-of-magnitude fluctuations in the galactic center $\gamma$-ray spectrum on Myr timescales. These factors affect the normalization and shape of the $\gamma$-ray spectrum, as well as each component of the spectrum (i.e. from neutral pion decay, relativistic Bremsstrahlung, and ICS), and related quantities like the proton-to-electron ratio and positron fraction. These fluctuations over time cannot be captured by models that fit $\gamma$-ray observations to static templates, which have been utilized by most studies of the Milky Way's galactic center. 

The methods we present, in which we use relatively simple prescriptions for $\gamma$-ray production from CR interactions, while simulating a great deal of astrophysical complexity, can serve as an important complement to existing $\gamma$-ray propagation codes like GALPROP and DRAGON2, which have more sophisticated modeling of CR particle physics but necessarily make more simplistic assumptions for the astrophysical features of the Milky Way than a full cosmological simulation. While we have focused on the galactic center for this first paper, we can utilize the infrastructure for modeling $\gamma$-ray emission in live-MHD galaxy formation simulations with multi-bin, multi-species cosmic rays to study other astrophysical systems (for instance, to place bounds on $\gamma$-ray emission from dwarf spheroidal satellite galaxies, or to constrain CR transport with CGM observables). We have shown in this work that the inverse Compton scattering of CR leptons produces bi-lobed $\gamma$-ray features above and below the plane of the disk; whether this emission is sufficient to explain the Fermi bubbles will be the focus of a companion paper. We will also conduct a more detailed study of the galactic center $\gamma$-ray excess detected by Fermi-LAT in an upcoming paper. 

There are additional sources that may contribute to GC $\gamma$-ray emission that we have not modeled here, including AGN, millisecond pulsar populations, and annihilating dark matter. In the future, we intend to add these sources to our $\gamma$-ray modeling; the machinery already exists to study the former, although models for AGN cosmic ray injection (and thus $\gamma$-ray emission) are highly unconstrained \cite{Wellons_2023, ponnada2024hookslinessinkersagn, sivasankaran2024agnfeedbackisolatedgalaxies, ponnada2025timedependentcosmicrayhalos, su2025modelingcosmicraysagn}. Modeling these more exotic sources of galactic $\gamma$-rays will be essential to elucidating our understanding of the galactic center; however, it is clear that the interpretation of the GC $\gamma$-ray excess (as well as other CR and $\gamma$-ray anomalies like the positron excess and 511 keV line) requires taking into account the full range of potential variations in CR physics identified in this work. Ultimately, the order-of-magnitude fluctuations in diffuse $\gamma$-ray emission within the GC in these simulations may indicate that this region is not an ideal, ``clean'' environment in which to search for more exotic origins of $\gamma$-ray emission, such as processes related to the interactions of elusive DM particles.

\vspace{1 cm}
The numerical calculations in this paper were run on the Texas Advanced Computing Center (TACC) allocation AST21010. P.F.H. was supported by a Simons Investigator Grant.

\bibliography{apssamp}% Produces the bibliography via BibTeX.

\begin{thebibliography}{105}%
\makeatletter
\providecommand \@ifxundefined [1]{%
 \@ifx{#1\undefined}
}%
\providecommand \@ifnum [1]{%
 \ifnum #1\expandafter \@firstoftwo
 \else \expandafter \@secondoftwo
 \fi
}%
\providecommand \@ifx [1]{%
 \ifx #1\expandafter \@firstoftwo
 \else \expandafter \@secondoftwo
 \fi
}%
\providecommand \natexlab [1]{#1}%
\providecommand \enquote  [1]{``#1''}%
\providecommand \bibnamefont  [1]{#1}%
\providecommand \bibfnamefont [1]{#1}%
\providecommand \citenamefont [1]{#1}%
\providecommand \href@noop [0]{\@secondoftwo}%
\providecommand \href [0]{\begingroup \@sanitize@url \@href}%
\providecommand \@href[1]{\@@startlink{#1}\@@href}%
\providecommand \@@href[1]{\endgroup#1\@@endlink}%
\providecommand \@sanitize@url [0]{\catcode `\\12\catcode `\$12\catcode `\&12\catcode `\#12\catcode `\^12\catcode `\_12\catcode `\%12\relax}%
\providecommand \@@startlink[1]{}%
\providecommand \@@endlink[0]{}%
\providecommand \url  [0]{\begingroup\@sanitize@url \@url }%
\providecommand \@url [1]{\endgroup\@href {#1}{\urlprefix }}%
\providecommand \urlprefix  [0]{URL }%
\providecommand \Eprint [0]{\href }%
\providecommand \doibase [0]{https://doi.org/}%
\providecommand \selectlanguage [0]{\@gobble}%
\providecommand \bibinfo  [0]{\@secondoftwo}%
\providecommand \bibfield  [0]{\@secondoftwo}%
\providecommand \translation [1]{[#1]}%
\providecommand \BibitemOpen [0]{}%
\providecommand \bibitemStop [0]{}%
\providecommand \bibitemNoStop [0]{.\EOS\space}%
\providecommand \EOS [0]{\spacefactor3000\relax}%
\providecommand \BibitemShut  [1]{\csname bibitem#1\endcsname}%
\let\auto@bib@innerbib\@empty
%</preamble>
\bibitem [{\citenamefont {Strong}\ \emph {et~al.}(2000)\citenamefont {Strong}, \citenamefont {Moskalenko},\ and\ \citenamefont {Reimer}}]{Strong_2000}%
  \BibitemOpen
  \bibfield  {author} {\bibinfo {author} {\bibfnamefont {A.~W.}\ \bibnamefont {Strong}}, \bibinfo {author} {\bibfnamefont {I.~V.}\ \bibnamefont {Moskalenko}},\ and\ \bibinfo {author} {\bibfnamefont {O.}~\bibnamefont {Reimer}},\ }\bibfield  {title} {\bibinfo {title} {Diffuse continuum gamma rays from the galaxy},\ }\href {https://doi.org/10.1086/309038} {\bibfield  {journal} {\bibinfo  {journal} {The Astrophysical Journal}\ }\textbf {\bibinfo {volume} {537}},\ \bibinfo {pages} {763–784} (\bibinfo {year} {2000})}\BibitemShut {NoStop}%
\bibitem [{\citenamefont {BLUMENTHAL}\ and\ \citenamefont {GOULD}(1970)}]{BlumenthalRevModPhys.42.237}%
  \BibitemOpen
  \bibfield  {author} {\bibinfo {author} {\bibfnamefont {G.~R.}\ \bibnamefont {BLUMENTHAL}}\ and\ \bibinfo {author} {\bibfnamefont {R.~J.}\ \bibnamefont {GOULD}},\ }\bibfield  {title} {\bibinfo {title} {Bremsstrahlung, synchrotron radiation, and compton scattering of high-energy electrons traversing dilute gases},\ }\href {https://doi.org/10.1103/RevModPhys.42.237} {\bibfield  {journal} {\bibinfo  {journal} {Rev. Mod. Phys.}\ }\textbf {\bibinfo {volume} {42}},\ \bibinfo {pages} {237} (\bibinfo {year} {1970})}\BibitemShut {NoStop}%
\bibitem [{\citenamefont {{Mannheim}}\ and\ \citenamefont {{Schlickeiser}}(1994)}]{1994A&A...286..983M}%
  \BibitemOpen
  \bibfield  {author} {\bibinfo {author} {\bibfnamefont {K.}~\bibnamefont {{Mannheim}}}\ and\ \bibinfo {author} {\bibfnamefont {R.}~\bibnamefont {{Schlickeiser}}},\ }\bibfield  {title} {\bibinfo {title} {{Interactions of cosmic ray nuclei}},\ }\href {https://doi.org/10.48550/arXiv.astro-ph/9402042} {\bibfield  {journal} {\bibinfo  {journal} {Astronomy \& Astrophysics}\ }\textbf {\bibinfo {volume} {286}},\ \bibinfo {pages} {983} (\bibinfo {year} {1994})},\ \Eprint {https://arxiv.org/abs/astro-ph/9402042} {arXiv:astro-ph/9402042 [astro-ph]} \BibitemShut {NoStop}%
\bibitem [{\citenamefont {Ackermann}\ \emph {et~al.}(2017)\citenamefont {Ackermann} \emph {et~al.}}]{Ackermann_2017}%
  \BibitemOpen
  \bibfield  {author} {\bibinfo {author} {\bibfnamefont {M.}~\bibnamefont {Ackermann}} \emph {et~al.},\ }\bibfield  {title} {\bibinfo {title} {The fermi galactic center gev excess and implications for dark matter},\ }\href {https://doi.org/10.3847/1538-4357/aa6cab} {\bibfield  {journal} {\bibinfo  {journal} {The Astrophysical Journal}\ }\textbf {\bibinfo {volume} {840}},\ \bibinfo {pages} {43} (\bibinfo {year} {2017})}\BibitemShut {NoStop}%
\bibitem [{\citenamefont {Genolini}\ \emph {et~al.}(2015)\citenamefont {Genolini}, \citenamefont {Putze}, \citenamefont {Salati},\ and\ \citenamefont {Serpico}}]{Genolini_2015}%
  \BibitemOpen
  \bibfield  {author} {\bibinfo {author} {\bibfnamefont {Y.}~\bibnamefont {Genolini}}, \bibinfo {author} {\bibfnamefont {A.}~\bibnamefont {Putze}}, \bibinfo {author} {\bibfnamefont {P.}~\bibnamefont {Salati}},\ and\ \bibinfo {author} {\bibfnamefont {P.~D.}\ \bibnamefont {Serpico}},\ }\bibfield  {title} {\bibinfo {title} {Theoretical uncertainties in extracting cosmic-ray diffusion parameters: the boron-to-carbon ratio},\ }\href {https://doi.org/10.1051/0004-6361/201526344} {\bibfield  {journal} {\bibinfo  {journal} {Astronomy \& Astrophysics}\ }\textbf {\bibinfo {volume} {580}},\ \bibinfo {pages} {A9} (\bibinfo {year} {2015})}\BibitemShut {NoStop}%
\bibitem [{\citenamefont {Amato}\ and\ \citenamefont {Blasi}(2018)}]{Amato_2018}%
  \BibitemOpen
  \bibfield  {author} {\bibinfo {author} {\bibfnamefont {E.}~\bibnamefont {Amato}}\ and\ \bibinfo {author} {\bibfnamefont {P.}~\bibnamefont {Blasi}},\ }\bibfield  {title} {\bibinfo {title} {Cosmic ray transport in the galaxy: A review},\ }\href {https://doi.org/10.1016/j.asr.2017.04.019} {\bibfield  {journal} {\bibinfo  {journal} {Advances in Space Research}\ }\textbf {\bibinfo {volume} {62}},\ \bibinfo {pages} {2731–2749} (\bibinfo {year} {2018})}\BibitemShut {NoStop}%
\bibitem [{\citenamefont {Bryant}\ and\ \citenamefont {Krabbe}(2021)}]{BRYANT2021101630}%
  \BibitemOpen
  \bibfield  {author} {\bibinfo {author} {\bibfnamefont {A.}~\bibnamefont {Bryant}}\ and\ \bibinfo {author} {\bibfnamefont {A.}~\bibnamefont {Krabbe}},\ }\bibfield  {title} {\bibinfo {title} {The episodic and multiscale galactic centre},\ }\href {https://doi.org/https://doi.org/10.1016/j.newar.2021.101630} {\bibfield  {journal} {\bibinfo  {journal} {New Astronomy Reviews}\ }\textbf {\bibinfo {volume} {93}},\ \bibinfo {pages} {101630} (\bibinfo {year} {2021})}\BibitemShut {NoStop}%
\bibitem [{\citenamefont {Figer}\ \emph {et~al.}(2004)\citenamefont {Figer}, \citenamefont {Rich}, \citenamefont {Kim}, \citenamefont {Morris},\ and\ \citenamefont {Serabyn}}]{Figer_2004}%
  \BibitemOpen
  \bibfield  {author} {\bibinfo {author} {\bibfnamefont {D.~F.}\ \bibnamefont {Figer}}, \bibinfo {author} {\bibfnamefont {R.~M.}\ \bibnamefont {Rich}}, \bibinfo {author} {\bibfnamefont {S.~S.}\ \bibnamefont {Kim}}, \bibinfo {author} {\bibfnamefont {M.}~\bibnamefont {Morris}},\ and\ \bibinfo {author} {\bibfnamefont {E.}~\bibnamefont {Serabyn}},\ }\bibfield  {title} {\bibinfo {title} {An extended star formation history for the galactic center fromhubble space telescopenicmos observations},\ }\href {https://doi.org/10.1086/380392} {\bibfield  {journal} {\bibinfo  {journal} {The Astrophysical Journal}\ }\textbf {\bibinfo {volume} {601}},\ \bibinfo {pages} {319–339} (\bibinfo {year} {2004})}\BibitemShut {NoStop}%
\bibitem [{\citenamefont {{Nogueras-Lara}}\ \emph {et~al.}(2020)\citenamefont {{Nogueras-Lara}}, \citenamefont {{Sch{\"o}del}}, \citenamefont {{Gallego-Calvente}}, \citenamefont {{Gallego-Cano}}, \citenamefont {{Shahzamanian}}, \citenamefont {{Dong}}, \citenamefont {{Neumayer}}, \citenamefont {{Hilker}}, \citenamefont {{Najarro}}, \citenamefont {{Nishiyama}}, \citenamefont {{Feldmeier-Krause}}, \citenamefont {{Girard}},\ and\ \citenamefont {{Cassisi}}}]{2020NatAs...4..377N}%
  \BibitemOpen
  \bibfield  {author} {\bibinfo {author} {\bibfnamefont {F.}~\bibnamefont {{Nogueras-Lara}}}, \bibinfo {author} {\bibfnamefont {R.}~\bibnamefont {{Sch{\"o}del}}}, \bibinfo {author} {\bibfnamefont {A.~T.}\ \bibnamefont {{Gallego-Calvente}}}, \bibinfo {author} {\bibfnamefont {E.}~\bibnamefont {{Gallego-Cano}}}, \bibinfo {author} {\bibfnamefont {B.}~\bibnamefont {{Shahzamanian}}}, \bibinfo {author} {\bibfnamefont {H.}~\bibnamefont {{Dong}}}, \bibinfo {author} {\bibfnamefont {N.}~\bibnamefont {{Neumayer}}}, \bibinfo {author} {\bibfnamefont {M.}~\bibnamefont {{Hilker}}}, \bibinfo {author} {\bibfnamefont {F.}~\bibnamefont {{Najarro}}}, \bibinfo {author} {\bibfnamefont {S.}~\bibnamefont {{Nishiyama}}}, \bibinfo {author} {\bibfnamefont {A.}~\bibnamefont {{Feldmeier-Krause}}}, \bibinfo {author} {\bibfnamefont {J.~H.~V.}\ \bibnamefont {{Girard}}},\ and\ \bibinfo {author} {\bibfnamefont {S.}~\bibnamefont {{Cassisi}}},\ }\bibfield  {title} {\bibinfo {title} {{Early formation and recent starburst activity in the nuclear
  disk of the Milky Way}},\ }\href {https://doi.org/10.1038/s41550-019-0967-9} {\bibfield  {journal} {\bibinfo  {journal} {Nature Astronomy}\ }\textbf {\bibinfo {volume} {4}},\ \bibinfo {pages} {377} (\bibinfo {year} {2020})},\ \Eprint {https://arxiv.org/abs/1910.06968} {arXiv:1910.06968 [astro-ph.GA]} \BibitemShut {NoStop}%
\bibitem [{\citenamefont {Nogueras-Lara}\ \emph {et~al.}(2020)\citenamefont {Nogueras-Lara}, \citenamefont {Schödel}, \citenamefont {Neumayer}, \citenamefont {Gallego-Cano}, \citenamefont {Shahzamanian}, \citenamefont {Gallego-Calvente},\ and\ \citenamefont {Najarro}}]{Nogueras_Lara_2020}%
  \BibitemOpen
  \bibfield  {author} {\bibinfo {author} {\bibfnamefont {F.}~\bibnamefont {Nogueras-Lara}}, \bibinfo {author} {\bibfnamefont {R.}~\bibnamefont {Schödel}}, \bibinfo {author} {\bibfnamefont {N.}~\bibnamefont {Neumayer}}, \bibinfo {author} {\bibfnamefont {E.}~\bibnamefont {Gallego-Cano}}, \bibinfo {author} {\bibfnamefont {B.}~\bibnamefont {Shahzamanian}}, \bibinfo {author} {\bibfnamefont {A.~T.}\ \bibnamefont {Gallego-Calvente}},\ and\ \bibinfo {author} {\bibfnamefont {F.}~\bibnamefont {Najarro}},\ }\bibfield  {title} {\bibinfo {title} {Galacticnucleus: A high angular-resolution jhks imaging survey of the galactic centre: Iii. evidence for wavelength-dependence of the extinction curve in the near-infrared},\ }\href {https://doi.org/10.1051/0004-6361/202038606} {\bibfield  {journal} {\bibinfo  {journal} {Astronomy \& Astrophysics}\ }\textbf {\bibinfo {volume} {641}},\ \bibinfo {pages} {A141} (\bibinfo {year} {2020})}\BibitemShut {NoStop}%
\bibitem [{\citenamefont {{Genzel}}\ \emph {et~al.}(2010)\citenamefont {{Genzel}}, \citenamefont {{Eisenhauer}},\ and\ \citenamefont {{Gillessen}}}]{2010RvMP...82.3121G}%
  \BibitemOpen
  \bibfield  {author} {\bibinfo {author} {\bibfnamefont {R.}~\bibnamefont {{Genzel}}}, \bibinfo {author} {\bibfnamefont {F.}~\bibnamefont {{Eisenhauer}}},\ and\ \bibinfo {author} {\bibfnamefont {S.}~\bibnamefont {{Gillessen}}},\ }\bibfield  {title} {\bibinfo {title} {{The Galactic Center massive black hole and nuclear star cluster}},\ }\href {https://doi.org/10.1103/RevModPhys.82.3121} {\bibfield  {journal} {\bibinfo  {journal} {Reviews of Modern Physics}\ }\textbf {\bibinfo {volume} {82}},\ \bibinfo {pages} {3121} (\bibinfo {year} {2010})},\ \Eprint {https://arxiv.org/abs/1006.0064} {arXiv:1006.0064 [astro-ph.GA]} \BibitemShut {NoStop}%
\bibitem [{\citenamefont {Chan}\ \emph {et~al.}(2019)\citenamefont {Chan}, \citenamefont {Kereš}, \citenamefont {Hopkins}, \citenamefont {Quataert}, \citenamefont {Su}, \citenamefont {Hayward},\ and\ \citenamefont {Faucher-Giguère}}]{Chan_2019}%
  \BibitemOpen
  \bibfield  {author} {\bibinfo {author} {\bibfnamefont {T.~K.}\ \bibnamefont {Chan}}, \bibinfo {author} {\bibfnamefont {D.}~\bibnamefont {Kereš}}, \bibinfo {author} {\bibfnamefont {P.~F.}\ \bibnamefont {Hopkins}}, \bibinfo {author} {\bibfnamefont {E.}~\bibnamefont {Quataert}}, \bibinfo {author} {\bibfnamefont {K.-Y.}\ \bibnamefont {Su}}, \bibinfo {author} {\bibfnamefont {C.~C.}\ \bibnamefont {Hayward}},\ and\ \bibinfo {author} {\bibfnamefont {C.-A.}\ \bibnamefont {Faucher-Giguère}},\ }\bibfield  {title} {\bibinfo {title} {Cosmic ray feedback in the fire simulations: constraining cosmic ray propagation with gev gamma-ray emission},\ }\href {https://doi.org/10.1093/mnras/stz1895} {\bibfield  {journal} {\bibinfo  {journal} {Monthly Notices of the Royal Astronomical Society}\ }\textbf {\bibinfo {volume} {488}},\ \bibinfo {pages} {3716–3744} (\bibinfo {year} {2019})}\BibitemShut {NoStop}%
\bibitem [{\citenamefont {Hopkins}\ \emph {et~al.}(2019)\citenamefont {Hopkins}, \citenamefont {Chan}, \citenamefont {Garrison-Kimmel}, \citenamefont {Ji}, \citenamefont {Su}, \citenamefont {Hummels}, \citenamefont {Kereš}, \citenamefont {Quataert},\ and\ \citenamefont {Faucher-Giguère}}]{Hopkins_2019}%
  \BibitemOpen
  \bibfield  {author} {\bibinfo {author} {\bibfnamefont {P.~F.}\ \bibnamefont {Hopkins}}, \bibinfo {author} {\bibfnamefont {T.~K.}\ \bibnamefont {Chan}}, \bibinfo {author} {\bibfnamefont {S.}~\bibnamefont {Garrison-Kimmel}}, \bibinfo {author} {\bibfnamefont {S.}~\bibnamefont {Ji}}, \bibinfo {author} {\bibfnamefont {K.-Y.}\ \bibnamefont {Su}}, \bibinfo {author} {\bibfnamefont {C.~B.}\ \bibnamefont {Hummels}}, \bibinfo {author} {\bibfnamefont {D.}~\bibnamefont {Kereš}}, \bibinfo {author} {\bibfnamefont {E.}~\bibnamefont {Quataert}},\ and\ \bibinfo {author} {\bibfnamefont {C.-A.}\ \bibnamefont {Faucher-Giguère}},\ }\bibfield  {title} {\bibinfo {title} {But what about...: cosmic rays, magnetic fields, conduction, and viscosity in galaxy formation},\ }\href {https://doi.org/10.1093/mnras/stz3321} {\bibfield  {journal} {\bibinfo  {journal} {Monthly Notices of the Royal Astronomical Society}\ }\textbf {\bibinfo {volume} {492}},\ \bibinfo {pages} {3465–3498} (\bibinfo {year} {2019})}\BibitemShut {NoStop}%
\bibitem [{\citenamefont {Hopkins}\ \emph {et~al.}(2020)\citenamefont {Hopkins}, \citenamefont {Squire}, \citenamefont {Chan}, \citenamefont {Quataert}, \citenamefont {Ji}, \citenamefont {Kereš},\ and\ \citenamefont {Faucher-Giguère}}]{Hopkins_2020}%
  \BibitemOpen
  \bibfield  {author} {\bibinfo {author} {\bibfnamefont {P.~F.}\ \bibnamefont {Hopkins}}, \bibinfo {author} {\bibfnamefont {J.}~\bibnamefont {Squire}}, \bibinfo {author} {\bibfnamefont {T.~K.}\ \bibnamefont {Chan}}, \bibinfo {author} {\bibfnamefont {E.}~\bibnamefont {Quataert}}, \bibinfo {author} {\bibfnamefont {S.}~\bibnamefont {Ji}}, \bibinfo {author} {\bibfnamefont {D.}~\bibnamefont {Kereš}},\ and\ \bibinfo {author} {\bibfnamefont {C.-A.}\ \bibnamefont {Faucher-Giguère}},\ }\bibfield  {title} {\bibinfo {title} {Testing physical models for cosmic ray transport coefficients on galactic scales: self-confinement and extrinsic turbulence at gev energies},\ }\href {https://doi.org/10.1093/mnras/staa3691} {\bibfield  {journal} {\bibinfo  {journal} {Monthly Notices of the Royal Astronomical Society}\ }\textbf {\bibinfo {volume} {501}},\ \bibinfo {pages} {4184–4213} (\bibinfo {year} {2020})}\BibitemShut {NoStop}%
\bibitem [{\citenamefont {Chan}\ \emph {et~al.}(2022)\citenamefont {Chan}, \citenamefont {Kereš}, \citenamefont {Gurvich}, \citenamefont {Hopkins}, \citenamefont {Trapp}, \citenamefont {Ji},\ and\ \citenamefont {Faucher-Giguère}}]{Chan_2022}%
  \BibitemOpen
  \bibfield  {author} {\bibinfo {author} {\bibfnamefont {T.~K.}\ \bibnamefont {Chan}}, \bibinfo {author} {\bibfnamefont {D.}~\bibnamefont {Kereš}}, \bibinfo {author} {\bibfnamefont {A.~B.}\ \bibnamefont {Gurvich}}, \bibinfo {author} {\bibfnamefont {P.~F.}\ \bibnamefont {Hopkins}}, \bibinfo {author} {\bibfnamefont {C.}~\bibnamefont {Trapp}}, \bibinfo {author} {\bibfnamefont {S.}~\bibnamefont {Ji}},\ and\ \bibinfo {author} {\bibfnamefont {C.-A.}\ \bibnamefont {Faucher-Giguère}},\ }\bibfield  {title} {\bibinfo {title} {The impact of cosmic rays on dynamical balance and disc–halo interaction in l\* disc galaxies},\ }\href {https://doi.org/10.1093/mnras/stac2236} {\bibfield  {journal} {\bibinfo  {journal} {Monthly Notices of the Royal Astronomical Society}\ }\textbf {\bibinfo {volume} {517}},\ \bibinfo {pages} {597–615} (\bibinfo {year} {2022})}\BibitemShut {NoStop}%
\bibitem [{\citenamefont {Lee}\ \emph {et~al.}(2016)\citenamefont {Lee}, \citenamefont {Lisanti}, \citenamefont {Safdi}, \citenamefont {Slatyer},\ and\ \citenamefont {Xue}}]{Lee_2016}%
  \BibitemOpen
  \bibfield  {author} {\bibinfo {author} {\bibfnamefont {S.~K.}\ \bibnamefont {Lee}}, \bibinfo {author} {\bibfnamefont {M.}~\bibnamefont {Lisanti}}, \bibinfo {author} {\bibfnamefont {B.~R.}\ \bibnamefont {Safdi}}, \bibinfo {author} {\bibfnamefont {T.~R.}\ \bibnamefont {Slatyer}},\ and\ \bibinfo {author} {\bibfnamefont {W.}~\bibnamefont {Xue}},\ }\bibfield  {title} {\bibinfo {title} {Evidence for unresolved gamma-ray point sources in the inner galaxy},\ }\bibfield  {journal} {\bibinfo  {journal} {Physical Review Letters}\ }\textbf {\bibinfo {volume} {116}},\ \href {https://doi.org/10.1103/physrevlett.116.051103} {10.1103/physrevlett.116.051103} (\bibinfo {year} {2016})\BibitemShut {NoStop}%
\bibitem [{\citenamefont {Bartels}\ \emph {et~al.}(2016)\citenamefont {Bartels}, \citenamefont {Krishnamurthy},\ and\ \citenamefont {Weniger}}]{Bartels_2016}%
  \BibitemOpen
  \bibfield  {author} {\bibinfo {author} {\bibfnamefont {R.}~\bibnamefont {Bartels}}, \bibinfo {author} {\bibfnamefont {S.}~\bibnamefont {Krishnamurthy}},\ and\ \bibinfo {author} {\bibfnamefont {C.}~\bibnamefont {Weniger}},\ }\bibfield  {title} {\bibinfo {title} {Strong support for the millisecond pulsar origin of the galactic center gev excess},\ }\bibfield  {journal} {\bibinfo  {journal} {Physical Review Letters}\ }\textbf {\bibinfo {volume} {116}},\ \href {https://doi.org/10.1103/physrevlett.116.051102} {10.1103/physrevlett.116.051102} (\bibinfo {year} {2016})\BibitemShut {NoStop}%
\bibitem [{\citenamefont {Hooper}\ and\ \citenamefont {Goodenough}(2011)}]{Hooper_2011}%
  \BibitemOpen
  \bibfield  {author} {\bibinfo {author} {\bibfnamefont {D.}~\bibnamefont {Hooper}}\ and\ \bibinfo {author} {\bibfnamefont {L.}~\bibnamefont {Goodenough}},\ }\bibfield  {title} {\bibinfo {title} {Dark matter annihilation in the galactic center as seen by the fermi gamma ray space telescope},\ }\href {https://doi.org/10.1016/j.physletb.2011.02.029} {\bibfield  {journal} {\bibinfo  {journal} {Physics Letters B}\ }\textbf {\bibinfo {volume} {697}},\ \bibinfo {pages} {412–428} (\bibinfo {year} {2011})}\BibitemShut {NoStop}%
\bibitem [{\citenamefont {{Daylan}}\ \emph {et~al.}(2016)\citenamefont {{Daylan}}, \citenamefont {{Finkbeiner}}, \citenamefont {{Hooper}}, \citenamefont {{Linden}}, \citenamefont {{Portillo}}, \citenamefont {{Rodd}},\ and\ \citenamefont {{Slatyer}}}]{2016PDU....12....1D}%
  \BibitemOpen
  \bibfield  {author} {\bibinfo {author} {\bibfnamefont {T.}~\bibnamefont {{Daylan}}}, \bibinfo {author} {\bibfnamefont {D.~P.}\ \bibnamefont {{Finkbeiner}}}, \bibinfo {author} {\bibfnamefont {D.}~\bibnamefont {{Hooper}}}, \bibinfo {author} {\bibfnamefont {T.}~\bibnamefont {{Linden}}}, \bibinfo {author} {\bibfnamefont {S.~K.~N.}\ \bibnamefont {{Portillo}}}, \bibinfo {author} {\bibfnamefont {N.~L.}\ \bibnamefont {{Rodd}}},\ and\ \bibinfo {author} {\bibfnamefont {T.~R.}\ \bibnamefont {{Slatyer}}},\ }\bibfield  {title} {\bibinfo {title} {{The characterization of the gamma-ray signal from the central Milky Way: A case for annihilating dark matter}},\ }\href {https://doi.org/10.1016/j.dark.2015.12.005} {\bibfield  {journal} {\bibinfo  {journal} {Physics of the Dark Universe}\ }\textbf {\bibinfo {volume} {12}},\ \bibinfo {pages} {1} (\bibinfo {year} {2016})},\ \Eprint {https://arxiv.org/abs/1402.6703} {arXiv:1402.6703 [astro-ph.HE]} \BibitemShut {NoStop}%
\bibitem [{\citenamefont {Hooper}(2023)}]{Hooper:2022bec}%
  \BibitemOpen
  \bibfield  {author} {\bibinfo {author} {\bibfnamefont {D.}~\bibnamefont {Hooper}},\ }\bibfield  {title} {\bibinfo {title} {{The status of the galactic center gamma-ray excess}},\ }\href {https://doi.org/10.21468/SciPostPhysProc.12.006} {\bibfield  {journal} {\bibinfo  {journal} {SciPost Phys. Proc.}\ }\textbf {\bibinfo {volume} {12}},\ \bibinfo {pages} {006} (\bibinfo {year} {2023})},\ \Eprint {https://arxiv.org/abs/2209.14370} {arXiv:2209.14370 [astro-ph.HE]} \BibitemShut {NoStop}%
\bibitem [{\citenamefont {Carlson}\ and\ \citenamefont {Profumo}(2014)}]{Carlson_2014}%
  \BibitemOpen
  \bibfield  {author} {\bibinfo {author} {\bibfnamefont {E.}~\bibnamefont {Carlson}}\ and\ \bibinfo {author} {\bibfnamefont {S.}~\bibnamefont {Profumo}},\ }\bibfield  {title} {\bibinfo {title} {Cosmic ray protons in the inner galaxy and the galactic center gamma-ray excess},\ }\bibfield  {journal} {\bibinfo  {journal} {Physical Review D}\ }\textbf {\bibinfo {volume} {90}},\ \href {https://doi.org/10.1103/physrevd.90.023015} {10.1103/physrevd.90.023015} (\bibinfo {year} {2014})\BibitemShut {NoStop}%
\bibitem [{\citenamefont {Petrović}\ \emph {et~al.}(2014)\citenamefont {Petrović}, \citenamefont {Serpico},\ and\ \citenamefont {Zaharijaš}}]{Petrovi__2014}%
  \BibitemOpen
  \bibfield  {author} {\bibinfo {author} {\bibfnamefont {J.}~\bibnamefont {Petrović}}, \bibinfo {author} {\bibfnamefont {P.~D.}\ \bibnamefont {Serpico}},\ and\ \bibinfo {author} {\bibfnamefont {G.}~\bibnamefont {Zaharijaš}},\ }\bibfield  {title} {\bibinfo {title} {Galactic center gamma-ray ``excess’’ from an active past of the galactic centre?},\ }\href {https://doi.org/10.1088/1475-7516/2014/10/052} {\bibfield  {journal} {\bibinfo  {journal} {Journal of Cosmology and Astroparticle Physics}\ }\textbf {\bibinfo {volume} {2014}}\bibinfo  {number} { (10)},\ \bibinfo {pages} {052–052}}\BibitemShut {NoStop}%
\bibitem [{\citenamefont {Gaggero}\ \emph {et~al.}(2015)\citenamefont {Gaggero}, \citenamefont {Taoso}, \citenamefont {Urbano}, \citenamefont {Valli},\ and\ \citenamefont {Ullio}}]{Gaggero_2015}%
  \BibitemOpen
\bibfield  {number} {  }\bibfield  {author} {\bibinfo {author} {\bibfnamefont {D.}~\bibnamefont {Gaggero}}, \bibinfo {author} {\bibfnamefont {M.}~\bibnamefont {Taoso}}, \bibinfo {author} {\bibfnamefont {A.}~\bibnamefont {Urbano}}, \bibinfo {author} {\bibfnamefont {M.}~\bibnamefont {Valli}},\ and\ \bibinfo {author} {\bibfnamefont {P.}~\bibnamefont {Ullio}},\ }\bibfield  {title} {\bibinfo {title} {Towards a realistic astrophysical interpretation of the gamma-ray galactic center excess},\ }\href {https://doi.org/10.1088/1475-7516/2015/12/056} {\bibfield  {journal} {\bibinfo  {journal} {Journal of Cosmology and Astroparticle Physics}\ }\textbf {\bibinfo {volume} {2015}}\bibinfo  {number} { (12)},\ \bibinfo {pages} {056–056}}\BibitemShut {NoStop}%
\bibitem [{\citenamefont {Su}\ \emph {et~al.}(2010)\citenamefont {Su}, \citenamefont {Slatyer},\ and\ \citenamefont {Finkbeiner}}]{Su_2010}%
  \BibitemOpen
\bibfield  {number} {  }\bibfield  {author} {\bibinfo {author} {\bibfnamefont {M.}~\bibnamefont {Su}}, \bibinfo {author} {\bibfnamefont {T.~R.}\ \bibnamefont {Slatyer}},\ and\ \bibinfo {author} {\bibfnamefont {D.~P.}\ \bibnamefont {Finkbeiner}},\ }\bibfield  {title} {\bibinfo {title} {Giant gamma-ray bubbles fromfermi-lat: Active galactic nucleus activity or bipolar galactic wind?},\ }\href {https://doi.org/10.1088/0004-637x/724/2/1044} {\bibfield  {journal} {\bibinfo  {journal} {The Astrophysical Journal}\ }\textbf {\bibinfo {volume} {724}},\ \bibinfo {pages} {1044–1082} (\bibinfo {year} {2010})}\BibitemShut {NoStop}%
\bibitem [{\citenamefont {Sarkar}(2024)}]{sarkar2024fermierositabubbleslooknuclear}%
  \BibitemOpen
  \bibfield  {author} {\bibinfo {author} {\bibfnamefont {K.~C.}\ \bibnamefont {Sarkar}},\ }\href {https://arxiv.org/abs/2403.09824} {\bibinfo {title} {The fermi/erosita bubbles: A look into the nuclear outflow from the milky way}} (\bibinfo {year} {2024}),\ \Eprint {https://arxiv.org/abs/2403.09824} {arXiv:2403.09824 [astro-ph.HE]} \BibitemShut {NoStop}%
\bibitem [{\citenamefont {Booth}\ \emph {et~al.}(2013)\citenamefont {Booth}, \citenamefont {Agertz}, \citenamefont {Kravtsov},\ and\ \citenamefont {Gnedin}}]{Booth_2013}%
  \BibitemOpen
  \bibfield  {author} {\bibinfo {author} {\bibfnamefont {C.~M.}\ \bibnamefont {Booth}}, \bibinfo {author} {\bibfnamefont {O.}~\bibnamefont {Agertz}}, \bibinfo {author} {\bibfnamefont {A.~V.}\ \bibnamefont {Kravtsov}},\ and\ \bibinfo {author} {\bibfnamefont {N.~Y.}\ \bibnamefont {Gnedin}},\ }\bibfield  {title} {\bibinfo {title} {Simulations of disk galaxies with cosmic ray driven galactic winds},\ }\href {https://doi.org/10.1088/2041-8205/777/1/l16} {\bibfield  {journal} {\bibinfo  {journal} {The Astrophysical Journal}\ }\textbf {\bibinfo {volume} {777}},\ \bibinfo {pages} {L16} (\bibinfo {year} {2013})}\BibitemShut {NoStop}%
\bibitem [{\citenamefont {Salem}\ and\ \citenamefont {Bryan}(2013)}]{Salem_2013}%
  \BibitemOpen
  \bibfield  {author} {\bibinfo {author} {\bibfnamefont {M.}~\bibnamefont {Salem}}\ and\ \bibinfo {author} {\bibfnamefont {G.~L.}\ \bibnamefont {Bryan}},\ }\bibfield  {title} {\bibinfo {title} {Cosmic ray driven outflows in global galaxy disc models},\ }\href {https://doi.org/10.1093/mnras/stt2121} {\bibfield  {journal} {\bibinfo  {journal} {Monthly Notices of the Royal Astronomical Society}\ }\textbf {\bibinfo {volume} {437}},\ \bibinfo {pages} {3312–3330} (\bibinfo {year} {2013})}\BibitemShut {NoStop}%
\bibitem [{\citenamefont {{Girichidis}}\ \emph {et~al.}(2016)\citenamefont {{Girichidis}}, \citenamefont {{Naab}}, \citenamefont {{Walch}}, \citenamefont {{Hanasz}}, \citenamefont {{Mac Low}}, \citenamefont {{Ostriker}}, \citenamefont {{Gatto}}, \citenamefont {{Peters}}, \citenamefont {{W{\"u}nsch}}, \citenamefont {{Glover}}, \citenamefont {{Klessen}}, \citenamefont {{Clark}},\ and\ \citenamefont {{Baczynski}}}]{2016ApJ...816L..19G}%
  \BibitemOpen
  \bibfield  {author} {\bibinfo {author} {\bibfnamefont {P.}~\bibnamefont {{Girichidis}}}, \bibinfo {author} {\bibfnamefont {T.}~\bibnamefont {{Naab}}}, \bibinfo {author} {\bibfnamefont {S.}~\bibnamefont {{Walch}}}, \bibinfo {author} {\bibfnamefont {M.}~\bibnamefont {{Hanasz}}}, \bibinfo {author} {\bibfnamefont {M.-M.}\ \bibnamefont {{Mac Low}}}, \bibinfo {author} {\bibfnamefont {J.~P.}\ \bibnamefont {{Ostriker}}}, \bibinfo {author} {\bibfnamefont {A.}~\bibnamefont {{Gatto}}}, \bibinfo {author} {\bibfnamefont {T.}~\bibnamefont {{Peters}}}, \bibinfo {author} {\bibfnamefont {R.}~\bibnamefont {{W{\"u}nsch}}}, \bibinfo {author} {\bibfnamefont {S.~C.~O.}\ \bibnamefont {{Glover}}}, \bibinfo {author} {\bibfnamefont {R.~S.}\ \bibnamefont {{Klessen}}}, \bibinfo {author} {\bibfnamefont {P.~C.}\ \bibnamefont {{Clark}}},\ and\ \bibinfo {author} {\bibfnamefont {C.}~\bibnamefont {{Baczynski}}},\ }\bibfield  {title} {\bibinfo {title} {{Launching Cosmic-Ray-driven Outflows from the Magnetized Interstellar Medium}},\ }\href
  {https://doi.org/10.3847/2041-8205/816/2/L19} {\bibfield  {journal} {\bibinfo  {journal} {The Astrophysical Journall}\ }\textbf {\bibinfo {volume} {816}},\ \bibinfo {eid} {L19} (\bibinfo {year} {2016})},\ \Eprint {https://arxiv.org/abs/1509.07247} {arXiv:1509.07247 [astro-ph.GA]} \BibitemShut {NoStop}%
\bibitem [{\citenamefont {Butsky}\ and\ \citenamefont {Quinn}(2018)}]{Butsky_2018}%
  \BibitemOpen
  \bibfield  {author} {\bibinfo {author} {\bibfnamefont {I.~S.}\ \bibnamefont {Butsky}}\ and\ \bibinfo {author} {\bibfnamefont {T.~R.}\ \bibnamefont {Quinn}},\ }\bibfield  {title} {\bibinfo {title} {The role of cosmic-ray transport in shaping the simulated circumgalactic medium},\ }\href {https://doi.org/10.3847/1538-4357/aaeac2} {\bibfield  {journal} {\bibinfo  {journal} {The Astrophysical Journal}\ }\textbf {\bibinfo {volume} {868}},\ \bibinfo {pages} {108} (\bibinfo {year} {2018})}\BibitemShut {NoStop}%
\bibitem [{\citenamefont {Buck}\ \emph {et~al.}(2020)\citenamefont {Buck}, \citenamefont {Pfrommer}, \citenamefont {Pakmor}, \citenamefont {Grand},\ and\ \citenamefont {Springel}}]{Buck_2020}%
  \BibitemOpen
  \bibfield  {author} {\bibinfo {author} {\bibfnamefont {T.}~\bibnamefont {Buck}}, \bibinfo {author} {\bibfnamefont {C.}~\bibnamefont {Pfrommer}}, \bibinfo {author} {\bibfnamefont {R.}~\bibnamefont {Pakmor}}, \bibinfo {author} {\bibfnamefont {R.~J.~J.}\ \bibnamefont {Grand}},\ and\ \bibinfo {author} {\bibfnamefont {V.}~\bibnamefont {Springel}},\ }\bibfield  {title} {\bibinfo {title} {The effects of cosmic rays on the formation of milky way-mass galaxies in a cosmological context},\ }\href {https://doi.org/10.1093/mnras/staa1960} {\bibfield  {journal} {\bibinfo  {journal} {Monthly Notices of the Royal Astronomical Society}\ }\textbf {\bibinfo {volume} {497}},\ \bibinfo {pages} {1712–1737} (\bibinfo {year} {2020})}\BibitemShut {NoStop}%
\bibitem [{\citenamefont {Werhahn}\ \emph {et~al.}(2021)\citenamefont {Werhahn}, \citenamefont {Pfrommer}, \citenamefont {Girichidis}, \citenamefont {Puchwein},\ and\ \citenamefont {Pakmor}}]{Werhahn_2021}%
  \BibitemOpen
  \bibfield  {author} {\bibinfo {author} {\bibfnamefont {M.}~\bibnamefont {Werhahn}}, \bibinfo {author} {\bibfnamefont {C.}~\bibnamefont {Pfrommer}}, \bibinfo {author} {\bibfnamefont {P.}~\bibnamefont {Girichidis}}, \bibinfo {author} {\bibfnamefont {E.}~\bibnamefont {Puchwein}},\ and\ \bibinfo {author} {\bibfnamefont {R.}~\bibnamefont {Pakmor}},\ }\bibfield  {title} {\bibinfo {title} {Cosmic rays and non-thermal emission in simulated galaxies i. electron and proton spectra compared to voyager 1 data},\ }\href {https://doi.org/10.1093/mnras/stab1324} {\bibfield  {journal} {\bibinfo  {journal} {Monthly Notices of the Royal Astronomical Society}\ }\textbf {\bibinfo {volume} {505}},\ \bibinfo {pages} {3273–3294} (\bibinfo {year} {2021})}\BibitemShut {NoStop}%
\bibitem [{\citenamefont {Pfrommer}\ \emph {et~al.}(2022)\citenamefont {Pfrommer}, \citenamefont {Werhahn}, \citenamefont {Pakmor}, \citenamefont {Girichidis},\ and\ \citenamefont {Simpson}}]{Pfrommer_2022}%
  \BibitemOpen
  \bibfield  {author} {\bibinfo {author} {\bibfnamefont {C.}~\bibnamefont {Pfrommer}}, \bibinfo {author} {\bibfnamefont {M.}~\bibnamefont {Werhahn}}, \bibinfo {author} {\bibfnamefont {R.}~\bibnamefont {Pakmor}}, \bibinfo {author} {\bibfnamefont {P.}~\bibnamefont {Girichidis}},\ and\ \bibinfo {author} {\bibfnamefont {C.~M.}\ \bibnamefont {Simpson}},\ }\bibfield  {title} {\bibinfo {title} {Simulating radio synchrotron emission in star-forming galaxies: small-scale magnetic dynamo and the origin of the far-infrared–radio correlation},\ }\href {https://doi.org/10.1093/mnras/stac1808} {\bibfield  {journal} {\bibinfo  {journal} {Monthly Notices of the Royal Astronomical Society}\ }\textbf {\bibinfo {volume} {515}},\ \bibinfo {pages} {4229–4264} (\bibinfo {year} {2022})}\BibitemShut {NoStop}%
\bibitem [{\citenamefont {Farcy}\ \emph {et~al.}(2022)\citenamefont {Farcy}, \citenamefont {Rosdahl}, \citenamefont {Dubois}, \citenamefont {Blaizot},\ and\ \citenamefont {Martin-Alvarez}}]{Farcy_2022}%
  \BibitemOpen
  \bibfield  {author} {\bibinfo {author} {\bibfnamefont {M.}~\bibnamefont {Farcy}}, \bibinfo {author} {\bibfnamefont {J.}~\bibnamefont {Rosdahl}}, \bibinfo {author} {\bibfnamefont {Y.}~\bibnamefont {Dubois}}, \bibinfo {author} {\bibfnamefont {J.}~\bibnamefont {Blaizot}},\ and\ \bibinfo {author} {\bibfnamefont {S.}~\bibnamefont {Martin-Alvarez}},\ }\bibfield  {title} {\bibinfo {title} {Radiation-magnetohydrodynamics simulations of cosmic ray feedback in disc galaxies},\ }\href {https://doi.org/10.1093/mnras/stac1196} {\bibfield  {journal} {\bibinfo  {journal} {Monthly Notices of the Royal Astronomical Society}\ }\textbf {\bibinfo {volume} {513}},\ \bibinfo {pages} {5000–5019} (\bibinfo {year} {2022})}\BibitemShut {NoStop}%
\bibitem [{\citenamefont {Thomas}\ \emph {et~al.}(2022)\citenamefont {Thomas}, \citenamefont {Pfrommer},\ and\ \citenamefont {Pakmor}}]{thomas2022cosmic}%
  \BibitemOpen
  \bibfield  {author} {\bibinfo {author} {\bibfnamefont {T.}~\bibnamefont {Thomas}}, \bibinfo {author} {\bibfnamefont {C.}~\bibnamefont {Pfrommer}},\ and\ \bibinfo {author} {\bibfnamefont {R.}~\bibnamefont {Pakmor}},\ }\href@noop {} {\bibinfo {title} {Cosmic ray-driven galactic winds: transport modes of cosmic rays and alfvén-wave dark regions}} (\bibinfo {year} {2022}),\ \Eprint {https://arxiv.org/abs/2203.12029} {arXiv:2203.12029 [astro-ph.GA]} \BibitemShut {NoStop}%
\bibitem [{\citenamefont {Hopkins}\ \emph {et~al.}(2022)\citenamefont {Hopkins}, \citenamefont {Butsky}, \citenamefont {Panopoulou}, \citenamefont {Ji}, \citenamefont {Quataert}, \citenamefont {Faucher-Giguère},\ and\ \citenamefont {Kereš}}]{Hopkins_CR}%
  \BibitemOpen
  \bibfield  {author} {\bibinfo {author} {\bibfnamefont {P.~F.}\ \bibnamefont {Hopkins}}, \bibinfo {author} {\bibfnamefont {I.~S.}\ \bibnamefont {Butsky}}, \bibinfo {author} {\bibfnamefont {G.~V.}\ \bibnamefont {Panopoulou}}, \bibinfo {author} {\bibfnamefont {S.}~\bibnamefont {Ji}}, \bibinfo {author} {\bibfnamefont {E.}~\bibnamefont {Quataert}}, \bibinfo {author} {\bibfnamefont {C.-A.}\ \bibnamefont {Faucher-Giguère}},\ and\ \bibinfo {author} {\bibfnamefont {D.}~\bibnamefont {Kereš}},\ }\bibfield  {title} {\bibinfo {title} {First predicted cosmic ray spectra, primary-to-secondary ratios, and ionization rates from mhd galaxy formation simulations},\ }\href {https://doi.org/10.1093/mnras/stac1791} {\bibfield  {journal} {\bibinfo  {journal} {Monthly Notices of the Royal Astronomical Society}\ }\textbf {\bibinfo {volume} {516}},\ \bibinfo {pages} {3470–3514} (\bibinfo {year} {2022})}\BibitemShut {NoStop}%
\bibitem [{\citenamefont {Ponnada}\ \emph {et~al.}(2023)\citenamefont {Ponnada}, \citenamefont {Panopoulou}, \citenamefont {Butsky}, \citenamefont {Hopkins}, \citenamefont {Skalidis}, \citenamefont {Hummels}, \citenamefont {Quataert}, \citenamefont {Kereš}, \citenamefont {Faucher-Giguère},\ and\ \citenamefont {Su}}]{Ponnada_2023}%
  \BibitemOpen
  \bibfield  {author} {\bibinfo {author} {\bibfnamefont {S.~B.}\ \bibnamefont {Ponnada}}, \bibinfo {author} {\bibfnamefont {G.~V.}\ \bibnamefont {Panopoulou}}, \bibinfo {author} {\bibfnamefont {I.~S.}\ \bibnamefont {Butsky}}, \bibinfo {author} {\bibfnamefont {P.~F.}\ \bibnamefont {Hopkins}}, \bibinfo {author} {\bibfnamefont {R.}~\bibnamefont {Skalidis}}, \bibinfo {author} {\bibfnamefont {C.}~\bibnamefont {Hummels}}, \bibinfo {author} {\bibfnamefont {E.}~\bibnamefont {Quataert}}, \bibinfo {author} {\bibfnamefont {D.}~\bibnamefont {Kereš}}, \bibinfo {author} {\bibfnamefont {C.-A.}\ \bibnamefont {Faucher-Giguère}},\ and\ \bibinfo {author} {\bibfnamefont {K.-Y.}\ \bibnamefont {Su}},\ }\bibfield  {title} {\bibinfo {title} {Synchrotron emission on fire: equipartition estimators of magnetic fields in simulated galaxies with spectrally resolved cosmic rays},\ }\href {https://doi.org/10.1093/mnras/stad3978} {\bibfield  {journal} {\bibinfo  {journal} {Monthly Notices of the Royal Astronomical Society}\ }\textbf
  {\bibinfo {volume} {527}},\ \bibinfo {pages} {11707–11718} (\bibinfo {year} {2023})}\BibitemShut {NoStop}%
\bibitem [{\citenamefont {Ponnada}\ \emph {et~al.}(2024)\citenamefont {Ponnada}, \citenamefont {Cochrane}, \citenamefont {Hopkins}, \citenamefont {Butsky}, \citenamefont {Wellons}, \citenamefont {Sanchez}, \citenamefont {Hummels}, \citenamefont {Lu}, \citenamefont {Kereš},\ and\ \citenamefont {Hayward}}]{ponnada2024hookslinessinkersagn}%
  \BibitemOpen
  \bibfield  {author} {\bibinfo {author} {\bibfnamefont {S.~B.}\ \bibnamefont {Ponnada}}, \bibinfo {author} {\bibfnamefont {R.~K.}\ \bibnamefont {Cochrane}}, \bibinfo {author} {\bibfnamefont {P.~F.}\ \bibnamefont {Hopkins}}, \bibinfo {author} {\bibfnamefont {I.~S.}\ \bibnamefont {Butsky}}, \bibinfo {author} {\bibfnamefont {S.}~\bibnamefont {Wellons}}, \bibinfo {author} {\bibfnamefont {N.~N.}\ \bibnamefont {Sanchez}}, \bibinfo {author} {\bibfnamefont {C.}~\bibnamefont {Hummels}}, \bibinfo {author} {\bibfnamefont {Y.~S.}\ \bibnamefont {Lu}}, \bibinfo {author} {\bibfnamefont {D.}~\bibnamefont {Kereš}},\ and\ \bibinfo {author} {\bibfnamefont {C.~C.}\ \bibnamefont {Hayward}},\ }\href {https://arxiv.org/abs/2410.02971} {\bibinfo {title} {Hooks, lines, and sinkers: How agn feedback and cosmic-ray transport shape the far infrared-radio correlation of galaxies}} (\bibinfo {year} {2024}),\ \Eprint {https://arxiv.org/abs/2410.02971} {arXiv:2410.02971 [astro-ph.GA]} \BibitemShut {NoStop}%
\bibitem [{\citenamefont {Hopkins}\ \emph {et~al.}(2018)\citenamefont {Hopkins} \emph {et~al.}}]{Hopkins_2018}%
  \BibitemOpen
  \bibfield  {author} {\bibinfo {author} {\bibfnamefont {P.~F.}\ \bibnamefont {Hopkins}} \emph {et~al.},\ }\bibfield  {title} {\bibinfo {title} {Fire-2 simulations: physics versus numerics in galaxy formation},\ }\href {https://doi.org/10.1093/mnras/sty1690} {\bibfield  {journal} {\bibinfo  {journal} {Monthly Notices of the Royal Astronomical Society}\ }\textbf {\bibinfo {volume} {480}},\ \bibinfo {pages} {800–863} (\bibinfo {year} {2018})}\BibitemShut {NoStop}%
\bibitem [{\citenamefont {Bisschoff}\ \emph {et~al.}(2019)\citenamefont {Bisschoff}, \citenamefont {Potgieter},\ and\ \citenamefont {Aslam}}]{Bisschoff_2019}%
  \BibitemOpen
  \bibfield  {author} {\bibinfo {author} {\bibfnamefont {D.}~\bibnamefont {Bisschoff}}, \bibinfo {author} {\bibfnamefont {M.~S.}\ \bibnamefont {Potgieter}},\ and\ \bibinfo {author} {\bibfnamefont {O.~P.~M.}\ \bibnamefont {Aslam}},\ }\bibfield  {title} {\bibinfo {title} {New very local interstellar spectra for electrons, positrons, protons, and light cosmic ray nuclei},\ }\href {https://doi.org/10.3847/1538-4357/ab1e4a} {\bibfield  {journal} {\bibinfo  {journal} {The Astrophysical Journal}\ }\textbf {\bibinfo {volume} {878}},\ \bibinfo {pages} {59} (\bibinfo {year} {2019})}\BibitemShut {NoStop}%
\bibitem [{\citenamefont {Hopkins}\ \emph {et~al.}(2021)\citenamefont {Hopkins}, \citenamefont {Squire},\ and\ \citenamefont {Butsky}}]{Hopkins_2021a}%
  \BibitemOpen
  \bibfield  {author} {\bibinfo {author} {\bibfnamefont {P.~F.}\ \bibnamefont {Hopkins}}, \bibinfo {author} {\bibfnamefont {J.}~\bibnamefont {Squire}},\ and\ \bibinfo {author} {\bibfnamefont {I.~S.}\ \bibnamefont {Butsky}},\ }\bibfield  {title} {\bibinfo {title} {A consistent reduced-speed-of-light formulation of cosmic ray transport valid in weak- and strong-scattering regimes},\ }\href {https://doi.org/10.1093/mnras/stab2635} {\bibfield  {journal} {\bibinfo  {journal} {Monthly Notices of the Royal Astronomical Society}\ }\textbf {\bibinfo {volume} {509}},\ \bibinfo {pages} {3779–3797} (\bibinfo {year} {2021})}\BibitemShut {NoStop}%
\bibitem [{\citenamefont {{Zweibel}}(2013)}]{Zwei13}%
  \BibitemOpen
  \bibfield  {author} {\bibinfo {author} {\bibfnamefont {E.~G.}\ \bibnamefont {{Zweibel}}},\ }\bibfield  {title} {\bibinfo {title} {{The microphysics and macrophysics of cosmic rays}},\ }\href {https://doi.org/10.1063/1.4807033} {\bibfield  {journal} {\bibinfo  {journal} {Physics of Plasmas}\ }\textbf {\bibinfo {volume} {20}},\ \bibinfo {eid} {055501} (\bibinfo {year} {2013})}\BibitemShut {NoStop}%
\bibitem [{\citenamefont {{Thomas}}\ and\ \citenamefont {{Pfrommer}}(2019)}]{thomas.pfrommer.18:alfven.reg.cr.transport}%
  \BibitemOpen
  \bibfield  {author} {\bibinfo {author} {\bibfnamefont {T.}~\bibnamefont {{Thomas}}}\ and\ \bibinfo {author} {\bibfnamefont {C.}~\bibnamefont {{Pfrommer}}},\ }\bibfield  {title} {\bibinfo {title} {{Cosmic-ray hydrodynamics: Alfv{\'e}n-wave regulated transport of cosmic rays}},\ }\href {https://doi.org/10.1093/mnras/stz263} {\bibfield  {journal} {\bibinfo  {journal} {Monthly Notices of the Royal Astronomical Society}\ }\textbf {\bibinfo {volume} {485}},\ \bibinfo {pages} {2977} (\bibinfo {year} {2019})},\ \Eprint {https://arxiv.org/abs/1805.11092} {arXiv:1805.11092 [astro-ph.HE]} \BibitemShut {NoStop}%
\bibitem [{\citenamefont {De~La Torre~Luque}\ \emph {et~al.}(2021)\citenamefont {De~La Torre~Luque}, \citenamefont {Mazziotta}, \citenamefont {Loparco}, \citenamefont {Gargano},\ and\ \citenamefont {Serini}}]{De_La_Torre_Luque_2021}%
  \BibitemOpen
  \bibfield  {author} {\bibinfo {author} {\bibfnamefont {P.}~\bibnamefont {De~La Torre~Luque}}, \bibinfo {author} {\bibfnamefont {M.}~\bibnamefont {Mazziotta}}, \bibinfo {author} {\bibfnamefont {F.}~\bibnamefont {Loparco}}, \bibinfo {author} {\bibfnamefont {F.}~\bibnamefont {Gargano}},\ and\ \bibinfo {author} {\bibfnamefont {D.}~\bibnamefont {Serini}},\ }\bibfield  {title} {\bibinfo {title} {Implications of current nuclear cross sections on secondary cosmic rays with the upcoming dragon2 code},\ }\href {https://doi.org/10.1088/1475-7516/2021/03/099} {\bibfield  {journal} {\bibinfo  {journal} {Journal of Cosmology and Astroparticle Physics}\ }\textbf {\bibinfo {volume} {2021}}\bibinfo  {number} { (03)},\ \bibinfo {pages} {099}}\BibitemShut {NoStop}%
\bibitem [{\citenamefont {Korsmeier}\ and\ \citenamefont {Cuoco}(2021)}]{Korsmeier_2021}%
  \BibitemOpen
\bibfield  {number} {  }\bibfield  {author} {\bibinfo {author} {\bibfnamefont {M.}~\bibnamefont {Korsmeier}}\ and\ \bibinfo {author} {\bibfnamefont {A.}~\bibnamefont {Cuoco}},\ }\bibfield  {title} {\bibinfo {title} {Implications of lithium to oxygen ams-02 spectra on our understanding of cosmic-ray diffusion},\ }\bibfield  {journal} {\bibinfo  {journal} {Physical Review D}\ }\textbf {\bibinfo {volume} {103}},\ \href {https://doi.org/10.1103/physrevd.103.103016} {10.1103/physrevd.103.103016} (\bibinfo {year} {2021})\BibitemShut {NoStop}%
\bibitem [{\citenamefont {Kafexhiu}\ \emph {et~al.}(2014)\citenamefont {Kafexhiu}, \citenamefont {Aharonian}, \citenamefont {Taylor},\ and\ \citenamefont {Vila}}]{Kafexhiu_2014}%
  \BibitemOpen
  \bibfield  {author} {\bibinfo {author} {\bibfnamefont {E.}~\bibnamefont {Kafexhiu}}, \bibinfo {author} {\bibfnamefont {F.}~\bibnamefont {Aharonian}}, \bibinfo {author} {\bibfnamefont {A.~M.}\ \bibnamefont {Taylor}},\ and\ \bibinfo {author} {\bibfnamefont {G.~S.}\ \bibnamefont {Vila}},\ }\bibfield  {title} {\bibinfo {title} {Parametrization of gamma-ray production cross sections for interactions in a broad proton energy range from the kinematic threshold to pev energies},\ }\bibfield  {journal} {\bibinfo  {journal} {Physical Review D}\ }\textbf {\bibinfo {volume} {90}},\ \href {https://doi.org/10.1103/physrevd.90.123014} {10.1103/physrevd.90.123014} (\bibinfo {year} {2014})\BibitemShut {NoStop}%
\bibitem [{\citenamefont {Agostinelli}\ \emph {et~al.}(2003)\citenamefont {Agostinelli} \emph {et~al.}}]{GEANT4:2002zbu}%
  \BibitemOpen
  \bibfield  {author} {\bibinfo {author} {\bibfnamefont {S.}~\bibnamefont {Agostinelli}} \emph {et~al.} (\bibinfo {collaboration} {GEANT4}),\ }\bibfield  {title} {\bibinfo {title} {{GEANT4 - A Simulation Toolkit}},\ }\href {https://doi.org/10.1016/S0168-9002(03)01368-8} {\bibfield  {journal} {\bibinfo  {journal} {Nucl. Instrum. Meth. A}\ }\textbf {\bibinfo {volume} {506}},\ \bibinfo {pages} {250} (\bibinfo {year} {2003})}\BibitemShut {NoStop}%
\bibitem [{\citenamefont {Allison}\ \emph {et~al.}(2006)\citenamefont {Allison} \emph {et~al.}}]{Allison_06}%
  \BibitemOpen
  \bibfield  {author} {\bibinfo {author} {\bibfnamefont {J.}~\bibnamefont {Allison}} \emph {et~al.},\ }\bibfield  {title} {\bibinfo {title} {Geant4 developments and applications},\ }\href {https://doi.org/10.1109/TNS.2006.869826} {\bibfield  {journal} {\bibinfo  {journal} {IEEE Transactions on Nuclear Science}\ }\textbf {\bibinfo {volume} {53}},\ \bibinfo {pages} {270} (\bibinfo {year} {2006})}\BibitemShut {NoStop}%
\bibitem [{\citenamefont {Allison}\ \emph {et~al.}(2016)\citenamefont {Allison} \emph {et~al.}}]{Allison:2016lfl}%
  \BibitemOpen
  \bibfield  {author} {\bibinfo {author} {\bibfnamefont {J.}~\bibnamefont {Allison}} \emph {et~al.},\ }\bibfield  {title} {\bibinfo {title} {{Recent developments in Geant4}},\ }\href {https://doi.org/10.1016/j.nima.2016.06.125} {\bibfield  {journal} {\bibinfo  {journal} {Nucl. Instrum. Meth. A}\ }\textbf {\bibinfo {volume} {835}},\ \bibinfo {pages} {186} (\bibinfo {year} {2016})}\BibitemShut {NoStop}%
\bibitem [{\citenamefont {Longair}(1992)}]{Longair:1992ze}%
  \BibitemOpen
  \bibinfo {editor} {\bibfnamefont {M.~S.}\ \bibnamefont {Longair}},\ ed.,\ \href@noop {} {\emph {\bibinfo {title} {{High-energy astrophysics. Vol. 1: Particles, photons and their detection}}}}\ (\bibinfo {year} {1992})\BibitemShut {NoStop}%
\bibitem [{\citenamefont {{Hopkins}}\ \emph {et~al.}(2020)\citenamefont {{Hopkins}}, \citenamefont {{Grudi{\'c}}}, \citenamefont {{Wetzel}}, \citenamefont {{Kere{\v{s}}}}, \citenamefont {{Faucher-Gigu{\`e}re}}, \citenamefont {{Ma}}, \citenamefont {{Murray}},\ and\ \citenamefont {{Butcher}}}]{2020MNRAS.491.3702H}%
  \BibitemOpen
  \bibfield  {author} {\bibinfo {author} {\bibfnamefont {P.~F.}\ \bibnamefont {{Hopkins}}}, \bibinfo {author} {\bibfnamefont {M.~Y.}\ \bibnamefont {{Grudi{\'c}}}}, \bibinfo {author} {\bibfnamefont {A.}~\bibnamefont {{Wetzel}}}, \bibinfo {author} {\bibfnamefont {D.}~\bibnamefont {{Kere{\v{s}}}}}, \bibinfo {author} {\bibfnamefont {C.-A.}\ \bibnamefont {{Faucher-Gigu{\`e}re}}}, \bibinfo {author} {\bibfnamefont {X.}~\bibnamefont {{Ma}}}, \bibinfo {author} {\bibfnamefont {N.}~\bibnamefont {{Murray}}},\ and\ \bibinfo {author} {\bibfnamefont {N.}~\bibnamefont {{Butcher}}},\ }\bibfield  {title} {\bibinfo {title} {{Radiative stellar feedback in galaxy formation: Methods and physics}},\ }\href {https://doi.org/10.1093/mnras/stz3129} {\bibfield  {journal} {\bibinfo  {journal} {Monthly Notices of the Royal Astronomical Society}\ }\textbf {\bibinfo {volume} {491}},\ \bibinfo {pages} {3702} (\bibinfo {year} {2020})},\ \Eprint {https://arxiv.org/abs/1811.12462} {arXiv:1811.12462 [astro-ph.GA]} \BibitemShut {NoStop}%
\bibitem [{\citenamefont {Dodelson}(2003)}]{Dodelson:2003ft}%
  \BibitemOpen
  \bibfield  {author} {\bibinfo {author} {\bibfnamefont {S.}~\bibnamefont {Dodelson}},\ }\href@noop {} {\emph {\bibinfo {title} {{Modern Cosmology}}}}\ (\bibinfo  {publisher} {Academic Press},\ \bibinfo {address} {Amsterdam},\ \bibinfo {year} {2003})\BibitemShut {NoStop}%
\bibitem [{\citenamefont {Acero}\ \emph {et~al.}(2016)\citenamefont {Acero} \emph {et~al.}}]{Acero_2016}%
  \BibitemOpen
  \bibfield  {author} {\bibinfo {author} {\bibfnamefont {F.}~\bibnamefont {Acero}} \emph {et~al.},\ }\bibfield  {title} {\bibinfo {title} {Development of the model of galactic interstellar emission for standard point-source analysis of fermi large area telescope data},\ }\href {https://doi.org/10.3847/0067-0049/223/2/26} {\bibfield  {journal} {\bibinfo  {journal} {The Astrophysical Journal Supplement Series}\ }\textbf {\bibinfo {volume} {223}},\ \bibinfo {pages} {26} (\bibinfo {year} {2016})}\BibitemShut {NoStop}%
\bibitem [{\citenamefont {{Torrey}}\ \emph {et~al.}(2017)\citenamefont {{Torrey}}, \citenamefont {{Hopkins}}, \citenamefont {{Faucher-Gigu{\`e}re}}, \citenamefont {{Vogelsberger}}, \citenamefont {{Quataert}}, \citenamefont {{Kere{\v{s}}}},\ and\ \citenamefont {{Murray}}}]{2017MNRAS.467.2301T}%
  \BibitemOpen
  \bibfield  {author} {\bibinfo {author} {\bibfnamefont {P.}~\bibnamefont {{Torrey}}}, \bibinfo {author} {\bibfnamefont {P.~F.}\ \bibnamefont {{Hopkins}}}, \bibinfo {author} {\bibfnamefont {C.-A.}\ \bibnamefont {{Faucher-Gigu{\`e}re}}}, \bibinfo {author} {\bibfnamefont {M.}~\bibnamefont {{Vogelsberger}}}, \bibinfo {author} {\bibfnamefont {E.}~\bibnamefont {{Quataert}}}, \bibinfo {author} {\bibfnamefont {D.}~\bibnamefont {{Kere{\v{s}}}}},\ and\ \bibinfo {author} {\bibfnamefont {N.}~\bibnamefont {{Murray}}},\ }\bibfield  {title} {\bibinfo {title} {{An instability of feedback-regulated star formation in galactic nuclei}},\ }\href {https://doi.org/10.1093/mnras/stx254} {\bibfield  {journal} {\bibinfo  {journal} {Monthly Notices of the Royal Astronomical Society}\ }\textbf {\bibinfo {volume} {467}},\ \bibinfo {pages} {2301} (\bibinfo {year} {2017})},\ \Eprint {https://arxiv.org/abs/1601.07186} {arXiv:1601.07186 [astro-ph.GA]} \BibitemShut {NoStop}%
\bibitem [{\citenamefont {{Orr}}\ \emph {et~al.}(2021)\citenamefont {{Orr}}, \citenamefont {{Hatchfield}}, \citenamefont {{Battersby}}, \citenamefont {{Hayward}}, \citenamefont {{Hopkins}}, \citenamefont {{Wetzel}}, \citenamefont {{Benincasa}}, \citenamefont {{Loebman}}, \citenamefont {{Sormani}},\ and\ \citenamefont {{Klessen}}}]{2021ApJ...908L..31O}%
  \BibitemOpen
  \bibfield  {author} {\bibinfo {author} {\bibfnamefont {M.~E.}\ \bibnamefont {{Orr}}}, \bibinfo {author} {\bibfnamefont {H.~P.}\ \bibnamefont {{Hatchfield}}}, \bibinfo {author} {\bibfnamefont {C.}~\bibnamefont {{Battersby}}}, \bibinfo {author} {\bibfnamefont {C.~C.}\ \bibnamefont {{Hayward}}}, \bibinfo {author} {\bibfnamefont {P.~F.}\ \bibnamefont {{Hopkins}}}, \bibinfo {author} {\bibfnamefont {A.}~\bibnamefont {{Wetzel}}}, \bibinfo {author} {\bibfnamefont {S.~M.}\ \bibnamefont {{Benincasa}}}, \bibinfo {author} {\bibfnamefont {S.~R.}\ \bibnamefont {{Loebman}}}, \bibinfo {author} {\bibfnamefont {M.~C.}\ \bibnamefont {{Sormani}}},\ and\ \bibinfo {author} {\bibfnamefont {R.~S.}\ \bibnamefont {{Klessen}}},\ }\bibfield  {title} {\bibinfo {title} {{Fiery Cores: Bursty and Smooth Star Formation Distributions across Galaxy Centers in Cosmological Zoom-in Simulations}},\ }\href {https://doi.org/10.3847/2041-8213/abdebd} {\bibfield  {journal} {\bibinfo  {journal} {The Astrophysical Journall}\ }\textbf {\bibinfo
  {volume} {908}},\ \bibinfo {eid} {L31} (\bibinfo {year} {2021})},\ \Eprint {https://arxiv.org/abs/2101.11034} {arXiv:2101.11034 [astro-ph.GA]} \BibitemShut {NoStop}%
\bibitem [{\citenamefont {{Engelmann}}\ \emph {et~al.}(1990)\citenamefont {{Engelmann}}, \citenamefont {{Ferrando}}, \citenamefont {{Soutoul}}, \citenamefont {{Goret}}, \citenamefont {{Juliusson}}, \citenamefont {{Koch-Miramond}}, \citenamefont {{Lund}}, \citenamefont {{Masse}}, \citenamefont {{Peters}}, \citenamefont {{Petrou}},\ and\ \citenamefont {{Rasmussen}}}]{1990A&A...233...96E}%
  \BibitemOpen
  \bibfield  {author} {\bibinfo {author} {\bibfnamefont {J.~J.}\ \bibnamefont {{Engelmann}}}, \bibinfo {author} {\bibfnamefont {P.}~\bibnamefont {{Ferrando}}}, \bibinfo {author} {\bibfnamefont {A.}~\bibnamefont {{Soutoul}}}, \bibinfo {author} {\bibfnamefont {P.}~\bibnamefont {{Goret}}}, \bibinfo {author} {\bibfnamefont {E.}~\bibnamefont {{Juliusson}}}, \bibinfo {author} {\bibfnamefont {L.}~\bibnamefont {{Koch-Miramond}}}, \bibinfo {author} {\bibfnamefont {N.}~\bibnamefont {{Lund}}}, \bibinfo {author} {\bibfnamefont {P.}~\bibnamefont {{Masse}}}, \bibinfo {author} {\bibfnamefont {B.}~\bibnamefont {{Peters}}}, \bibinfo {author} {\bibfnamefont {N.}~\bibnamefont {{Petrou}}},\ and\ \bibinfo {author} {\bibfnamefont {I.~L.}\ \bibnamefont {{Rasmussen}}},\ }\bibfield  {title} {\bibinfo {title} {{Charge composition and energy spectra of cosmic-ray nuclei for elements from Be to Ni - Results from HEAO-3-C2.}},\ }\href@noop {} {\bibfield  {journal} {\bibinfo  {journal} {Astronomy \& Astrophysics}\ }\textbf {\bibinfo
  {volume} {233}},\ \bibinfo {pages} {96} (\bibinfo {year} {1990})}\BibitemShut {NoStop}%
\bibitem [{\citenamefont {{Shikaze}}\ \emph {et~al.}(2007)\citenamefont {{Shikaze}} \emph {et~al.}}]{2007APh....28..154S}%
  \BibitemOpen
  \bibfield  {author} {\bibinfo {author} {\bibfnamefont {Y.}~\bibnamefont {{Shikaze}}} \emph {et~al.},\ }\bibfield  {title} {\bibinfo {title} {{Measurements of 0.2 20 GeV/n cosmic-ray proton and helium spectra from 1997 through 2002 with the BESS spectrometer}},\ }\href {https://doi.org/10.1016/j.astropartphys.2007.05.001} {\bibfield  {journal} {\bibinfo  {journal} {Astroparticle Physics}\ }\textbf {\bibinfo {volume} {28}},\ \bibinfo {pages} {154} (\bibinfo {year} {2007})},\ \Eprint {https://arxiv.org/abs/astro-ph/0611388} {arXiv:astro-ph/0611388 [astro-ph]} \BibitemShut {NoStop}%
\bibitem [{\citenamefont {{Boezio}}\ \emph {et~al.}(2000)\citenamefont {{Boezio}} \emph {et~al.}}]{2000ApJ...532..653B}%
  \BibitemOpen
  \bibfield  {author} {\bibinfo {author} {\bibfnamefont {M.}~\bibnamefont {{Boezio}}} \emph {et~al.},\ }\bibfield  {title} {\bibinfo {title} {{The Cosmic-Ray Electron and Positron Spectra Measured at 1 AU during Solar Minimum Activity}},\ }\href {https://doi.org/10.1086/308545} {\bibfield  {journal} {\bibinfo  {journal} {The Astrophysical Journal}\ }\textbf {\bibinfo {volume} {532}},\ \bibinfo {pages} {653} (\bibinfo {year} {2000})}\BibitemShut {NoStop}%
\bibitem [{\citenamefont {{Obermeier}}\ \emph {et~al.}(2011)\citenamefont {{Obermeier}}, \citenamefont {{Ave}}, \citenamefont {{Boyle}}, \citenamefont {{H{\"o}ppner}}, \citenamefont {{H{\"o}randel}},\ and\ \citenamefont {{M{\"u}ller}}}]{2011ApJ...742...14O}%
  \BibitemOpen
  \bibfield  {author} {\bibinfo {author} {\bibfnamefont {A.}~\bibnamefont {{Obermeier}}}, \bibinfo {author} {\bibfnamefont {M.}~\bibnamefont {{Ave}}}, \bibinfo {author} {\bibfnamefont {P.}~\bibnamefont {{Boyle}}}, \bibinfo {author} {\bibfnamefont {C.}~\bibnamefont {{H{\"o}ppner}}}, \bibinfo {author} {\bibfnamefont {J.}~\bibnamefont {{H{\"o}randel}}},\ and\ \bibinfo {author} {\bibfnamefont {D.}~\bibnamefont {{M{\"u}ller}}},\ }\bibfield  {title} {\bibinfo {title} {{Energy Spectra of Primary and Secondary Cosmic-Ray Nuclei Measured with TRACER}},\ }\href {https://doi.org/10.1088/0004-637X/742/1/14} {\bibfield  {journal} {\bibinfo  {journal} {The Astrophysical Journal}\ }\textbf {\bibinfo {volume} {742}},\ \bibinfo {eid} {14} (\bibinfo {year} {2011})},\ \Eprint {https://arxiv.org/abs/1108.4838} {arXiv:1108.4838 [astro-ph.HE]} \BibitemShut {NoStop}%
\bibitem [{\citenamefont {{Adriani}}\ \emph {et~al.}(2014)\citenamefont {{Adriani}} \emph {et~al.}}]{2014ApJ...791...93A}%
  \BibitemOpen
  \bibfield  {author} {\bibinfo {author} {\bibfnamefont {O.}~\bibnamefont {{Adriani}}} \emph {et~al.},\ }\bibfield  {title} {\bibinfo {title} {{Measurement of Boron and Carbon Fluxes in Cosmic Rays with the PAMELA Experiment}},\ }\href {https://doi.org/10.1088/0004-637X/791/2/93} {\bibfield  {journal} {\bibinfo  {journal} {The Astrophysical Journal}\ }\textbf {\bibinfo {volume} {791}},\ \bibinfo {eid} {93} (\bibinfo {year} {2014})},\ \Eprint {https://arxiv.org/abs/1407.1657} {arXiv:1407.1657 [astro-ph.HE]} \BibitemShut {NoStop}%
\bibitem [{\citenamefont {{Abdollahi}}\ \emph {et~al.}(2017)\citenamefont {{Abdollahi}} \emph {et~al.}}]{2017PhRvD..95h2007A}%
  \BibitemOpen
  \bibfield  {author} {\bibinfo {author} {\bibfnamefont {S.}~\bibnamefont {{Abdollahi}}} \emph {et~al.},\ }\bibfield  {title} {\bibinfo {title} {{Cosmic-ray electron-positron spectrum from 7 GeV to 2 TeV with the Fermi Large Area Telescope}},\ }\href {https://doi.org/10.1103/PhysRevD.95.082007} {\bibfield  {journal} {\bibinfo  {journal} {\prd}\ }\textbf {\bibinfo {volume} {95}},\ \bibinfo {eid} {082007} (\bibinfo {year} {2017})},\ \Eprint {https://arxiv.org/abs/1704.07195} {arXiv:1704.07195 [astro-ph.HE]} \BibitemShut {NoStop}%
\bibitem [{\citenamefont {{H.~E.~S.~S. Collaboration}:}\ \emph {et~al.}(2017)\citenamefont {{H.~E.~S.~S. Collaboration}:}, \citenamefont {{Abdalla}} \emph {et~al.}}]{2017arXiv170906442H}%
  \BibitemOpen
  \bibfield  {author} {\bibinfo {author} {\bibnamefont {{H.~E.~S.~S. Collaboration}:}}, \bibinfo {author} {\bibfnamefont {H.}~\bibnamefont {{Abdalla}}}, \emph {et~al.},\ }\bibfield  {title} {\bibinfo {title} {{Contributions of the High Energy Stereoscopic System (H.E.S.S.) to the 35th International Cosmic Ray Conference (ICRC), Busan, Korea}},\ }\href {https://doi.org/10.48550/arXiv.1709.06442} {\bibfield  {journal} {\bibinfo  {journal} {arXiv e-prints}\ ,\ \bibinfo {eid} {arXiv:1709.06442}} (\bibinfo {year} {2017})},\ \Eprint {https://arxiv.org/abs/1709.06442} {arXiv:1709.06442 [astro-ph.HE]} \BibitemShut {NoStop}%
\bibitem [{\citenamefont {{Yoon}}\ \emph {et~al.}(2017)\citenamefont {{Yoon}} \emph {et~al.}}]{2017ApJ...839....5Y}%
  \BibitemOpen
  \bibfield  {author} {\bibinfo {author} {\bibfnamefont {Y.~S.}\ \bibnamefont {{Yoon}}} \emph {et~al.},\ }\bibfield  {title} {\bibinfo {title} {{Proton and Helium Spectra from the CREAM-III Flight}},\ }\href {https://doi.org/10.3847/1538-4357/aa68e4} {\bibfield  {journal} {\bibinfo  {journal} {The Astrophysical Journal}\ }\textbf {\bibinfo {volume} {839}},\ \bibinfo {eid} {5} (\bibinfo {year} {2017})},\ \Eprint {https://arxiv.org/abs/1704.02512} {arXiv:1704.02512 [astro-ph.HE]} \BibitemShut {NoStop}%
\bibitem [{\citenamefont {{DAMPE Collaboration}:}\ \emph {et~al.}(2017)\citenamefont {{DAMPE Collaboration}:}, \citenamefont {{Ambrosi}} \emph {et~al.}}]{2017Natur.552...63D}%
  \BibitemOpen
  \bibfield  {author} {\bibinfo {author} {\bibnamefont {{DAMPE Collaboration}:}}, \bibinfo {author} {\bibfnamefont {G.}~\bibnamefont {{Ambrosi}}}, \emph {et~al.},\ }\bibfield  {title} {\bibinfo {title} {{Direct detection of a break in the teraelectronvolt cosmic-ray spectrum of electrons and positrons}},\ }\href {https://doi.org/10.1038/nature24475} {\bibfield  {journal} {\bibinfo  {journal} {\nat}\ }\textbf {\bibinfo {volume} {552}},\ \bibinfo {pages} {63} (\bibinfo {year} {2017})},\ \Eprint {https://arxiv.org/abs/1711.10981} {arXiv:1711.10981 [astro-ph.HE]} \BibitemShut {NoStop}%
\bibitem [{\citenamefont {{Adriani}}\ \emph {et~al.}(2018)\citenamefont {{Adriani}}, \citenamefont {{Calet Collaboration}} \emph {et~al.}}]{2018PhRvL.120z1102A}%
  \BibitemOpen
  \bibfield  {author} {\bibinfo {author} {\bibfnamefont {O.}~\bibnamefont {{Adriani}}}, \bibinfo {author} {\bibnamefont {{Calet Collaboration}}}, \emph {et~al.},\ }\bibfield  {title} {\bibinfo {title} {{Extended Measurement of the Cosmic-Ray Electron and Positron Spectrum from 11 GeV to 4.8 TeV with the Calorimetric Electron Telescope on the International Space Station}},\ }\href {https://doi.org/10.1103/PhysRevLett.120.261102} {\bibfield  {journal} {\bibinfo  {journal} {\prl}\ }\textbf {\bibinfo {volume} {120}},\ \bibinfo {eid} {261102} (\bibinfo {year} {2018})},\ \Eprint {https://arxiv.org/abs/1806.09728} {arXiv:1806.09728 [astro-ph.HE]} \BibitemShut {NoStop}%
\bibitem [{\citenamefont {{Atkin}}\ \emph {et~al.}(2019)\citenamefont {{Atkin}} \emph {et~al.}}]{2019ARep...63...66A}%
  \BibitemOpen
  \bibfield  {author} {\bibinfo {author} {\bibfnamefont {E.~V.}\ \bibnamefont {{Atkin}}} \emph {et~al.},\ }\bibfield  {title} {\bibinfo {title} {{Energy Spectra of Cosmic-Ray Protons and Nuclei Measured in the NUCLEON Experiment Using a New Method}},\ }\href {https://doi.org/10.1134/S1063772919010013} {\bibfield  {journal} {\bibinfo  {journal} {Astronomy Reports}\ }\textbf {\bibinfo {volume} {63}},\ \bibinfo {pages} {66} (\bibinfo {year} {2019})}\BibitemShut {NoStop}%
\bibitem [{\citenamefont {{Aguilar}}\ \emph {et~al.}(2018)\citenamefont {{Aguilar}} \emph {et~al.}}]{2018PhRvL.120b1101A}%
  \BibitemOpen
  \bibfield  {author} {\bibinfo {author} {\bibfnamefont {M.}~\bibnamefont {{Aguilar}}} \emph {et~al.},\ }\bibfield  {title} {\bibinfo {title} {{Observation of New Properties of Secondary Cosmic Rays Lithium, Beryllium, and Boron by the Alpha Magnetic Spectrometer on the International Space Station}},\ }\href {https://doi.org/10.1103/PhysRevLett.120.021101} {\bibfield  {journal} {\bibinfo  {journal} {\prl}\ }\textbf {\bibinfo {volume} {120}},\ \bibinfo {eid} {021101} (\bibinfo {year} {2018})}\BibitemShut {NoStop}%
\bibitem [{\citenamefont {{Aguilar}}\ \emph {et~al.}(2019{\natexlab{a}})\citenamefont {{Aguilar}} \emph {et~al.}}]{2019PhRvL.122d1102A}%
  \BibitemOpen
  \bibfield  {author} {\bibinfo {author} {\bibfnamefont {M.}~\bibnamefont {{Aguilar}}} \emph {et~al.},\ }\bibfield  {title} {\bibinfo {title} {{Towards Understanding the Origin of Cosmic-Ray Positrons}},\ }\href {https://doi.org/10.1103/PhysRevLett.122.041102} {\bibfield  {journal} {\bibinfo  {journal} {\prl}\ }\textbf {\bibinfo {volume} {122}},\ \bibinfo {eid} {041102} (\bibinfo {year} {2019}{\natexlab{a}})}\BibitemShut {NoStop}%
\bibitem [{\citenamefont {{Aguilar}}\ \emph {et~al.}(2019{\natexlab{b}})\citenamefont {{Aguilar}} \emph {et~al.}}]{2019PhRvL.122j1101A}%
  \BibitemOpen
  \bibfield  {author} {\bibinfo {author} {\bibfnamefont {M.}~\bibnamefont {{Aguilar}}} \emph {et~al.},\ }\bibfield  {title} {\bibinfo {title} {{Towards Understanding the Origin of Cosmic-Ray Electrons}},\ }\href {https://doi.org/10.1103/PhysRevLett.122.101101} {\bibfield  {journal} {\bibinfo  {journal} {\prl}\ }\textbf {\bibinfo {volume} {122}},\ \bibinfo {eid} {101101} (\bibinfo {year} {2019}{\natexlab{b}})}\BibitemShut {NoStop}%
\bibitem [{\citenamefont {Adriani}\ \emph {et~al.}(2009)\citenamefont {Adriani} \emph {et~al.}}]{PAMELA:2008gwm}%
  \BibitemOpen
  \bibfield  {author} {\bibinfo {author} {\bibfnamefont {O.}~\bibnamefont {Adriani}} \emph {et~al.} (\bibinfo {collaboration} {PAMELA}),\ }\bibfield  {title} {\bibinfo {title} {{An anomalous positron abundance in cosmic rays with energies 1.5-100 GeV}},\ }\href {https://doi.org/10.1038/nature07942} {\bibfield  {journal} {\bibinfo  {journal} {Nature}\ }\textbf {\bibinfo {volume} {458}},\ \bibinfo {pages} {607} (\bibinfo {year} {2009})},\ \Eprint {https://arxiv.org/abs/0810.4995} {arXiv:0810.4995 [astro-ph]} \BibitemShut {NoStop}%
\bibitem [{\citenamefont {Moskalenko}\ and\ \citenamefont {Strong}(1998)}]{Moskalenko_1998}%
  \BibitemOpen
  \bibfield  {author} {\bibinfo {author} {\bibfnamefont {I.~V.}\ \bibnamefont {Moskalenko}}\ and\ \bibinfo {author} {\bibfnamefont {A.~W.}\ \bibnamefont {Strong}},\ }\bibfield  {title} {\bibinfo {title} {Production and propagation of cosmic‐ray positrons and electrons},\ }\href {https://doi.org/10.1086/305152} {\bibfield  {journal} {\bibinfo  {journal} {The Astrophysical Journal}\ }\textbf {\bibinfo {volume} {493}},\ \bibinfo {pages} {694–707} (\bibinfo {year} {1998})}\BibitemShut {NoStop}%
\bibitem [{\citenamefont {Shaviv}\ \emph {et~al.}(2009)\citenamefont {Shaviv}, \citenamefont {Nakar},\ and\ \citenamefont {Piran}}]{Shaviv:2009bu}%
  \BibitemOpen
  \bibfield  {author} {\bibinfo {author} {\bibfnamefont {N.~J.}\ \bibnamefont {Shaviv}}, \bibinfo {author} {\bibfnamefont {E.}~\bibnamefont {Nakar}},\ and\ \bibinfo {author} {\bibfnamefont {T.}~\bibnamefont {Piran}},\ }\bibfield  {title} {\bibinfo {title} {{Natural explanation for the anomalous positron to electron ratio with supernova remnants as the sole cosmic ray source}},\ }\href {https://doi.org/10.1103/PhysRevLett.103.111302} {\bibfield  {journal} {\bibinfo  {journal} {Phys. Rev. Lett.}\ }\textbf {\bibinfo {volume} {103}},\ \bibinfo {pages} {111302} (\bibinfo {year} {2009})},\ \Eprint {https://arxiv.org/abs/0902.0376} {arXiv:0902.0376 [astro-ph.HE]} \BibitemShut {NoStop}%
\bibitem [{\citenamefont {{Prantzos}}\ \emph {et~al.}(2011)\citenamefont {{Prantzos}}, \citenamefont {{Boehm}}, \citenamefont {{Bykov}}, \citenamefont {{Diehl}}, \citenamefont {{Ferri{\`e}re}}, \citenamefont {{Guessoum}}, \citenamefont {{Jean}}, \citenamefont {{Knoedlseder}}, \citenamefont {{Marcowith}}, \citenamefont {{Moskalenko}}, \citenamefont {{Strong}},\ and\ \citenamefont {{Weidenspointner}}}]{2011RvMP...83.1001P}%
  \BibitemOpen
  \bibfield  {author} {\bibinfo {author} {\bibfnamefont {N.}~\bibnamefont {{Prantzos}}}, \bibinfo {author} {\bibfnamefont {C.}~\bibnamefont {{Boehm}}}, \bibinfo {author} {\bibfnamefont {A.~M.}\ \bibnamefont {{Bykov}}}, \bibinfo {author} {\bibfnamefont {R.}~\bibnamefont {{Diehl}}}, \bibinfo {author} {\bibfnamefont {K.}~\bibnamefont {{Ferri{\`e}re}}}, \bibinfo {author} {\bibfnamefont {N.}~\bibnamefont {{Guessoum}}}, \bibinfo {author} {\bibfnamefont {P.}~\bibnamefont {{Jean}}}, \bibinfo {author} {\bibfnamefont {J.}~\bibnamefont {{Knoedlseder}}}, \bibinfo {author} {\bibfnamefont {A.}~\bibnamefont {{Marcowith}}}, \bibinfo {author} {\bibfnamefont {I.~V.}\ \bibnamefont {{Moskalenko}}}, \bibinfo {author} {\bibfnamefont {A.}~\bibnamefont {{Strong}}},\ and\ \bibinfo {author} {\bibfnamefont {G.}~\bibnamefont {{Weidenspointner}}},\ }\bibfield  {title} {\bibinfo {title} {{The 511 keV emission from positron annihilation in the Galaxy}},\ }\href {https://doi.org/10.1103/RevModPhys.83.1001} {\bibfield  {journal} {\bibinfo
  {journal} {Reviews of Modern Physics}\ }\textbf {\bibinfo {volume} {83}},\ \bibinfo {pages} {1001} (\bibinfo {year} {2011})},\ \Eprint {https://arxiv.org/abs/1009.4620} {arXiv:1009.4620 [astro-ph.HE]} \BibitemShut {NoStop}%
\bibitem [{\citenamefont {Porter}\ \emph {et~al.}(2022)\citenamefont {Porter}, \citenamefont {Jóhannesson},\ and\ \citenamefont {Moskalenko}}]{Porter_2022}%
  \BibitemOpen
  \bibfield  {author} {\bibinfo {author} {\bibfnamefont {T.~A.}\ \bibnamefont {Porter}}, \bibinfo {author} {\bibfnamefont {G.}~\bibnamefont {Jóhannesson}},\ and\ \bibinfo {author} {\bibfnamefont {I.~V.}\ \bibnamefont {Moskalenko}},\ }\bibfield  {title} {\bibinfo {title} {The galprop cosmic-ray propagation and nonthermal emissions framework: Release v57},\ }\href {https://doi.org/10.3847/1538-4365/ac80f6} {\bibfield  {journal} {\bibinfo  {journal} {The Astrophysical Journal Supplement Series}\ }\textbf {\bibinfo {volume} {262}},\ \bibinfo {pages} {30} (\bibinfo {year} {2022})}\BibitemShut {NoStop}%
\bibitem [{\citenamefont {{Yang}}\ \emph {et~al.}(2018)\citenamefont {{Yang}}, \citenamefont {{Kafexhiu}},\ and\ \citenamefont {{Aharonian}}}]{2018A&A...615A.108Y}%
  \BibitemOpen
  \bibfield  {author} {\bibinfo {author} {\bibfnamefont {R.-z.}\ \bibnamefont {{Yang}}}, \bibinfo {author} {\bibfnamefont {E.}~\bibnamefont {{Kafexhiu}}},\ and\ \bibinfo {author} {\bibfnamefont {F.}~\bibnamefont {{Aharonian}}},\ }\bibfield  {title} {\bibinfo {title} {{Exploring the shape of the {\ensuremath{\gamma}}-ray spectrum around the ``{\ensuremath{\pi}}$^{0}$-bump''}},\ }\href {https://doi.org/10.1051/0004-6361/201730908} {\bibfield  {journal} {\bibinfo  {journal} {Astronomy \& Astrophysics}\ }\textbf {\bibinfo {volume} {615}},\ \bibinfo {eid} {A108} (\bibinfo {year} {2018})},\ \Eprint {https://arxiv.org/abs/1803.05072} {arXiv:1803.05072 [astro-ph.HE]} \BibitemShut {NoStop}%
\bibitem [{\citenamefont {Moreno}\ \emph {et~al.}(2021)\citenamefont {Moreno} \emph {et~al.}}]{Moreno:2020bxn}%
  \BibitemOpen
  \bibfield  {author} {\bibinfo {author} {\bibfnamefont {J.}~\bibnamefont {Moreno}} \emph {et~al.},\ }\bibfield  {title} {\bibinfo {title} {{Spatially resolved star formation and fuelling in galaxy interactions}},\ }\href {https://doi.org/10.1093/mnras/staa2952} {\bibfield  {journal} {\bibinfo  {journal} {Mon. Not. Roy. Astron. Soc.}\ }\textbf {\bibinfo {volume} {503}},\ \bibinfo {pages} {3113} (\bibinfo {year} {2021})},\ \Eprint {https://arxiv.org/abs/2009.11289} {arXiv:2009.11289 [astro-ph.GA]} \BibitemShut {NoStop}%
\bibitem [{\citenamefont {Li}\ \emph {et~al.}(2024)\citenamefont {Li}, \citenamefont {Li}, \citenamefont {Cui}, \citenamefont {Marinacci}, \citenamefont {Sales}, \citenamefont {Vogelsberger},\ and\ \citenamefont {Torrey}}]{Li:2024aoq}%
  \BibitemOpen
  \bibfield  {author} {\bibinfo {author} {\bibfnamefont {C.}~\bibnamefont {Li}}, \bibinfo {author} {\bibfnamefont {H.}~\bibnamefont {Li}}, \bibinfo {author} {\bibfnamefont {W.}~\bibnamefont {Cui}}, \bibinfo {author} {\bibfnamefont {F.}~\bibnamefont {Marinacci}}, \bibinfo {author} {\bibfnamefont {L.~V.}\ \bibnamefont {Sales}}, \bibinfo {author} {\bibfnamefont {M.}~\bibnamefont {Vogelsberger}},\ and\ \bibinfo {author} {\bibfnamefont {P.}~\bibnamefont {Torrey}},\ }\bibfield  {title} {\bibinfo {title} {{Evolution and distribution of superbubbles in simulated Milky Way-like galaxies}},\ }\href {https://doi.org/10.1093/mnras/stae797} {\bibfield  {journal} {\bibinfo  {journal} {Mon. Not. Roy. Astron. Soc.}\ }\textbf {\bibinfo {volume} {529}},\ \bibinfo {pages} {4073} (\bibinfo {year} {2024})},\ \Eprint {https://arxiv.org/abs/2403.12135} {arXiv:2403.12135 [astro-ph.GA]} \BibitemShut {NoStop}%
\bibitem [{\citenamefont {Debattista}\ \emph {et~al.}(2019)\citenamefont {Debattista}, \citenamefont {Gonzalez}, \citenamefont {Sanderson}, \citenamefont {El-Badry}, \citenamefont {Garrison-Kimmel}, \citenamefont {Wetzel}, \citenamefont {Faucher-Giguère},\ and\ \citenamefont {Hopkins}}]{Debattista_2019}%
  \BibitemOpen
  \bibfield  {author} {\bibinfo {author} {\bibfnamefont {V.~P.}\ \bibnamefont {Debattista}}, \bibinfo {author} {\bibfnamefont {O.~A.}\ \bibnamefont {Gonzalez}}, \bibinfo {author} {\bibfnamefont {R.~E.}\ \bibnamefont {Sanderson}}, \bibinfo {author} {\bibfnamefont {K.}~\bibnamefont {El-Badry}}, \bibinfo {author} {\bibfnamefont {S.}~\bibnamefont {Garrison-Kimmel}}, \bibinfo {author} {\bibfnamefont {A.}~\bibnamefont {Wetzel}}, \bibinfo {author} {\bibfnamefont {C.-A.}\ \bibnamefont {Faucher-Giguère}},\ and\ \bibinfo {author} {\bibfnamefont {P.~F.}\ \bibnamefont {Hopkins}},\ }\bibfield  {title} {\bibinfo {title} {Formation, vertex deviation, and age of the milky way’s bulge: input from a cosmological simulation with a late-forming bar},\ }\href {https://doi.org/10.1093/mnras/stz746} {\bibfield  {journal} {\bibinfo  {journal} {Monthly Notices of the Royal Astronomical Society}\ }\textbf {\bibinfo {volume} {485}},\ \bibinfo {pages} {5073–5085} (\bibinfo {year} {2019})}\BibitemShut {NoStop}%
\bibitem [{\citenamefont {Elia}\ \emph {et~al.}(2022)\citenamefont {Elia}, \citenamefont {Molinari}, \citenamefont {Schisano}, \citenamefont {Soler}, \citenamefont {Merello}, \citenamefont {Russeil}, \citenamefont {Veneziani}, \citenamefont {Zavagno}, \citenamefont {Noriega-Crespo}, \citenamefont {Olmi}, \citenamefont {Benedettini}, \citenamefont {Hennebelle}, \citenamefont {Klessen}, \citenamefont {Leurini}, \citenamefont {Paladini}, \citenamefont {Pezzuto}, \citenamefont {Traficante}, \citenamefont {Eden}, \citenamefont {Martin}, \citenamefont {Sormani}, \citenamefont {Coletta}, \citenamefont {Colman}, \citenamefont {Plume}, \citenamefont {Maruccia}, \citenamefont {Mininni},\ and\ \citenamefont {Liu}}]{Elia_2022}%
  \BibitemOpen
  \bibfield  {author} {\bibinfo {author} {\bibfnamefont {D.}~\bibnamefont {Elia}}, \bibinfo {author} {\bibfnamefont {S.}~\bibnamefont {Molinari}}, \bibinfo {author} {\bibfnamefont {E.}~\bibnamefont {Schisano}}, \bibinfo {author} {\bibfnamefont {J.~D.}\ \bibnamefont {Soler}}, \bibinfo {author} {\bibfnamefont {M.}~\bibnamefont {Merello}}, \bibinfo {author} {\bibfnamefont {D.}~\bibnamefont {Russeil}}, \bibinfo {author} {\bibfnamefont {M.}~\bibnamefont {Veneziani}}, \bibinfo {author} {\bibfnamefont {A.}~\bibnamefont {Zavagno}}, \bibinfo {author} {\bibfnamefont {A.}~\bibnamefont {Noriega-Crespo}}, \bibinfo {author} {\bibfnamefont {L.}~\bibnamefont {Olmi}}, \bibinfo {author} {\bibfnamefont {M.}~\bibnamefont {Benedettini}}, \bibinfo {author} {\bibfnamefont {P.}~\bibnamefont {Hennebelle}}, \bibinfo {author} {\bibfnamefont {R.~S.}\ \bibnamefont {Klessen}}, \bibinfo {author} {\bibfnamefont {S.}~\bibnamefont {Leurini}}, \bibinfo {author} {\bibfnamefont {R.}~\bibnamefont {Paladini}}, \bibinfo {author} {\bibfnamefont
  {S.}~\bibnamefont {Pezzuto}}, \bibinfo {author} {\bibfnamefont {A.}~\bibnamefont {Traficante}}, \bibinfo {author} {\bibfnamefont {D.~J.}\ \bibnamefont {Eden}}, \bibinfo {author} {\bibfnamefont {P.~G.}\ \bibnamefont {Martin}}, \bibinfo {author} {\bibfnamefont {M.}~\bibnamefont {Sormani}}, \bibinfo {author} {\bibfnamefont {A.}~\bibnamefont {Coletta}}, \bibinfo {author} {\bibfnamefont {T.}~\bibnamefont {Colman}}, \bibinfo {author} {\bibfnamefont {R.}~\bibnamefont {Plume}}, \bibinfo {author} {\bibfnamefont {Y.}~\bibnamefont {Maruccia}}, \bibinfo {author} {\bibfnamefont {C.}~\bibnamefont {Mininni}},\ and\ \bibinfo {author} {\bibfnamefont {S.~J.}\ \bibnamefont {Liu}},\ }\bibfield  {title} {\bibinfo {title} {The star formation rate of the milky way as seen by herschel},\ }\href {https://doi.org/10.3847/1538-4357/aca27d} {\bibfield  {journal} {\bibinfo  {journal} {The Astrophysical Journal}\ }\textbf {\bibinfo {volume} {941}},\ \bibinfo {pages} {162} (\bibinfo {year} {2022})}\BibitemShut {NoStop}%
\bibitem [{\citenamefont {Jaffe}\ \emph {et~al.}(2010)\citenamefont {Jaffe}, \citenamefont {Leahy}, \citenamefont {Banday}, \citenamefont {Leach}, \citenamefont {Lowe},\ and\ \citenamefont {Wilkinson}}]{Jaffe_2010}%
  \BibitemOpen
  \bibfield  {author} {\bibinfo {author} {\bibfnamefont {T.~R.}\ \bibnamefont {Jaffe}}, \bibinfo {author} {\bibfnamefont {J.~P.}\ \bibnamefont {Leahy}}, \bibinfo {author} {\bibfnamefont {A.~J.}\ \bibnamefont {Banday}}, \bibinfo {author} {\bibfnamefont {S.~M.}\ \bibnamefont {Leach}}, \bibinfo {author} {\bibfnamefont {S.~R.}\ \bibnamefont {Lowe}},\ and\ \bibinfo {author} {\bibfnamefont {A.}~\bibnamefont {Wilkinson}},\ }\bibfield  {title} {\bibinfo {title} {Modelling the galactic magnetic field on the plane in two dimensions},\ }\href {https://doi.org/10.1111/j.1365-2966.2009.15745.x} {\bibfield  {journal} {\bibinfo  {journal} {Monthly Notices of the Royal Astronomical Society}\ }\textbf {\bibinfo {volume} {401}},\ \bibinfo {pages} {1013–1028} (\bibinfo {year} {2010})}\BibitemShut {NoStop}%
\bibitem [{\citenamefont {Jaffe}(2019)}]{Jaffe_2019}%
  \BibitemOpen
  \bibfield  {author} {\bibinfo {author} {\bibfnamefont {T.}~\bibnamefont {Jaffe}},\ }\bibfield  {title} {\bibinfo {title} {Practical modeling of large-scale galactic magnetic fields: Status and prospects},\ }\href {https://doi.org/10.3390/galaxies7020052} {\bibfield  {journal} {\bibinfo  {journal} {Galaxies}\ }\textbf {\bibinfo {volume} {7}},\ \bibinfo {pages} {52} (\bibinfo {year} {2019})}\BibitemShut {NoStop}%
\bibitem [{\citenamefont {Battersby}\ \emph {et~al.}(2025)\citenamefont {Battersby}, \citenamefont {Walker}, \citenamefont {Barnes}, \citenamefont {Ginsburg}, \citenamefont {Lipman}, \citenamefont {Alboslani}, \citenamefont {Hatchfield}, \citenamefont {Bally}, \citenamefont {Glover}, \citenamefont {Henshaw}, \citenamefont {Immer}, \citenamefont {Klessen}, \citenamefont {Longmore}, \citenamefont {Mills}, \citenamefont {Molinari}, \citenamefont {Smith}, \citenamefont {Sormani}, \citenamefont {Tress},\ and\ \citenamefont {Zhang}}]{battersby20253dcmzicentral}%
  \BibitemOpen
  \bibfield  {author} {\bibinfo {author} {\bibfnamefont {C.}~\bibnamefont {Battersby}}, \bibinfo {author} {\bibfnamefont {D.~L.}\ \bibnamefont {Walker}}, \bibinfo {author} {\bibfnamefont {A.}~\bibnamefont {Barnes}}, \bibinfo {author} {\bibfnamefont {A.}~\bibnamefont {Ginsburg}}, \bibinfo {author} {\bibfnamefont {D.}~\bibnamefont {Lipman}}, \bibinfo {author} {\bibfnamefont {D.}~\bibnamefont {Alboslani}}, \bibinfo {author} {\bibfnamefont {H.~P.}\ \bibnamefont {Hatchfield}}, \bibinfo {author} {\bibfnamefont {J.}~\bibnamefont {Bally}}, \bibinfo {author} {\bibfnamefont {S.~C.~O.}\ \bibnamefont {Glover}}, \bibinfo {author} {\bibfnamefont {J.~D.}\ \bibnamefont {Henshaw}}, \bibinfo {author} {\bibfnamefont {K.}~\bibnamefont {Immer}}, \bibinfo {author} {\bibfnamefont {R.~S.}\ \bibnamefont {Klessen}}, \bibinfo {author} {\bibfnamefont {S.~N.}\ \bibnamefont {Longmore}}, \bibinfo {author} {\bibfnamefont {E.~A.~C.}\ \bibnamefont {Mills}}, \bibinfo {author} {\bibfnamefont {S.}~\bibnamefont {Molinari}}, \bibinfo {author}
  {\bibfnamefont {R.}~\bibnamefont {Smith}}, \bibinfo {author} {\bibfnamefont {M.~C.}\ \bibnamefont {Sormani}}, \bibinfo {author} {\bibfnamefont {R.~G.}\ \bibnamefont {Tress}},\ and\ \bibinfo {author} {\bibfnamefont {Q.}~\bibnamefont {Zhang}},\ }\href {https://arxiv.org/abs/2410.17334} {\bibinfo {title} {3-d cmz i: Central molecular zone overview}} (\bibinfo {year} {2025}),\ \Eprint {https://arxiv.org/abs/2410.17334} {arXiv:2410.17334 [astro-ph.GA]} \BibitemShut {NoStop}%
\bibitem [{\citenamefont {Walker}\ \emph {et~al.}(2024)\citenamefont {Walker}, \citenamefont {Battersby}, \citenamefont {Lipman}, \citenamefont {Sormani}, \citenamefont {Ginsburg}, \citenamefont {Glover}, \citenamefont {Henshaw}, \citenamefont {Longmore}, \citenamefont {Klessen}, \citenamefont {Immer}, \citenamefont {Alboslani}, \citenamefont {Bally}, \citenamefont {Barnes}, \citenamefont {Hatchfield}, \citenamefont {Mills}, \citenamefont {Smith}, \citenamefont {Tress},\ and\ \citenamefont {Zhang}}]{walker20243dcmziiiconstraining}%
  \BibitemOpen
  \bibfield  {author} {\bibinfo {author} {\bibfnamefont {D.~L.}\ \bibnamefont {Walker}}, \bibinfo {author} {\bibfnamefont {C.}~\bibnamefont {Battersby}}, \bibinfo {author} {\bibfnamefont {D.}~\bibnamefont {Lipman}}, \bibinfo {author} {\bibfnamefont {M.~C.}\ \bibnamefont {Sormani}}, \bibinfo {author} {\bibfnamefont {A.}~\bibnamefont {Ginsburg}}, \bibinfo {author} {\bibfnamefont {S.~C.~O.}\ \bibnamefont {Glover}}, \bibinfo {author} {\bibfnamefont {J.~D.}\ \bibnamefont {Henshaw}}, \bibinfo {author} {\bibfnamefont {S.~N.}\ \bibnamefont {Longmore}}, \bibinfo {author} {\bibfnamefont {R.~S.}\ \bibnamefont {Klessen}}, \bibinfo {author} {\bibfnamefont {K.}~\bibnamefont {Immer}}, \bibinfo {author} {\bibfnamefont {D.}~\bibnamefont {Alboslani}}, \bibinfo {author} {\bibfnamefont {J.}~\bibnamefont {Bally}}, \bibinfo {author} {\bibfnamefont {A.}~\bibnamefont {Barnes}}, \bibinfo {author} {\bibfnamefont {H.~P.}\ \bibnamefont {Hatchfield}}, \bibinfo {author} {\bibfnamefont {E.~A.~C.}\ \bibnamefont {Mills}}, \bibinfo {author}
  {\bibfnamefont {R.}~\bibnamefont {Smith}}, \bibinfo {author} {\bibfnamefont {R.~G.}\ \bibnamefont {Tress}},\ and\ \bibinfo {author} {\bibfnamefont {Q.}~\bibnamefont {Zhang}},\ }\href {https://arxiv.org/abs/2410.17320} {\bibinfo {title} {3-d cmz iii: Constraining the 3-d structure of the central molecular zone via molecular line emission and absorption}} (\bibinfo {year} {2024}),\ \Eprint {https://arxiv.org/abs/2410.17320} {arXiv:2410.17320 [astro-ph.GA]} \BibitemShut {NoStop}%
\bibitem [{\citenamefont {{Ackermann}}\ \emph {et~al.}(2012)\citenamefont {{Ackermann}} \emph {et~al.}}]{2012ApJ...750....3A}%
  \BibitemOpen
  \bibfield  {author} {\bibinfo {author} {\bibfnamefont {M.}~\bibnamefont {{Ackermann}}} \emph {et~al.},\ }\bibfield  {title} {\bibinfo {title} {{Fermi-LAT Observations of the Diffuse {\ensuremath{\gamma}}-Ray Emission: Implications for Cosmic Rays and the Interstellar Medium}},\ }\href {https://doi.org/10.1088/0004-637X/750/1/3} {\bibfield  {journal} {\bibinfo  {journal} {The Astrophysical Journal}\ }\textbf {\bibinfo {volume} {750}},\ \bibinfo {eid} {3} (\bibinfo {year} {2012})},\ \Eprint {https://arxiv.org/abs/1202.4039} {arXiv:1202.4039 [astro-ph.HE]} \BibitemShut {NoStop}%
\bibitem [{\citenamefont {Yusef-Zadeh}\ \emph {et~al.}(2012)\citenamefont {Yusef-Zadeh}, \citenamefont {Hewitt}, \citenamefont {Wardle}, \citenamefont {Tatischeff}, \citenamefont {Roberts}, \citenamefont {Cotton}, \citenamefont {Uchiyama}, \citenamefont {Nobukawa}, \citenamefont {Tsuru}, \citenamefont {Heinke},\ and\ \citenamefont {Royster}}]{Yusef_Zadeh_2012}%
  \BibitemOpen
  \bibfield  {author} {\bibinfo {author} {\bibfnamefont {F.}~\bibnamefont {Yusef-Zadeh}}, \bibinfo {author} {\bibfnamefont {J.~W.}\ \bibnamefont {Hewitt}}, \bibinfo {author} {\bibfnamefont {M.}~\bibnamefont {Wardle}}, \bibinfo {author} {\bibfnamefont {V.}~\bibnamefont {Tatischeff}}, \bibinfo {author} {\bibfnamefont {D.~A.}\ \bibnamefont {Roberts}}, \bibinfo {author} {\bibfnamefont {W.}~\bibnamefont {Cotton}}, \bibinfo {author} {\bibfnamefont {H.}~\bibnamefont {Uchiyama}}, \bibinfo {author} {\bibfnamefont {M.}~\bibnamefont {Nobukawa}}, \bibinfo {author} {\bibfnamefont {T.~G.}\ \bibnamefont {Tsuru}}, \bibinfo {author} {\bibfnamefont {C.}~\bibnamefont {Heinke}},\ and\ \bibinfo {author} {\bibfnamefont {M.}~\bibnamefont {Royster}},\ }\bibfield  {title} {\bibinfo {title} {Interacting cosmic rays with molecular clouds: A bremsstrahlung origin of diffuse high-energy emission from the inner 2°×1° of the galactic center},\ }\href {https://doi.org/10.1088/0004-637x/762/1/33} {\bibfield  {journal} {\bibinfo  {journal}
  {The Astrophysical Journal}\ }\textbf {\bibinfo {volume} {762}},\ \bibinfo {pages} {33} (\bibinfo {year} {2012})}\BibitemShut {NoStop}%
\bibitem [{\citenamefont {{Yan}}\ and\ \citenamefont {{Lazarian}}(2004)}]{yan.lazarian.04:cr.scattering.fast.modes}%
  \BibitemOpen
  \bibfield  {author} {\bibinfo {author} {\bibfnamefont {H.}~\bibnamefont {{Yan}}}\ and\ \bibinfo {author} {\bibfnamefont {A.}~\bibnamefont {{Lazarian}}},\ }\bibfield  {title} {\bibinfo {title} {{Cosmic-Ray Scattering and Streaming in Compressible Magnetohydrodynamic Turbulence}},\ }\href {https://doi.org/10.1086/423733} {\bibfield  {journal} {\bibinfo  {journal} {The Astrophysical Journal}\ }\textbf {\bibinfo {volume} {614}},\ \bibinfo {pages} {757} (\bibinfo {year} {2004})},\ \Eprint {https://arxiv.org/abs/astro-ph/0408172} {arXiv:astro-ph/0408172 [astro-ph]} \BibitemShut {NoStop}%
\bibitem [{\citenamefont {{Yan}}\ and\ \citenamefont {{Lazarian}}(2008)}]{yan.lazarian.2008:cr.propagation.with.streaming}%
  \BibitemOpen
  \bibfield  {author} {\bibinfo {author} {\bibfnamefont {H.}~\bibnamefont {{Yan}}}\ and\ \bibinfo {author} {\bibfnamefont {A.}~\bibnamefont {{Lazarian}}},\ }\bibfield  {title} {\bibinfo {title} {{Cosmic-Ray Propagation: Nonlinear Diffusion Parallel and Perpendicular to Mean Magnetic Field}},\ }\href {https://doi.org/10.1086/524771} {\bibfield  {journal} {\bibinfo  {journal} {The Astrophysical Journal}\ }\textbf {\bibinfo {volume} {673}},\ \bibinfo {eid} {942-953} (\bibinfo {year} {2008})},\ \Eprint {https://arxiv.org/abs/0710.2617} {arXiv:0710.2617} \BibitemShut {NoStop}%
\bibitem [{\citenamefont {{Lazarian}}(2016)}]{lazarian:2016.cr.wave.damping}%
  \BibitemOpen
  \bibfield  {author} {\bibinfo {author} {\bibfnamefont {A.}~\bibnamefont {{Lazarian}}},\ }\bibfield  {title} {\bibinfo {title} {{Damping of Alfv{\'e}n Waves by Turbulence and Its Consequences: From Cosmic-ray Streaming to Launching Winds}},\ }\href {https://doi.org/10.3847/1538-4357/833/2/131} {\bibfield  {journal} {\bibinfo  {journal} {The Astrophysical Journal}\ }\textbf {\bibinfo {volume} {833}},\ \bibinfo {eid} {131} (\bibinfo {year} {2016})},\ \Eprint {https://arxiv.org/abs/1607.02042} {arXiv:1607.02042 [astro-ph.HE]} \BibitemShut {NoStop}%
\bibitem [{\citenamefont {{Zweibel}}(2017)}]{zweibel:cr.feedback.review}%
  \BibitemOpen
  \bibfield  {author} {\bibinfo {author} {\bibfnamefont {E.~G.}\ \bibnamefont {{Zweibel}}},\ }\bibfield  {title} {\bibinfo {title} {{The basis for cosmic ray feedback: Written on the wind}},\ }\href {https://doi.org/10.1063/1.4984017} {\bibfield  {journal} {\bibinfo  {journal} {Physics of Plasmas}\ }\textbf {\bibinfo {volume} {24}},\ \bibinfo {eid} {055402} (\bibinfo {year} {2017})}\BibitemShut {NoStop}%
\bibitem [{\citenamefont {{Ruszkowski}}\ \emph {et~al.}(2017)\citenamefont {{Ruszkowski}}, \citenamefont {{Yang}},\ and\ \citenamefont {{Zweibel}}}]{Rusz17}%
  \BibitemOpen
  \bibfield  {author} {\bibinfo {author} {\bibfnamefont {M.}~\bibnamefont {{Ruszkowski}}}, \bibinfo {author} {\bibfnamefont {H.-Y.~K.}\ \bibnamefont {{Yang}}},\ and\ \bibinfo {author} {\bibfnamefont {E.}~\bibnamefont {{Zweibel}}},\ }\bibfield  {title} {\bibinfo {title} {{Global Simulations of Galactic Winds Including Cosmic-ray Streaming}},\ }\href {https://doi.org/10.3847/1538-4357/834/2/208} {\bibfield  {journal} {\bibinfo  {journal} {The Astrophysical Journal}\ }\textbf {\bibinfo {volume} {834}},\ \bibinfo {eid} {208} (\bibinfo {year} {2017})},\ \Eprint {https://arxiv.org/abs/1602.04856} {arXiv:1602.04856} \BibitemShut {NoStop}%
\bibitem [{\citenamefont {{Farber}}\ \emph {et~al.}(2018)\citenamefont {{Farber}}, \citenamefont {{Ruszkowski}}, \citenamefont {{Yang}},\ and\ \citenamefont {{Zweibel}}}]{farber:decoupled.crs.in.neutral.gas}%
  \BibitemOpen
  \bibfield  {author} {\bibinfo {author} {\bibfnamefont {R.}~\bibnamefont {{Farber}}}, \bibinfo {author} {\bibfnamefont {M.}~\bibnamefont {{Ruszkowski}}}, \bibinfo {author} {\bibfnamefont {H.-Y.~K.}\ \bibnamefont {{Yang}}},\ and\ \bibinfo {author} {\bibfnamefont {E.~G.}\ \bibnamefont {{Zweibel}}},\ }\bibfield  {title} {\bibinfo {title} {{Impact of Cosmic-Ray Transport on Galactic Winds}},\ }\href {https://doi.org/10.3847/1538-4357/aab26d} {\bibfield  {journal} {\bibinfo  {journal} {The Astrophysical Journal}\ }\textbf {\bibinfo {volume} {856}},\ \bibinfo {eid} {112} (\bibinfo {year} {2018})},\ \Eprint {https://arxiv.org/abs/1707.04579} {arXiv:1707.04579 [astro-ph.HE]} \BibitemShut {NoStop}%
\bibitem [{\citenamefont {{Kempski}}\ \emph {et~al.}(2020)\citenamefont {{Kempski}}, \citenamefont {{Quataert}},\ and\ \citenamefont {{Squire}}}]{kempski:2020.cr.soundwave.instabilities.highbeta.plasmas.resemble.perseus.density.fluctuation.power.spectra}%
  \BibitemOpen
  \bibfield  {author} {\bibinfo {author} {\bibfnamefont {P.}~\bibnamefont {{Kempski}}}, \bibinfo {author} {\bibfnamefont {E.}~\bibnamefont {{Quataert}}},\ and\ \bibinfo {author} {\bibfnamefont {J.}~\bibnamefont {{Squire}}},\ }\bibfield  {title} {\bibinfo {title} {{Sound-wave instabilities in dilute plasmas with cosmic rays: implications for cosmic ray confinement and the Perseus X-ray ripples}},\ }\href {https://doi.org/10.1093/mnras/staa535} {\bibfield  {journal} {\bibinfo  {journal} {Monthly Notices of the Royal Astronomical Society}\ }\textbf {\bibinfo {volume} {493}},\ \bibinfo {pages} {5323} (\bibinfo {year} {2020})},\ \Eprint {https://arxiv.org/abs/1911.06328} {arXiv:1911.06328 [astro-ph.GA]} \BibitemShut {NoStop}%
\bibitem [{\citenamefont {{Kempski}}\ \emph {et~al.}(2023)\citenamefont {{Kempski}}, \citenamefont {{Fielding}}, \citenamefont {{Quataert}}, \citenamefont {{Galishnikova}}, \citenamefont {{Kunz}}, \citenamefont {{Philippov}},\ and\ \citenamefont {{Ripperda}}}]{kempski:2023.large.amplitude.fluctuations.and.cr.scattering}%
  \BibitemOpen
  \bibfield  {author} {\bibinfo {author} {\bibfnamefont {P.}~\bibnamefont {{Kempski}}}, \bibinfo {author} {\bibfnamefont {D.~B.}\ \bibnamefont {{Fielding}}}, \bibinfo {author} {\bibfnamefont {E.}~\bibnamefont {{Quataert}}}, \bibinfo {author} {\bibfnamefont {A.~K.}\ \bibnamefont {{Galishnikova}}}, \bibinfo {author} {\bibfnamefont {M.~W.}\ \bibnamefont {{Kunz}}}, \bibinfo {author} {\bibfnamefont {A.~A.}\ \bibnamefont {{Philippov}}},\ and\ \bibinfo {author} {\bibfnamefont {B.}~\bibnamefont {{Ripperda}}},\ }\bibfield  {title} {\bibinfo {title} {{Cosmic ray transport in large-amplitude turbulence with small-scale field reversals}},\ }\href {https://doi.org/10.48550/arXiv.2304.12335} {\bibfield  {journal} {\bibinfo  {journal} {arXiv e-prints}\ ,\ \bibinfo {eid} {arXiv:2304.12335}} (\bibinfo {year} {2023})},\ \Eprint {https://arxiv.org/abs/2304.12335} {arXiv:2304.12335 [astro-ph.HE]} \BibitemShut {NoStop}%
\bibitem [{\citenamefont {{Kempski}}\ \emph {et~al.}(2024)\citenamefont {{Kempski}}, \citenamefont {{Li}}, \citenamefont {{Fielding}}, \citenamefont {{Quataert}}, \citenamefont {{Phinney}}, \citenamefont {{Kunz}}, \citenamefont {{Jow}},\ and\ \citenamefont {{Philippov}}}]{kempski.li.2024:unified.cr.scattering.plasma.scattering.from.strong.field.curvature.intermittent.ism.structure.explained.together}%
  \BibitemOpen
  \bibfield  {author} {\bibinfo {author} {\bibfnamefont {P.}~\bibnamefont {{Kempski}}}, \bibinfo {author} {\bibfnamefont {D.}~\bibnamefont {{Li}}}, \bibinfo {author} {\bibfnamefont {D.~B.}\ \bibnamefont {{Fielding}}}, \bibinfo {author} {\bibfnamefont {E.}~\bibnamefont {{Quataert}}}, \bibinfo {author} {\bibfnamefont {E.~S.}\ \bibnamefont {{Phinney}}}, \bibinfo {author} {\bibfnamefont {M.~W.}\ \bibnamefont {{Kunz}}}, \bibinfo {author} {\bibfnamefont {D.~L.}\ \bibnamefont {{Jow}}},\ and\ \bibinfo {author} {\bibfnamefont {A.~A.}\ \bibnamefont {{Philippov}}},\ }\bibfield  {title} {\bibinfo {title} {{A Unified Model of Cosmic Ray Propagation and Radio Extreme Scattering Events from Intermittent Interstellar Structures}},\ }\href {https://doi.org/10.48550/arXiv.2412.03649} {\bibfield  {journal} {\bibinfo  {journal} {arXiv e-prints}\ ,\ \bibinfo {eid} {arXiv:2412.03649}} (\bibinfo {year} {2024})},\ \Eprint {https://arxiv.org/abs/2412.03649} {arXiv:2412.03649 [astro-ph.HE]} \BibitemShut {NoStop}%
\bibitem [{\citenamefont {{Krumholz}}\ \emph {et~al.}(2020)\citenamefont {{Krumholz}}, \citenamefont {{Crocker}}, \citenamefont {{Xu}}, \citenamefont {{Lazarian}}, \citenamefont {{Rosevear}},\ and\ \citenamefont {{Bedwell-Wilson}}}]{krumholz:2020.cr.transport.starbursts.upper.limit.kappa.gamma.rays}%
  \BibitemOpen
  \bibfield  {author} {\bibinfo {author} {\bibfnamefont {M.~R.}\ \bibnamefont {{Krumholz}}}, \bibinfo {author} {\bibfnamefont {R.~M.}\ \bibnamefont {{Crocker}}}, \bibinfo {author} {\bibfnamefont {S.}~\bibnamefont {{Xu}}}, \bibinfo {author} {\bibfnamefont {A.}~\bibnamefont {{Lazarian}}}, \bibinfo {author} {\bibfnamefont {M.~T.}\ \bibnamefont {{Rosevear}}},\ and\ \bibinfo {author} {\bibfnamefont {J.}~\bibnamefont {{Bedwell-Wilson}}},\ }\bibfield  {title} {\bibinfo {title} {{Cosmic ray transport in starburst galaxies}},\ }\href {https://doi.org/10.1093/mnras/staa493} {\bibfield  {journal} {\bibinfo  {journal} {Monthly Notices of the Royal Astronomical Society}\ }\textbf {\bibinfo {volume} {493}},\ \bibinfo {pages} {2817} (\bibinfo {year} {2020})},\ \Eprint {https://arxiv.org/abs/1911.09774} {arXiv:1911.09774 [astro-ph.HE]} \BibitemShut {NoStop}%
\bibitem [{\citenamefont {{Hopkins}}\ \emph {et~al.}(2022{\natexlab{a}})\citenamefont {{Hopkins}}, \citenamefont {{Squire}},\ and\ \citenamefont {{Butsky}}}]{hopkins:m1.cr.closure}%
  \BibitemOpen
  \bibfield  {author} {\bibinfo {author} {\bibfnamefont {P.~F.}\ \bibnamefont {{Hopkins}}}, \bibinfo {author} {\bibfnamefont {J.}~\bibnamefont {{Squire}}},\ and\ \bibinfo {author} {\bibfnamefont {I.~S.}\ \bibnamefont {{Butsky}}},\ }\bibfield  {title} {\bibinfo {title} {{A consistent reduced-speed-of-light formulation of cosmic ray transport valid in weak- and strong-scattering regimes}},\ }\href {https://doi.org/10.1093/mnras/stab2635} {\bibfield  {journal} {\bibinfo  {journal} {Monthly Notices of the Royal Astronomical Society}\ }\textbf {\bibinfo {volume} {509}},\ \bibinfo {pages} {3779} (\bibinfo {year} {2022}{\natexlab{a}})},\ \Eprint {https://arxiv.org/abs/2103.10443} {arXiv:2103.10443 [astro-ph.HE]} \BibitemShut {NoStop}%
\bibitem [{\citenamefont {{Ji}}\ \emph {et~al.}(2022)\citenamefont {{Ji}}, \citenamefont {{Squire}},\ and\ \citenamefont {{Hopkins}}}]{ji:2021.cr.mhd.pic.dust.sims}%
  \BibitemOpen
  \bibfield  {author} {\bibinfo {author} {\bibfnamefont {S.}~\bibnamefont {{Ji}}}, \bibinfo {author} {\bibfnamefont {J.}~\bibnamefont {{Squire}}},\ and\ \bibinfo {author} {\bibfnamefont {P.~F.}\ \bibnamefont {{Hopkins}}},\ }\bibfield  {title} {\bibinfo {title} {{Numerical study of cosmic ray confinement through dust resonant drag instabilities}},\ }\href {https://doi.org/10.1093/mnras/stac895} {\bibfield  {journal} {\bibinfo  {journal} {Monthly Notices of the Royal Astronomical Society}\ }\textbf {\bibinfo {volume} {513}},\ \bibinfo {pages} {282} (\bibinfo {year} {2022})},\ \Eprint {https://arxiv.org/abs/2112.00752} {arXiv:2112.00752 [astro-ph.HE]} \BibitemShut {NoStop}%
\bibitem [{\citenamefont {{Hopkins}}\ \emph {et~al.}(2022{\natexlab{b}})\citenamefont {{Hopkins}}, \citenamefont {{Squire}}, \citenamefont {{Butsky}},\ and\ \citenamefont {{Ji}}}]{hopkins:2021.sc.et.models.incompatible.obs}%
  \BibitemOpen
  \bibfield  {author} {\bibinfo {author} {\bibfnamefont {P.~F.}\ \bibnamefont {{Hopkins}}}, \bibinfo {author} {\bibfnamefont {J.}~\bibnamefont {{Squire}}}, \bibinfo {author} {\bibfnamefont {I.~S.}\ \bibnamefont {{Butsky}}},\ and\ \bibinfo {author} {\bibfnamefont {S.}~\bibnamefont {{Ji}}},\ }\bibfield  {title} {\bibinfo {title} {{Standard self-confinement and extrinsic turbulence models for cosmic ray transport are fundamentally incompatible with observations}},\ }\href {https://doi.org/10.1093/mnras/stac2909} {\bibfield  {journal} {\bibinfo  {journal} {Monthly Notices of the Royal Astronomical Society}\ }\textbf {\bibinfo {volume} {517}},\ \bibinfo {pages} {5413} (\bibinfo {year} {2022}{\natexlab{b}})},\ \Eprint {https://arxiv.org/abs/2112.02153} {arXiv:2112.02153 [astro-ph.HE]} \BibitemShut {NoStop}%
\bibitem [{\citenamefont {{Squire}}\ \emph {et~al.}(2021)\citenamefont {{Squire}}, \citenamefont {{Hopkins}}, \citenamefont {{Quataert}},\ and\ \citenamefont {{Kempski}}}]{squire:2021.dust.cr.confinement.damping}%
  \BibitemOpen
  \bibfield  {author} {\bibinfo {author} {\bibfnamefont {J.}~\bibnamefont {{Squire}}}, \bibinfo {author} {\bibfnamefont {P.~F.}\ \bibnamefont {{Hopkins}}}, \bibinfo {author} {\bibfnamefont {E.}~\bibnamefont {{Quataert}}},\ and\ \bibinfo {author} {\bibfnamefont {P.}~\bibnamefont {{Kempski}}},\ }\bibfield  {title} {\bibinfo {title} {{The impact of astrophysical dust grains on the confinement of cosmic rays}},\ }\href {https://doi.org/10.1093/mnras/stab179} {\bibfield  {journal} {\bibinfo  {journal} {Monthly Notices of the Royal Astronomical Society}\ }\textbf {\bibinfo {volume} {502}},\ \bibinfo {pages} {2630} (\bibinfo {year} {2021})},\ \Eprint {https://arxiv.org/abs/2011.02497} {arXiv:2011.02497 [astro-ph.HE]} \BibitemShut {NoStop}%
\bibitem [{\citenamefont {{Butsky}}\ \emph {et~al.}(2024)\citenamefont {{Butsky}}, \citenamefont {{Hopkins}}, \citenamefont {{Kempski}}, \citenamefont {{Ponnada}}, \citenamefont {{Quataert}},\ and\ \citenamefont {{Squire}}}]{butsky:2023.cosmic.ray.scattering.patchy.ism.structures}%
  \BibitemOpen
  \bibfield  {author} {\bibinfo {author} {\bibfnamefont {I.~S.}\ \bibnamefont {{Butsky}}}, \bibinfo {author} {\bibfnamefont {P.~F.}\ \bibnamefont {{Hopkins}}}, \bibinfo {author} {\bibfnamefont {P.}~\bibnamefont {{Kempski}}}, \bibinfo {author} {\bibfnamefont {S.~B.}\ \bibnamefont {{Ponnada}}}, \bibinfo {author} {\bibfnamefont {E.}~\bibnamefont {{Quataert}}},\ and\ \bibinfo {author} {\bibfnamefont {J.}~\bibnamefont {{Squire}}},\ }\bibfield  {title} {\bibinfo {title} {{Galactic cosmic-ray scattering due to intermittent structures}},\ }\href {https://doi.org/10.1093/mnras/stae276} {\bibfield  {journal} {\bibinfo  {journal} {Monthly Notices of the Royal Astronomical Society}\ }\textbf {\bibinfo {volume} {528}},\ \bibinfo {pages} {4245} (\bibinfo {year} {2024})},\ \Eprint {https://arxiv.org/abs/2308.06316} {arXiv:2308.06316 [astro-ph.HE]} \BibitemShut {NoStop}%
\bibitem [{\citenamefont {{Fitz Axen}}\ \emph {et~al.}(2024)\citenamefont {{Fitz Axen}}, \citenamefont {{Offner}}, \citenamefont {{Hopkins}}, \citenamefont {{Krumholz}},\ and\ \citenamefont {{Grudi{\'c}}}}]{fitzaxen:2024.cr.transport.into.gmcs.suppressed.starforge}%
  \BibitemOpen
  \bibfield  {author} {\bibinfo {author} {\bibfnamefont {M.}~\bibnamefont {{Fitz Axen}}}, \bibinfo {author} {\bibfnamefont {S.}~\bibnamefont {{Offner}}}, \bibinfo {author} {\bibfnamefont {P.~F.}\ \bibnamefont {{Hopkins}}}, \bibinfo {author} {\bibfnamefont {M.~R.}\ \bibnamefont {{Krumholz}}},\ and\ \bibinfo {author} {\bibfnamefont {M.~Y.}\ \bibnamefont {{Grudi{\'c}}}},\ }\bibfield  {title} {\bibinfo {title} {{Suppressed Cosmic-Ray Energy Densities in Molecular Clouds from Streaming Instability-regulated Transport}},\ }\href {https://doi.org/10.3847/1538-4357/ad675a} {\bibfield  {journal} {\bibinfo  {journal} {The Astrophysical Journal}\ }\textbf {\bibinfo {volume} {973}},\ \bibinfo {eid} {16} (\bibinfo {year} {2024})},\ \Eprint {https://arxiv.org/abs/2407.17597} {arXiv:2407.17597 [astro-ph.GA]} \BibitemShut {NoStop}%
\bibitem [{\citenamefont {{Barreto-Mota}}\ \emph {et~al.}(2024)\citenamefont {{Barreto-Mota}}, \citenamefont {{de Gouveia Dal Pino}}, \citenamefont {{Xu}},\ and\ \citenamefont {{Lazarian}}}]{barretomota:2024.mirror.scattering.ism.crs}%
  \BibitemOpen
  \bibfield  {author} {\bibinfo {author} {\bibfnamefont {L.}~\bibnamefont {{Barreto-Mota}}}, \bibinfo {author} {\bibfnamefont {E.~M.}\ \bibnamefont {{de Gouveia Dal Pino}}}, \bibinfo {author} {\bibfnamefont {S.}~\bibnamefont {{Xu}}},\ and\ \bibinfo {author} {\bibfnamefont {A.}~\bibnamefont {{Lazarian}}},\ }\bibfield  {title} {\bibinfo {title} {{Cosmic Ray Diffusion in the Turbulent Interstellar Medium: Effects of Mirror Diffusion and Pitch Angle Scattering}},\ }\href {https://doi.org/10.48550/arXiv.2405.12146} {\bibfield  {journal} {\bibinfo  {journal} {arXiv e-prints}\ ,\ \bibinfo {eid} {arXiv:2405.12146}} (\bibinfo {year} {2024})},\ \Eprint {https://arxiv.org/abs/2405.12146} {arXiv:2405.12146 [astro-ph.HE]} \BibitemShut {NoStop}%
\bibitem [{\citenamefont {Wellons}\ \emph {et~al.}(2023)\citenamefont {Wellons}, \citenamefont {Faucher-Giguère}, \citenamefont {Hopkins}, \citenamefont {Quataert}, \citenamefont {Anglés-Alcázar}, \citenamefont {Feldmann}, \citenamefont {Hayward}, \citenamefont {Kereš}, \citenamefont {Su},\ and\ \citenamefont {Wetzel}}]{Wellons_2023}%
  \BibitemOpen
  \bibfield  {author} {\bibinfo {author} {\bibfnamefont {S.}~\bibnamefont {Wellons}}, \bibinfo {author} {\bibfnamefont {C.-A.}\ \bibnamefont {Faucher-Giguère}}, \bibinfo {author} {\bibfnamefont {P.~F.}\ \bibnamefont {Hopkins}}, \bibinfo {author} {\bibfnamefont {E.}~\bibnamefont {Quataert}}, \bibinfo {author} {\bibfnamefont {D.}~\bibnamefont {Anglés-Alcázar}}, \bibinfo {author} {\bibfnamefont {R.}~\bibnamefont {Feldmann}}, \bibinfo {author} {\bibfnamefont {C.~C.}\ \bibnamefont {Hayward}}, \bibinfo {author} {\bibfnamefont {D.}~\bibnamefont {Kereš}}, \bibinfo {author} {\bibfnamefont {K.-Y.}\ \bibnamefont {Su}},\ and\ \bibinfo {author} {\bibfnamefont {A.}~\bibnamefont {Wetzel}},\ }\bibfield  {title} {\bibinfo {title} {Exploring supermassive black hole physics and galaxy quenching across halo mass in fire cosmological zoom simulations},\ }\href {https://doi.org/10.1093/mnras/stad511} {\bibfield  {journal} {\bibinfo  {journal} {Monthly Notices of the Royal Astronomical Society}\ }\textbf {\bibinfo {volume}
  {520}},\ \bibinfo {pages} {5394–5412} (\bibinfo {year} {2023})}\BibitemShut {NoStop}%
\bibitem [{\citenamefont {Sivasankaran}\ \emph {et~al.}(2024)\citenamefont {Sivasankaran}, \citenamefont {Blecha}, \citenamefont {Torrey}, \citenamefont {Kelley}, \citenamefont {Bhowmick}, \citenamefont {Vogelsberger}, \citenamefont {Hernquist}, \citenamefont {Marinacci},\ and\ \citenamefont {Sales}}]{sivasankaran2024agnfeedbackisolatedgalaxies}%
  \BibitemOpen
  \bibfield  {author} {\bibinfo {author} {\bibfnamefont {A.}~\bibnamefont {Sivasankaran}}, \bibinfo {author} {\bibfnamefont {L.}~\bibnamefont {Blecha}}, \bibinfo {author} {\bibfnamefont {P.}~\bibnamefont {Torrey}}, \bibinfo {author} {\bibfnamefont {L.~Z.}\ \bibnamefont {Kelley}}, \bibinfo {author} {\bibfnamefont {A.}~\bibnamefont {Bhowmick}}, \bibinfo {author} {\bibfnamefont {M.}~\bibnamefont {Vogelsberger}}, \bibinfo {author} {\bibfnamefont {L.}~\bibnamefont {Hernquist}}, \bibinfo {author} {\bibfnamefont {F.}~\bibnamefont {Marinacci}},\ and\ \bibinfo {author} {\bibfnamefont {L.~V.}\ \bibnamefont {Sales}},\ }\href {https://arxiv.org/abs/2402.15240} {\bibinfo {title} {Agn feedback in isolated galaxies with a smuggle multiphase ism}} (\bibinfo {year} {2024}),\ \Eprint {https://arxiv.org/abs/2402.15240} {arXiv:2402.15240 [astro-ph.GA]} \BibitemShut {NoStop}%
\bibitem [{\citenamefont {Ponnada}(2025)}]{ponnada2025timedependentcosmicrayhalos}%
  \BibitemOpen
  \bibfield  {author} {\bibinfo {author} {\bibfnamefont {S.~B.}\ \bibnamefont {Ponnada}},\ }\href {https://arxiv.org/abs/2509.02697} {\bibinfo {title} {Time-dependent cosmic ray halos from bursty star formation and active galactic nuclei: Semi-analytic formalism and galaxy formation implications}} (\bibinfo {year} {2025}),\ \Eprint {https://arxiv.org/abs/2509.02697} {arXiv:2509.02697 [astro-ph.GA]} \BibitemShut {NoStop}%
\bibitem [{\citenamefont {Su}\ \emph {et~al.}(2025)\citenamefont {Su}, \citenamefont {Bryan}, \citenamefont {Hopkins}, \citenamefont {Natarajan}, \citenamefont {Ponnada}, \citenamefont {Emami},\ and\ \citenamefont {Lu}}]{su2025modelingcosmicraysagn}%
  \BibitemOpen
  \bibfield  {author} {\bibinfo {author} {\bibfnamefont {K.-Y.}\ \bibnamefont {Su}}, \bibinfo {author} {\bibfnamefont {G.~L.}\ \bibnamefont {Bryan}}, \bibinfo {author} {\bibfnamefont {P.~F.}\ \bibnamefont {Hopkins}}, \bibinfo {author} {\bibfnamefont {P.}~\bibnamefont {Natarajan}}, \bibinfo {author} {\bibfnamefont {S.~B.}\ \bibnamefont {Ponnada}}, \bibinfo {author} {\bibfnamefont {R.}~\bibnamefont {Emami}},\ and\ \bibinfo {author} {\bibfnamefont {Y.~S.}\ \bibnamefont {Lu}},\ }\href {https://arxiv.org/abs/2502.00927} {\bibinfo {title} {Modeling cosmic rays at agn jet-driven shock fronts}} (\bibinfo {year} {2025}),\ \Eprint {https://arxiv.org/abs/2502.00927} {arXiv:2502.00927 [astro-ph.GA]} \BibitemShut {NoStop}%
\end{thebibliography}%

\end{document}